# Nonlinear light-matter interaction at terahertz frequencies

**Daniele Nicoletti,** [1,*] **Andrea Cavalleri** [1,2]

[1] *Max Planck Institute for the Structure and Dynamics of Matter, 22761 Hamburg, Germany*
[2] *Department of Physics, Oxford University, Clarendon Laboratory, Oxford OX1 3PU, UK*
[*] *Corresponding author: daniele.nicoletti@mpsd.mpg.de*

Strong optical pulses at mid-infrared and terahertz frequencies have recently emerged as a powerful tool to manipulate and control the solid state and especially complex condensed matter systems with strongly correlated electrons. The recent developments in high-power sources in the 0.1-30 THz frequency range, both from table-top laser systems and Free-Electron Lasers, has provided access to excitations of molecules and solids, which can be stimulated at their resonance frequencies. Amongst these, we discuss free electrons in metals, superconducting gaps and Josephson plasmons in layered superconductors, vibrational modes of the crystal lattice (phonons), as well as magnetic excitations. This Review provides an overview and illustrative examples of how intense THz transients can be used to resonantly control matter, with particular focus on strongly correlated electron systems and high-temperature superconductors.

## 1. Introduction

In the past few decades intense electromagnetic radiation has been widely employed to control the behavior of matter in its different phases. Ultrashort laser pulses at near-infrared and visible photon energies (0.5 – 3 eV, i.e. $10^2$ – $10^3$ THz frequency) are nowadays routinely used to perturb (or prepare) samples on sub-picosecond time scales. In most experiments performed so far on condensed matter systems, such pump pulses create a highly non-thermal electron distribution, which primarily relaxes by electron-electron scattering [1,2], followed by thermalization through coupling with other degrees of freedom [3]. In addition, coherence effects, like excitation of coherent phonons or magnons [4,5,6,7,8], can also be observed due to the impulsive nature of the initial photoexcitation [9].

In the particular case of strongly correlated materials, the delicate balance between different degrees of freedom brings such systems on the edge of stability between different phases, and to very large susceptibilities to arbitrarily small external perturbations [10]. In this context, the role played by optical stimulation is far more effective, leading to observation of a broad variety of dynamical phenomena. Among these, it is worth mentioning photo-induced insulator-to-metal transitions achieved by “photo-doping” Mott insulators [11,12,13,14], optical melting of magnetic order [15] or excitation of coherent orbital waves [16].

However, stimulation with high energy (~eV) photons can lead to strong limitations in terms of control capability. These constraints originate from the fact that above gap charge excitation typically results in highly incoherent dynamics, with a huge transfer of entropy in the system. Such photon energies are indeed at least one order of magnitude higher than all energy scales of relevant low-lying excitations in correlated materials, as electron plasmas in metals, superconducting gaps, vibrational modes of the crystal lattice, or magnetic excitations [10].

Only the most recent developments in generation of intense laser pulses of sub-picosecond duration at mid-infrared (MIR) and terahertz (THz) frequencies [17] have made it possible to achieve resonant excitation of these modes with strong transient fields, thus driving them into the nonlinear regime and accessing coherent dynamics, with no relevant increase in entropy.

In Fig. 1 we report a schematic classification of fundamental excitations in complex solids along with the corresponding energy scales. While infrared-active phonons and local vibrational modes of molecular solids can be typically reached with MIR pulses (5-20 μm wavelength, 50-250 meV photon energy, 15-50 THz frequency), all collective modes from broken symmetry states lie in the single-THz range (~300 μm wavelength, ~few meV photon energy).

In this Review Article we report the state of the art of light control experiments on strongly correlated solids performed by stimulation with intense THz field transients. A wide spectrum of striking dynamical phenomena has been observed in the recent years, including highly nonlinear lattice dynamics, magnetism control, modulation of electron correlations, insulator-to-metal transitions, as well as control and enhancement of superconductivity.

In Section 2 we introduce the concept of mode-selective control of the crystal lattice upon MIR excitation. We discuss how, via the so-called "nonlinear phononics" mechanism, anharmonic coupling to other (Raman-active) modes can be achieved. This corresponds to a net displacement of the crystal, which can drive electronic and magnetic phase transitions in many complex oxides [18,19].

The implication that resonant lattice excitation has on the control of superconductivity in high-$T_c$ copper oxides is discussed in Section 3. We report here how stimulation of Cu-O stretching modes in single-layer cuprates can lead to transient melting of charge and spin order and enhancement of superconductivity [20,21]. We also discuss experiments performed on the bilayer cuprate $YBa_2Cu_3O_x$, which showed that nonlinear lattice excitation can drive the system into a transient structure with superconducting-like optical properties up to room temperature [22,23].

In Section 4 we review the most recent results achieved by stimulating local vibrational modes in molecular solids, also located in the MIR spectral range. Resonant excitation of these modes allowed to access direct modulation of local electron correlations in a one-dimensional Mott insulator [24,25], as well as control of superconductivity in an alkali-doped fulleride [26].

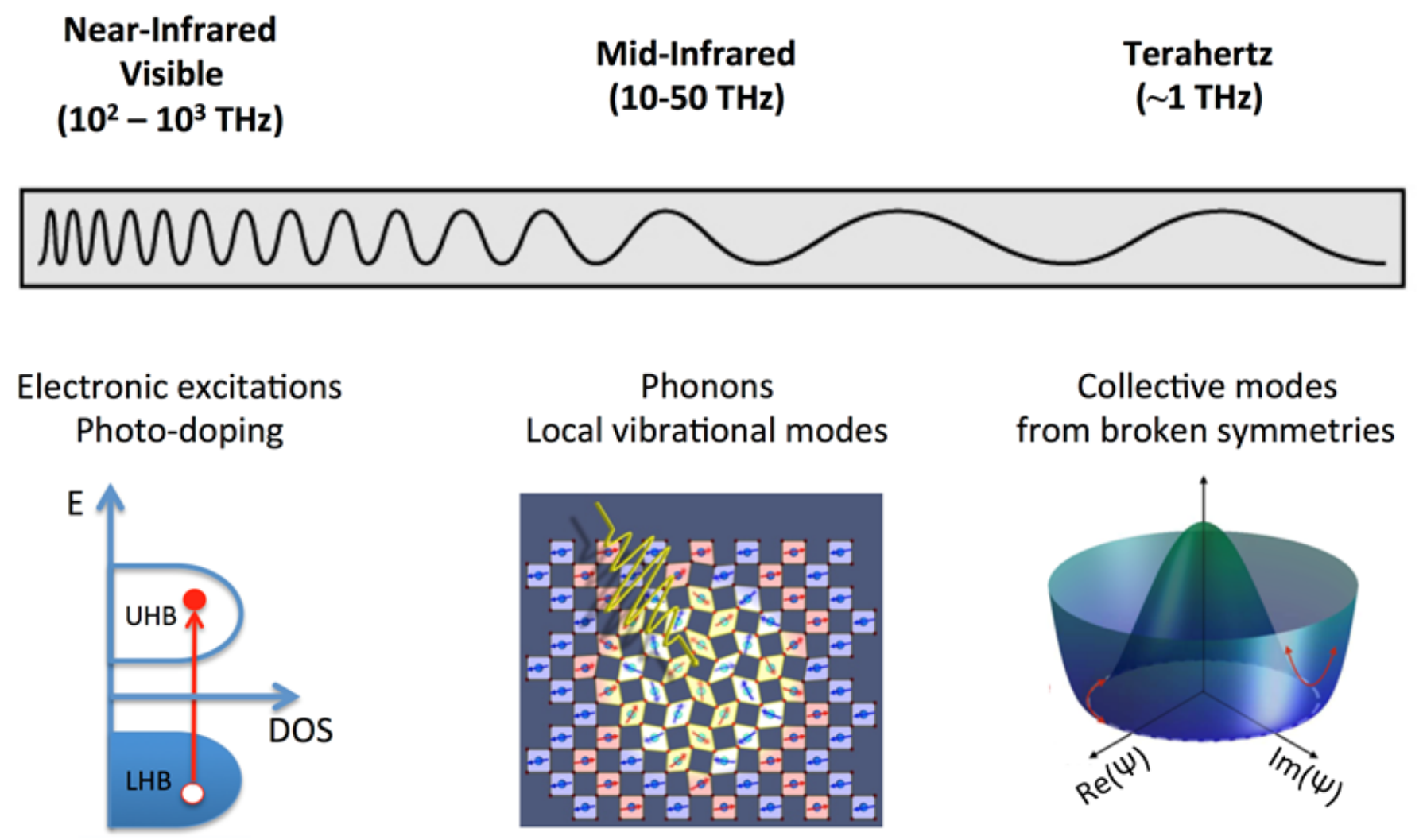

Fig. 1. Energy scales of the main fundamental excitations of solids in the visible and infrared. While stimulation in the near-infrared and visible range ($10^2$-$10^3$ THz) acts mainly on the electronic degrees of freedom, excitation of phonons and collective modes of broken symmetry states is achieved at MIR and THz frequencies.

In Section 5 we report experiments performed at longer excitation wavelengths, in the single- and sub-THz range. We summarize relevant results regarding the control of magnetism [27] and Cooper pairs in superconductors [28,29], along with examples of non-resonant, strong field control.

Finally, in Section 6 we focus on the most recent highlights in the field of the so-called "Josephson plasmonics". These were achieved by stimulating high-$T_c$ cuprates with strong THz pulses polarized in the direction perpendicular to the superconducting planes, which directly drove the interlayer Josephson current [30,31,32].

## 2. Selective control of the crystal lattice at mid-infrared wavelengths

Optical fields at mid-infrared frequencies can be used to excite the lattice of complex materials on a sub-picosecond timescale. New transient crystal structures, not stable at equilibrium, can be induced, thus introducing new effective Hamiltonians and driving the emergence of novel low-energy collective properties and functionalities. Here, we first analyze how anharmonic coupling between different modes of the lattice can transform the oscillatory motion of a resonantly driven mode to an average displacement of the crystal structure. We then discuss highlight experiments, in which insulators are turned into metals by selectively changing a specific bond angle. These effects are accompanied by changes in charge, orbital and magnetic order.

### 2.1. Nonlinear phononics

We introduce here the concept of nonlinear control of the crystal lattice. Nowadays, thanks to the most recent developments in nonlinear optics, strong transient fields in excess of $\sim$100 kV/cm can be generated at MIR and THz wavelengths [17]. These can be employed to excite collective modes in solids, like lattice phonons, provided that these modes are at small momentum and infrared (IR) active (*i.e.*, a change in their normal-mode coordinate implies a change in dipole moment). Lattice vibrations can be selectively driven to amplitudes as high as several percent of unit cell interatomic distances, thus giving rise to dynamical states which may have radically different average properties than those at equilibrium.

The linear response of a crystal lattice to a light field made resonant with an IR-active phonon mode is described by the potential energy term $H_{lin} = \frac{1}{2}\omega_{IR}^2 Q_{IR}^2$. In this expression, $Q_{IR}$ is the normal mode coordinate and $\omega_{IR}$ its eigenfrequency. When resonantly driven by a pulsed field $G(t)\sin(\omega_{IR}t)$ (being $G(t)$ a Gaussian envelope), the dynamics can be described by the equation of motion of a damped harmonic oscillator

$$\ddot{Q}_{IR} + 2\gamma\dot{Q}_{IR} + \omega_{IR}^2 Q_{IR} = G(t)\sin(\omega_{IR}t),$$

where $\gamma$ is the damping constant. After stimulation, the atoms start to oscillate about their equilibrium positions along the normal mode coordinate and then relax over a timescale set by either the envelope duration or by the decay time $1/\gamma$.

Upon increase of the driving electric field, anharmonic coupling to other modes with generic coordinate $Q_R$ becomes relevant. To lowest order, the lattice Hamiltonian describing the nonlinear interaction reads

$$H = \frac{1}{2}\omega_R^2 Q_R^2 - a_{12}Q_{IR}Q_R^2 - a_{21}Q_{IR}^2 Q_R,$$

where $a_{ij}$ are anharmonic coupling constants. In the case of centrosymmetric crystals, for which IR-active phonons have odd parity while Raman-active modes are even, the first coupling term $a_{12}Q_{IR}Q_R^2$ is always zero (for symmetry reasons) and the second term $a_{21}Q_{IR}^2 Q_R$ is nonzero only if $Q_R$ is a Raman mode. For a finite static displacement $Q_{IR}$, a shift of the Raman mode's energy potential $V_R$ along its coordinate $Q_R$ is induced, as depicted in Fig. 2(a). This phenomenon was analyzed theoretically in the 1970s within the framework of Ionic Raman scattering [33,34,35].

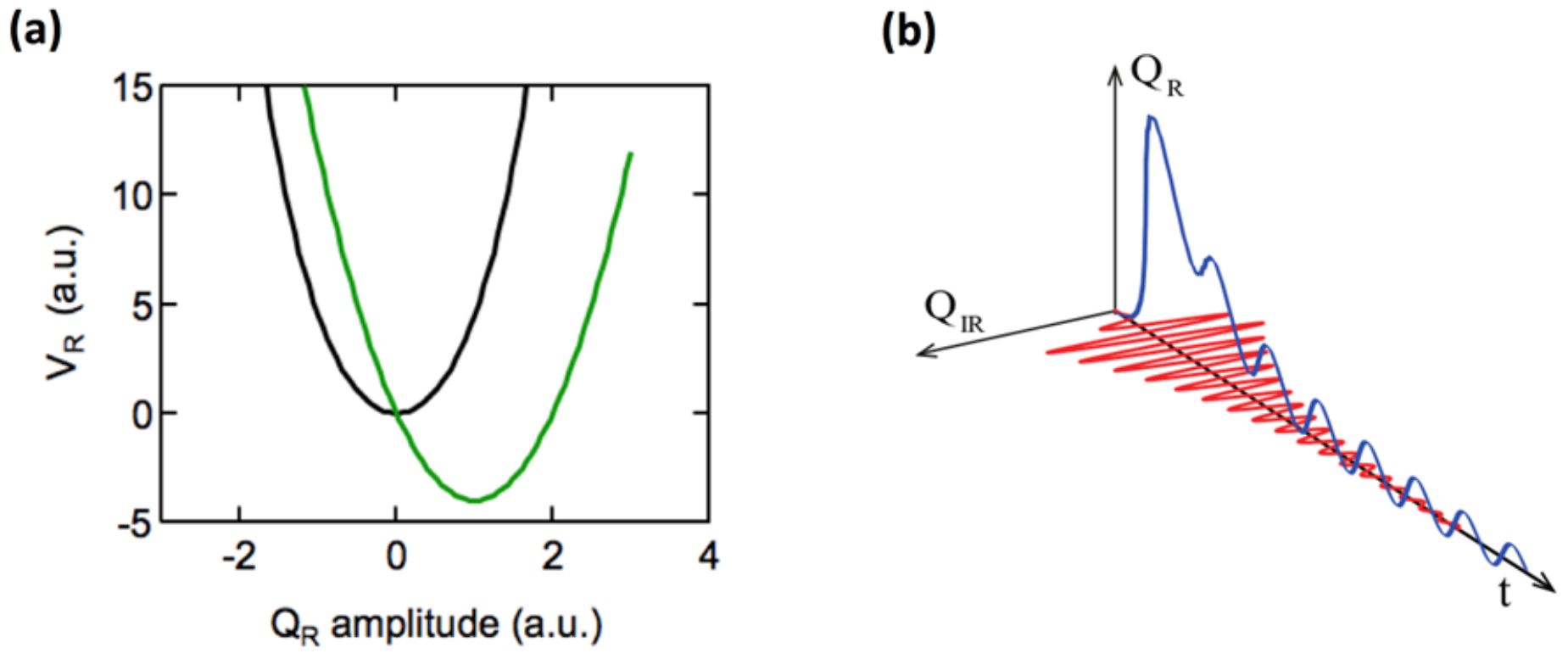

Fig. 2. (a) Parabolic energy potential $V_R$ of a Raman-active phonon mode (black curve). Within cubic coupling, this energy potential shifts towards a new position (green curve) for a finite static displacement of a coupled IR-active mode $Q_{IR}$. (b) Dynamical response of the two coupled modes. The ions oscillate coherently along the IR-active mode coordinate $Q_{IR}$ (red) and undergo simultaneously a directional displacement along $Q_R$, which scales as $Q_{IR}^2$. If optical excitation is fast compared to the Raman phonon period, coherent oscillations along $Q_R$ take place. Adapted from [18,19].

The coupled dynamics following excitation of the IR-active mode are described by the two equations of motion

$$\ddot{Q}_{IR} + 2\gamma_{IR}\dot{Q}_{IR} + \omega_{IR}^2 Q_{IR} = 2a_{21}Q_{IR}Q_R + G(t)\sin(\omega_{IR}t),$$

$$\ddot{Q}_R + 2\gamma_R\dot{Q}_R + \omega_R^2 Q_R = a_{21}Q_{IR}^2.$$

The anharmonically coupled Raman mode $Q_R$ undergoes the driving force $a_{21}Q_{IR}^2$, with its direction being independent of the sign of $Q_{IR}$. Hence, the atoms of the crystal lattice not only oscillate along the IR coordinate $Q_{IR}$, but are simultaneously displaced along the Raman coordinate $Q_R$, as schematically depicted in Fig. 2(b). This effect is equivalent to rectification in nonlinear optics. In addition, when the optical excitation is short compared to the Raman mode period, $Q_R$ exhibits coherent oscillations (blue line in Fig. 2(b)). These are also characteristic of the well-known coherent response of crystal lattices to pulsed excitation in the near-IR or visible [36,37,38]. However, in that case the lattice displacement and oscillations are driven by electron-phonon coupling and not, as it is the case of nonlinear phononics, by lattice anharmonicities. Note also that the excitation through nonlinear phononics is more selective and dissipation is highly suppressed thanks to the low photon energy, which reduces the number of accessible relaxation channels.

### 2.2. First experimental observation of nonlinear phonon coupling

Nonlinear phononics has been firstly observed in the rhombohedrally-distorted perovskite $La_{0.7}Sr_{0.3}MnO_3$, a double-exchange ferromagnet with low electrical conductivity below $T_c \sim 350$ K. An IR-active phonon mode of $E_u$ symmetry [39], which involves a Mn-O stretching motion, was resonantly excited to large amplitudes with MIR pump pulses tuned to ~18 THz (75 meV photon energy, 14 µm wavelength).

The time-dependent anisotropic reflectivity changes induced by this phonon excitation were measured at near-infrared (800 nm) wavelengths [40] and are displayed in Fig. 3(a). Strikingly, the ultrafast response exhibits oscillations at 1.2 THz. Because of their frequency and anisotropic character, these can be attributed to coherent oscillations of a Raman mode of $E_g$ symmetry, which corresponds to a rotation of the oxygen octahedra around the Mn ions [41,42]. Furthermore, as shown in Fig. 3(b), the pump photon energy dependent amplitude of the coherent $E_g$ Raman mode oscillations perfectly followed the resonance profile of the IR-active Mn-O stretching mode.

While the above experiment allowed a clear identification of the frequency and symmetry of the coherently driven Raman mode, the lattice displacement predicted by the nonlinear phononics theory could not be directly proven by this all-optical experiment.

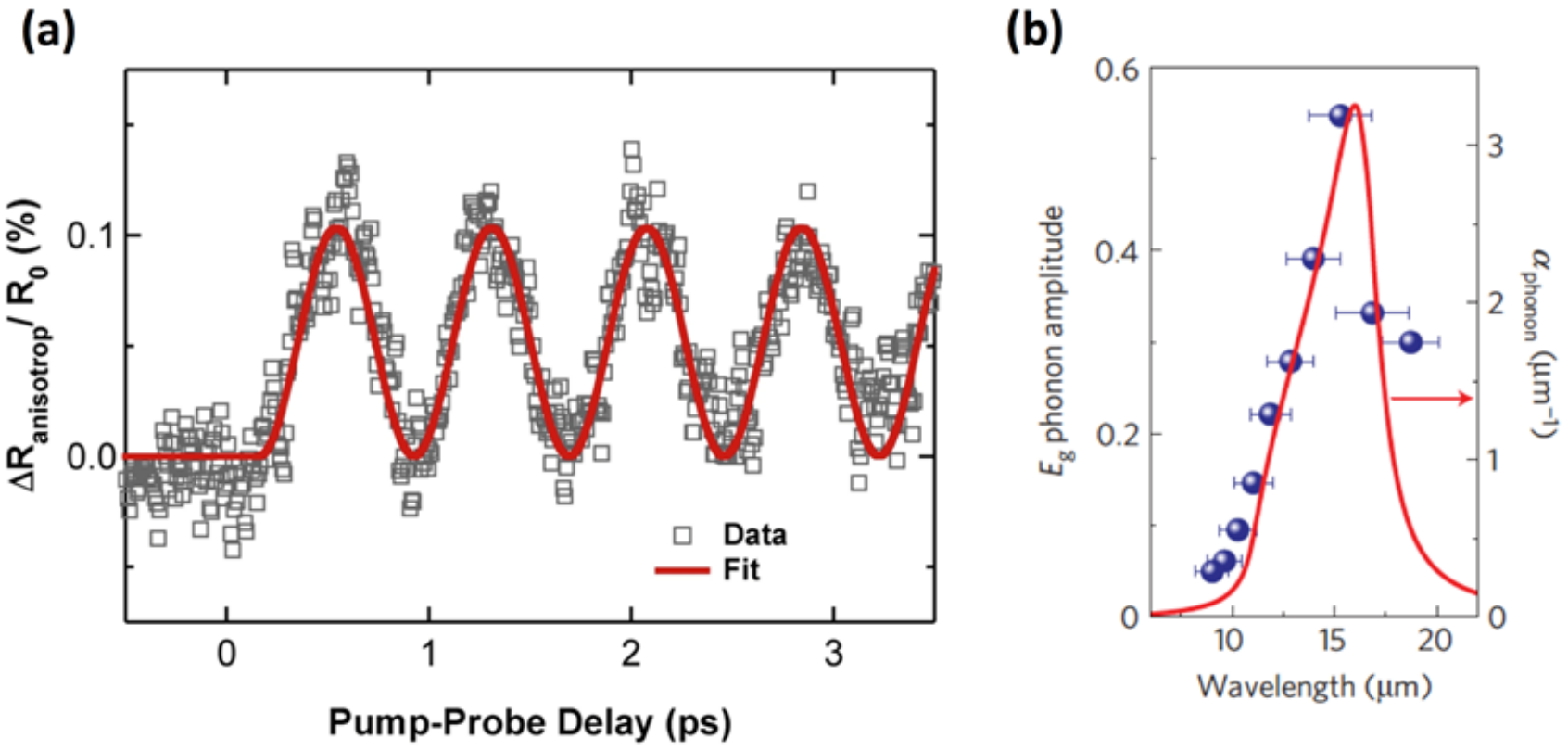

Fig. 3. (a) Time-resolved anisotropic reflectivity changes induced in $La_{0.7}Sr_{0.3}MnO_3$ by resonant excitation of the IR-active $E_u$ stretching mode at 14 μm. (b) Amplitude of the coherent $E_g$ Raman oscillation (blue circles) and linear absorption coefficient (red line) of $La_{0.7}Sr_{0.3}MnO_3$ as function of pump wavelength. Panel (a) adapted from [43]. Panel (b) reprinted from [40].

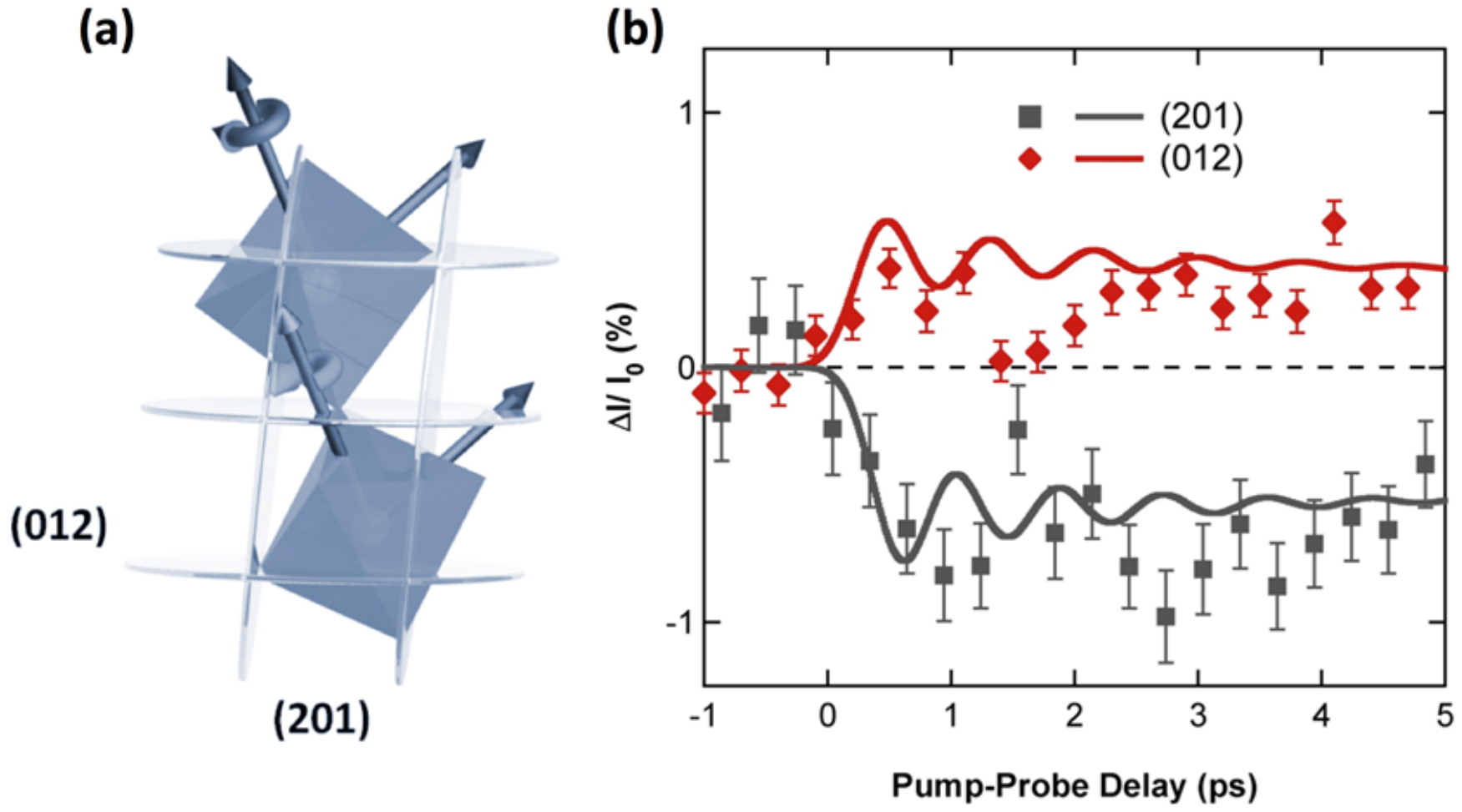

Fig. 4. (a) Rotational motions of the $MnO_6$ octahedra associated with the $E_g$ Raman mode. Diffraction planes are also displayed. (b) Relative x-ray intensity changes of the (201) and the (012) Bragg peaks in $La_{0.7}Sr_{0.3}MnO_3$ induced by MIR excitation. Solid lines are fits based on the coupled equations of motions for nonlinear phononics and the $E_g$ phonon induced structure factor changes. The anharmonic coupling constant is the only free fitting parameter. Adapted from [43].

The transient modifications of the crystal structure of $La_{0.7}Sr_{0.3}MnO_3$ after the same MIR excitation were then studied with femtosecond hard x-ray diffraction [43]. The atomic motion along phonon coordinates modulates the x-ray structure factor, which determines the intensity of specific Bragg peaks [44,45]. Motion along the $E_g$ Raman mode coordinates involves anti-phase rotations of the two $MnO_6$ octahedra of the $La_{0.7}Sr_{0.3}MnO_3$ unit cell [41,42], as illustrated in Fig. 4(a).

Figure 4(b) summarizes the results of this femtosecond x-ray diffraction experiment carried out at the LCLS Free-Electron Laser [46]. The transient intensity changes of the (201) and (012) structural Bragg peaks upon the vibrational excitation were measured with 70-fs x-ray pulses at 6 keV photon energy.

Two displacive responses were observed in the transient Bragg intensities, with different amplitudes and opposite signs for the two measured peaks. These curves were fitted by a numerical solution of the coupled equations of motion and taking also into account the calculated relative intensity changes ($\Delta I/I_0$) for a motion of the oxygen atoms with amplitude $Q_R$ along the coordinate of the rotational Raman mode. The anharmonic coupling constant $a_{21}$ was the only free fitting parameter. These fits revealed a rotation of the $MnO_6$ octahedra of ~4 mrad for the chosen excitation fluence of ~1 mJ/cm$^2$.

## 2.3. Vibrationally-induced phase transitions in manganites

The rotation of the $MnO_6$ octahedra, activated via IR-phonon excitation through nonlinear phononics, is expected to be strongly coupled to the collective electronic and magnetic properties of manganites, thus opening perspectives for phase control in these materials. Such rotation modifies indeed the Mn-O-Mn bond angles, thus affecting the spatial overlap of the orbital wavefunctions along these bonds and controlling the sign of the exchange interactions [47,48]. For example, it is known that the electron hopping between adjacent Mn sites is maximum for a bond angle of 180° (cubic lattice) and decreases with decreasing angle, resulting in maximum electron bandwidth for straight bonds.

The magnetic and structural properties of manganites can be understood in more detail by considering the Goodenough-Kanamori-Anderson rules for semi-covalent bonds [49], which explain the exchange interactions between neighboring Mn and O ions. In the case of doped manganites, both ferromagnetic metallic and antiferromagnetic insulating behaviors are possible. For asymmetric bonds of type $Mn^{3+}$ – $O^{2-}$ – $Mn^{4+}$, hopping occurs and the magnetic coupling is ferromagnetic as long as the bond is "straight", while a "bent" bond (with an angle $\ll$ 180°) corresponds to an insulating antiferromagnetic phase.

Due to its strongly distorted structure, $Pr_{0.7}Ca_{0.3}MnO_3$ is an insulator. However, its insulating character is unstable against various forms of external perturbation, including electric or magnetic fields, optical stimulation, as well as static external pressure [50,51,52,53,54,55].

The dynamical lattice control of the electronic phase of a manganite was first reported by Rini *et al.* [56], who measured the electric transport properties in a $Pr_{0.7}Ca_{0.3}MnO_3$ after resonant excitation of the Mn-O stretching mode with femtosecond pulses tuned to ~17 μm wavelength. The light-induced changes in electric transport are displayed in Fig. 5(a), where the transition from the insulating ground state into a metastable metallic state is evidenced by a five-order-of-magnitude increase in electrical conductivity within a few nanoseconds. This photo-induced phase transition can be uniquely attributed to large-amplitude crystal lattice distortions, as the effect completely disappears when the pump pulses are tuned away from the phonon resonance [56].

A microscopic theory for the vibrationally driven ultrafast phase control was recently developed [57]. Such theory, which predicts the dynamical path taken by the crystal lattice and its effect on the electronic material properties, was applied to compute the coupling strength of the resonantly driven Mn-O stretching mode to other phonon modes in $PrMnO_3$. In Fig. 5(b) the energy potential of a strongly coupled, $A_g$ symmetry Raman mode is displayed for various amplitudes of the directly-driven, IR-active Mn-O stretching mode ($B_{1u}$ symmetry). Finite displacement of the crystal lattice along the $B_{1u}$ coordinate shifts the parabolic potential of the Raman mode to a new minimum position, with this displacement varying quadratically with the driving amplitude, as expected for the cubic coupling $Q_{B1u}^2 Q_{Ag}$.

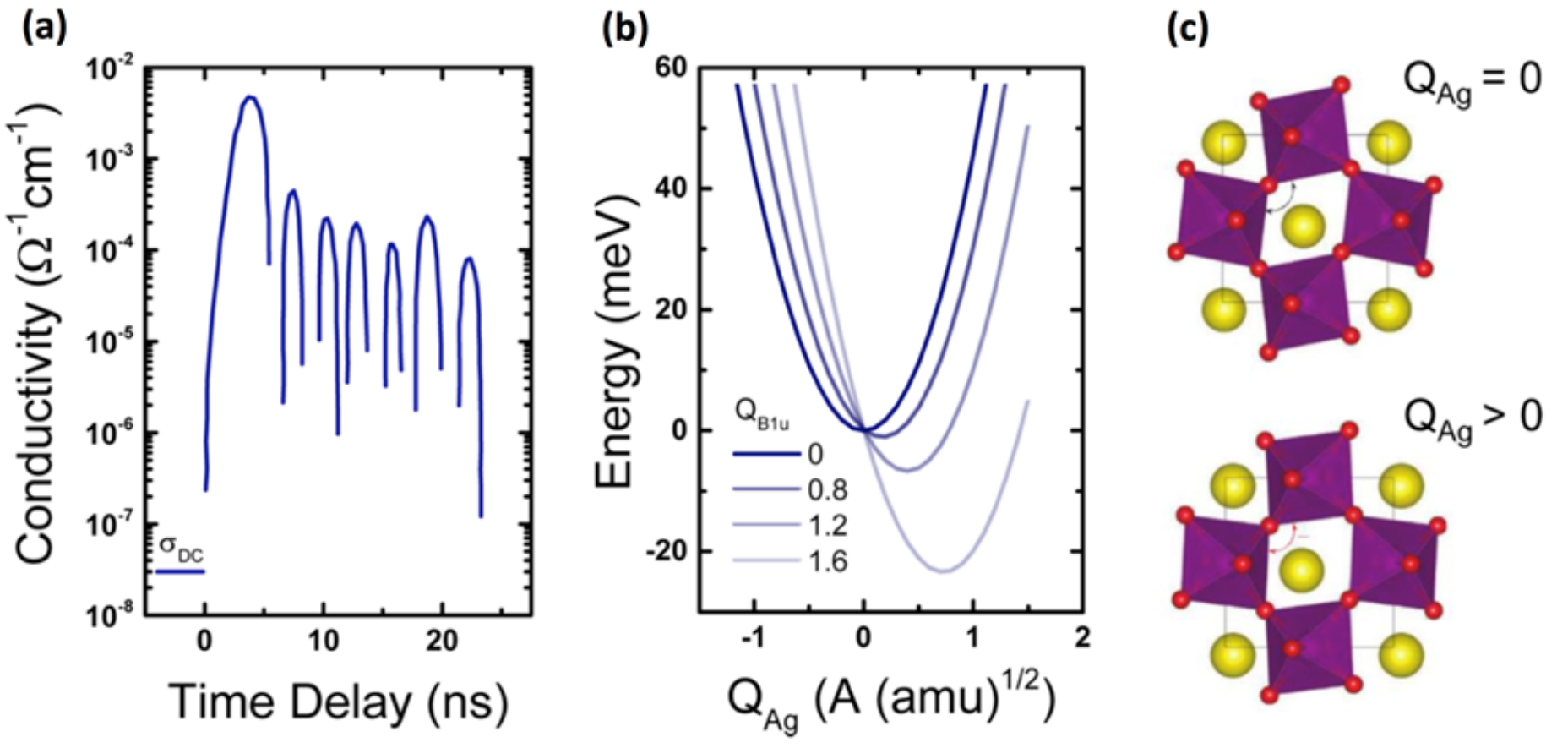

Fig. 5. (a) Time-dependent conductivity of $Pr_{0.7}Ca_{0.3}MnO_3$ following resonant excitation of the Mn-O stretching vibration. The time resolution of this experiment was 4 ns, and the ringing at later time delays resulted from the lack of impedance matching in the measurement. (b) Calculated energy potential of the $A_g$ rotational Raman mode in $PrMnO_3$ for different static displacements of the IR-active $B_{1u}$ mode. (c) Sketches of atomic positions of the distorted perovskite at equilibrium (top) and after excitation (bottom). Reprinted with permission from [18]. Copyright 2015 American Chemical Society.

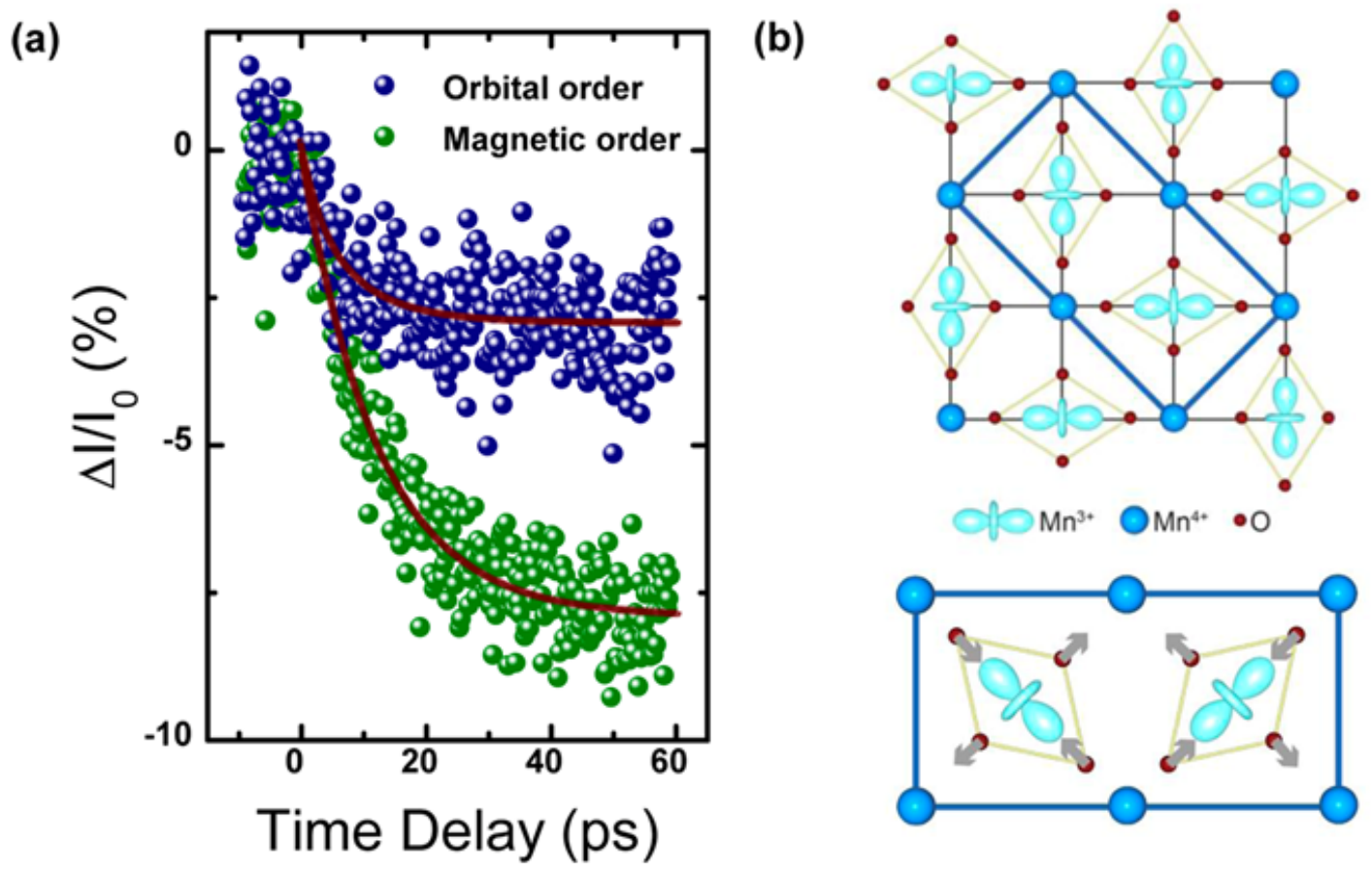

Fig. 6. (a) Lattice-driven melting of antiferromagnetic and charge/orbital order in the layered manganite $La_{0.5}Sr_{1.5}MnO_4$, measured via time-resolved resonant soft x-ray diffraction. (b) Sketch of the equilibrium in-plane charge and orbital order. The thick blue line (top) indicates the orbital ordering unit cell. Displacement of the oxygen atoms associated with the nonlinearly coupled Jahn–Teller mode. The sketched relaxation of the Jahn–Teller distortions is also displayed (bottom). Reprinted with permission from [18]. Copyright 2015 American Chemical Society.

As shown in Fig. 5(c), the atomic motion associated with the $A_g$ Raman displacement tends to reduce the rotation of the $MnO_6$ octahedra, hence bringing the crystal structure closer to the cubic perovskite and straightening the Mn-O-Mn bonds. In order to prove that this motion is responsible for driving the insulator-to-metal-transitions, the electronic density of states of the Mn *3d* orbitals was also calculated [57]. For the equilibrium crystal structure, these calculations yield a gap at the Fermi energy, characteristic of the insulating state. This gap is then shown to close in the transiently displaced state. These results support the hypothesis that a nonlinear coupling of the resonantly driven Mn-O stretching mode to the rotational Raman mode drives the observed insulator-to-metal transition in $Pr_{0.7}Ca_{0.3}MnO_3$ [56].

The ultrafast response of the orbital and magnetic order to vibrational excitation was also investigated by means of time-resolved resonant soft x-ray diffraction at the LCLS Free Electron Laser. In the related layered manganite $La_{0.5}Sr_{1.5}MnO_4$ [58,59], a reduction of both orders following resonant excitation of the IR-active in-plane Mn−O stretching mode was observed by measuring the transient intensity changes of selected superlattice reflections for photon energies made resonant with the Mn $L$-edge (see Fig. 6(a)) and thus sensitive to the arrangement of Mn 3$d$ electrons [60]. This response was understood by considering that, according to selection rules, the driven $B_{2u}$ mode can couple nonlinearly to a Raman-active $A_g$ mode, associated to a Jahn−Teller distortion (see Fig. 6(b)). At equilibrium, Jahn−Teller distortions lift the degeneracy of the 3$d$ orbitals to lower the energy of the system, thus stabilizing orbital ordering and also affecting the exchange interaction to give rise to magnetic order. The proposed scenario is that the applied stimulation causes a relaxation of these distortions, thus melting the ordering of orbitals and spins in the driven state.

One should note that the mechanism behind this lattice-driven insulator-to-metal transition in manganites is different from excitation at near-infrared or visible wavelengths. Indeed, in the latter case the phase transition is triggered by charge transfer between different orbitals or adjacent sites, subsequently releasing Jahn−Teller distortions and modifying the exchange interaction [15,61,62].

## 2.4. Vibrational control at heterostructure interfaces

As discussed in Section 2.3, many of the functional properties of $ABO_3$ perovskite oxides are extremely sensitive to rotation and tilting of the oxygen octahedra. Hence, designing or actively controlling such distortions paves the way for engineering the electronic and magnetic properties [63,64,65].

One class of materials where this approach has been successfully demonstrated is that of rare earth nickelates [66]. At equilibrium, these compounds display a sharp transition from a high-temperature metallic to a low-temperature insulating state [67,68], accompanied by an increased Ni-O-Ni bond bending and the appearance of a charge density wave. Additionally, charge disproportionation between adjacent Ni sites is associated with different Ni-O bond lengths [69]. At low temperatures, the nickelates also possess an unusual antiferromagnetic spin arrangement [70]. The transition temperature depends crucially on epitaxial strain, being therefore sensitive to lattice distortions [71].

A first MIR pump – optical and THz probe experiment [72] was performed on 33-nm thick (100 unit cells) $NdNiO_3$ epitaxial thin films deposited on (001)-oriented $LaAlO_3$ [73]. This substrate provided a compressive strain to the material and reduced the equilibrium metal-insulator transition temperature $T_{MI}$ from a bulk value of ~200 K to about 130 K.

In Fig. 7(a) time-resolved 800-nm reflectivity changes induced by 15-μm MIR excitation are displayed for different base temperatures. At room T, where $NdNiO_3$ is metallic, only a modest, short-lived increase in reflectivity was observed, likely associated to electronic excitations near the Fermi level. Below the metal-insulator transition temperature, however, a long-lived ~20% reduction in reflectivity was measured, indicative of the formation of a metastable electronic phase. The time-dependent long-lived increase in the low-frequency THz reflectivity (Fig. 7(b)) suggested that this transient phase was indeed metallic [74]. This interpretation was confirmed by frequency-resolving the THz response for the unperturbed heterostructure and for the photo-induced state, from which a five orders of magnitude increase in DC conductivity was extracted for the $NdNiO_3$ film [72].

This optical response was also analyzed as function of pump wavelength, as shown in Fig. 7(c). The effective photo-susceptibility was found to follow the absorption of a phonon mode of the $LaAlO_3$ substrate, rather than that of $NdNiO_3$. This all-optical experiment evidenced that the insulator-to-metal transition in the functional $NdNiO_3$ film is driven across the interface upon exciting the vibrational mode of the substrate.

Time-resolved resonant soft x-ray diffraction was then applied to measure the simultaneous magnetic response in the $NdNiO_3$ film with nanometer spatial and femtosecond temporal resolution [75]. 15-μm wavelength pulses were used to excite a ~30 nm thick compressively strained $NdNiO_3$ film deposited on a (111) $LaAlO_3$ crystal. The dynamics of the antiferromagnetic order on the Ni sublattice was measured by detecting x-ray pulses tuned to the Ni $L_3$ edge and diffracted at the corresponding pseudo-cubic (¼ ¼ ¼) wave vector [70].

In Fig. 8(a) we show the 80% drop in diffraction intensity, which develops within ~2 ps after vibrational excitation (blue dots). This time scale is significantly longer than the drop time following direct interband excitation in the same material [76], while it is very similar to the phonon-induced metallization time, as estimated by transient THz reflectivity (green dots). This observation indicates a tight connection between the insulator-to-metal transition and the melting of magnetic order.

In the same experiment, the spatial distribution of the melting front of magnetic order was also analyzed by studying the width of the diffraction peak and the amplitude of Laue oscillations throughout the vibrationally-induced dynamics. A model assuming a soliton-like demagnetization front propagating from the hetero-interface into the thin film could reproduce well the data.

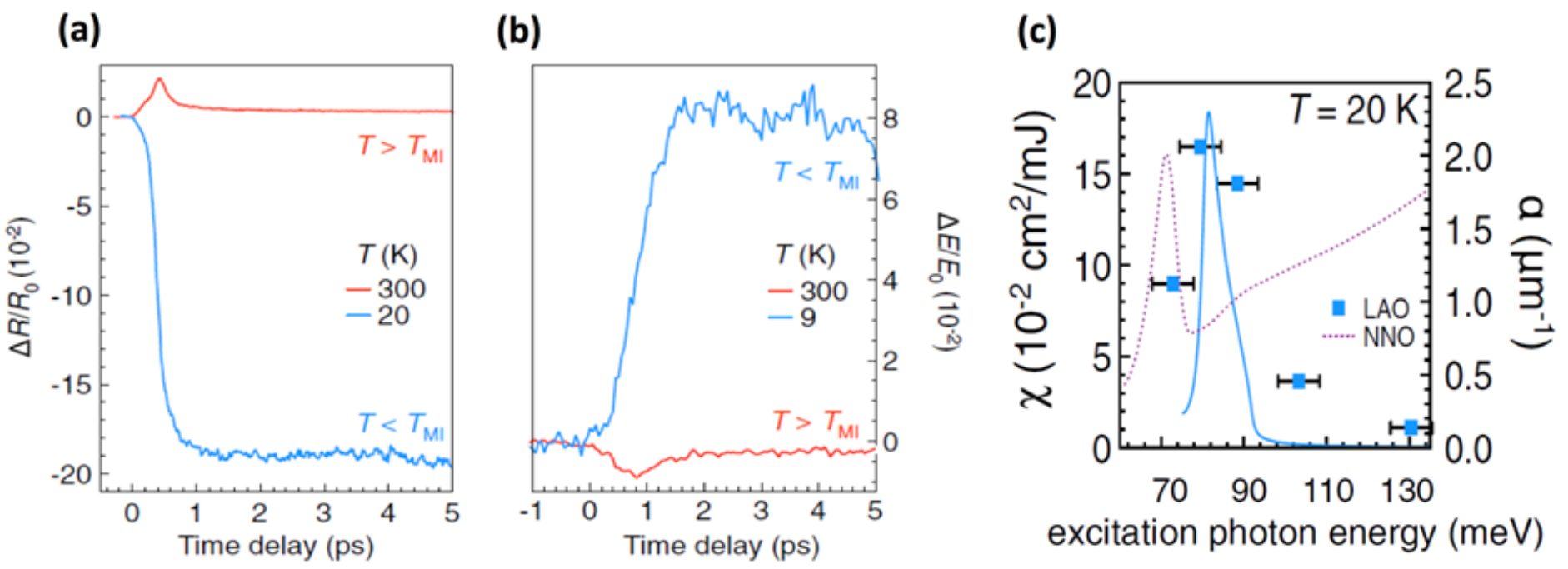

Fig. 7. (a) Transient reflectivity changes at 800 nm following vibrational excitation of the $NdNiO_3/LaAlO_3$ heterostructure. Data are shown at two temperatures, below and above the insulator-metal transition. (b) Corresponding changes in the low-frequency THz reflectivity. (c) Photo-susceptibility $\chi$ of the vibrationally driven insulator-to-metal transition as a function of pump wavelength. The solid blue line shows the linear extinction coefficient of the IR-active phonon of $LaAlO_3$ [77], while the dashed lines is the linear absorption of bulk $NdNiO_3$ [74]. Reprinted with permission from [72]. Copyright 2012 by the American Physical Society.

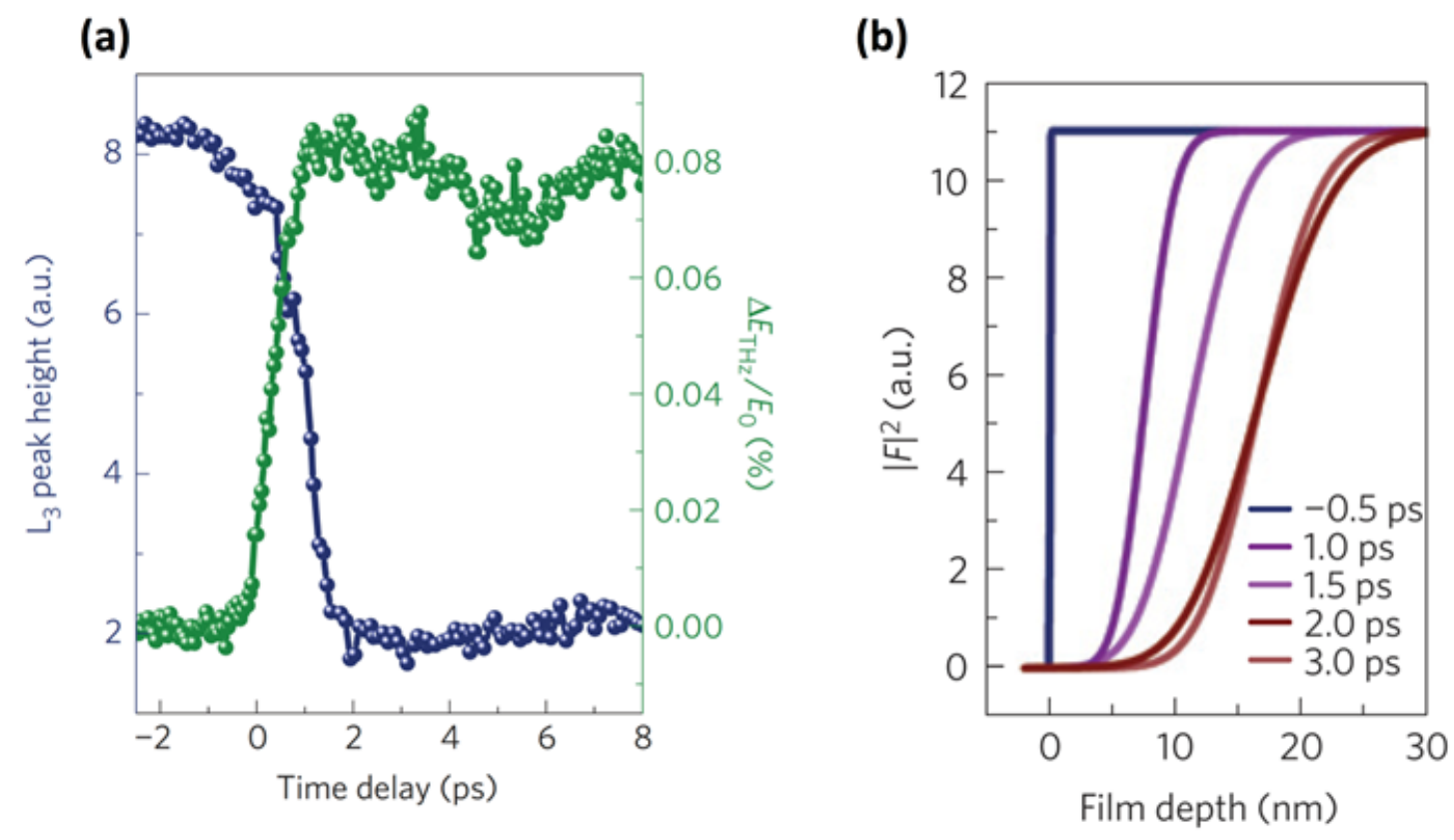

Fig. 8. (a) Intensity changes of the $NdNiO_3$ antiferromagnetic order related (¼ ¼ ¼) diffraction peak at the 852-eV Ni $L_3$ edge (blue dots) following MIR excitation. Green dots indicate the changes in THz reflectivity after the same excitation (see also Fig. 7(b)). (b) Spatiotemporal dynamics of the magnetic order along the sample growth direction, extracted from the numerical fits to the data. Reprinted from [75].

Fig. 8(b) shows the early time evolution of the space-dependent magnetic order, extracted by numerically fitting the diffraction peaks at various time delays using this model. At equilibrium, the $NdNiO_3$ film is homogenously ordered, while at positive time delays heterogeneous melting takes place, with a demagnetizing phase front that propagates halfway into the film and leaves a magnetically disordered region behind. Analysis of the time-dependent position of the phase front (see Ref. [75]) suggested that the magnetic melt front propagates at a speed comparable with the $NdNiO_3$ longitudinal sound velocity [78].

These experimental results can be explained by considering that the direct excitation of the IR-active substrate phonon induces octahedral distortions across the interface, which locally act on the electronic and magnetic order of the $NdNiO_3$ film. A similar scenario has indeed been predicted for static deformations in perovskite heterostructures [63], which were proposed to locally create itinerant charge carriers. These would then melt the antiferromagnetic order while diffusing into the nickelate film [79], thus explaining the observed stalling of the phase front. This scenario, which was also supported by a model Hamiltonian description [75], is compatible with the similar time scales observed for the insulator-to-metal transition and the magnetic order melting.

## 3. Phonon-mediated control of high-temperature superconductivity

In Section 2 we have introduced the concept of nonlinear lattice dynamics and shown how selective excitation of particular phonon modes at MIR wavelengths has been used to induce phase transitions and control charge, orbital, and magnetic order in different families of complex oxides. Here we turn our attention to high-$T_c$ cuprates and review the most recent results regarding the control of superconducting properties in these materials via nonlinear phonon excitation. In particular, we show how stimulation of the crystal lattice can serve to remove charge and spin orders in single-layer cuprates, thus enhancing their superconducting response on ultrafast timescales. We also review the most recent experiments performed on the bilayer cuprate $YBa_2Cu_3O_x$, where signatures of light-induced, transient superconductivity have been observed up to room temperature.

### 3.1. Melting of charge stripes and light-induced superconductivity in single-layer cuprates

High-$T_c$ superconductivity is achieved in single-layer cuprates when a few percent excess charges are injected in the $CuO_2$ planes of the antiferromagnetic Mott insulating parent compounds (*e.g.*, $La_2CuO_4$). These dopants are typically holes, which are introduced via chemical substitution (for example in $La_{2-x}(Ba/Sr)_xCuO_4$). While superconductivity is found for concentrations $x > 0.05$, reaching the maximum $T_c$ at $x \simeq 0.16$, the region of the phase diagram around $x = 1/8$ is also characterized by the presence of periodic one-dimensional modulation of charge and spin density (the so-called *stripes*) [80,81], which causes a reduction in $T_c$ [82,83]. In $La_{2-x}Ba_xCuO_4$ these stripes become static at 1/8-doping, thanks to a periodic buckling of the $CuO_2$ planes in a low-temperature tetragonal (LTT) phase [84,85,86].

Static stripes, LTT phases and suppressed superconductivity are also detected in other single-layer cuprates, such as $La_{1.8-x}Eu_{0.2}Sr_xCuO_4$ [87,88]. A typical crystal structure of these compounds is displayed in Fig. 9(a), depicting $CuO_2$ layers stacked along the *c* axis. The superconducting critical temperature of $La_{1.8-x}Eu_{0.2}Sr_xCuO_4$ is strongly reduced for all doping values $x < 1/8$ and completely suppressed at $x = 1/8$, as illustrated in the phase diagram plotted in Fig. 9(b) [89,90].

The coexistence and competition between superconductivity, stripe order, and structural distortion are a fundamental feature of the cuprate phase diagram, which is not fully understood yet. While stripes are typically pinned by the LTT distorted structure and compete with superconductivity, measurements performed under external pressure have revealed the presence of stripes without any LTT distortion [91,92], thus indicating that charge ordering alone may be driving the microscopic physics [93]. Recent experiments [94,95] suggest that the individual striped planes would be in fact be in a highly coherent paired state, a so-called *pair density wave*, in which the stripes (see Fig. 9(c)) modulate the superconducting order parameter in plane and frustrate the interlayer coherent transport.

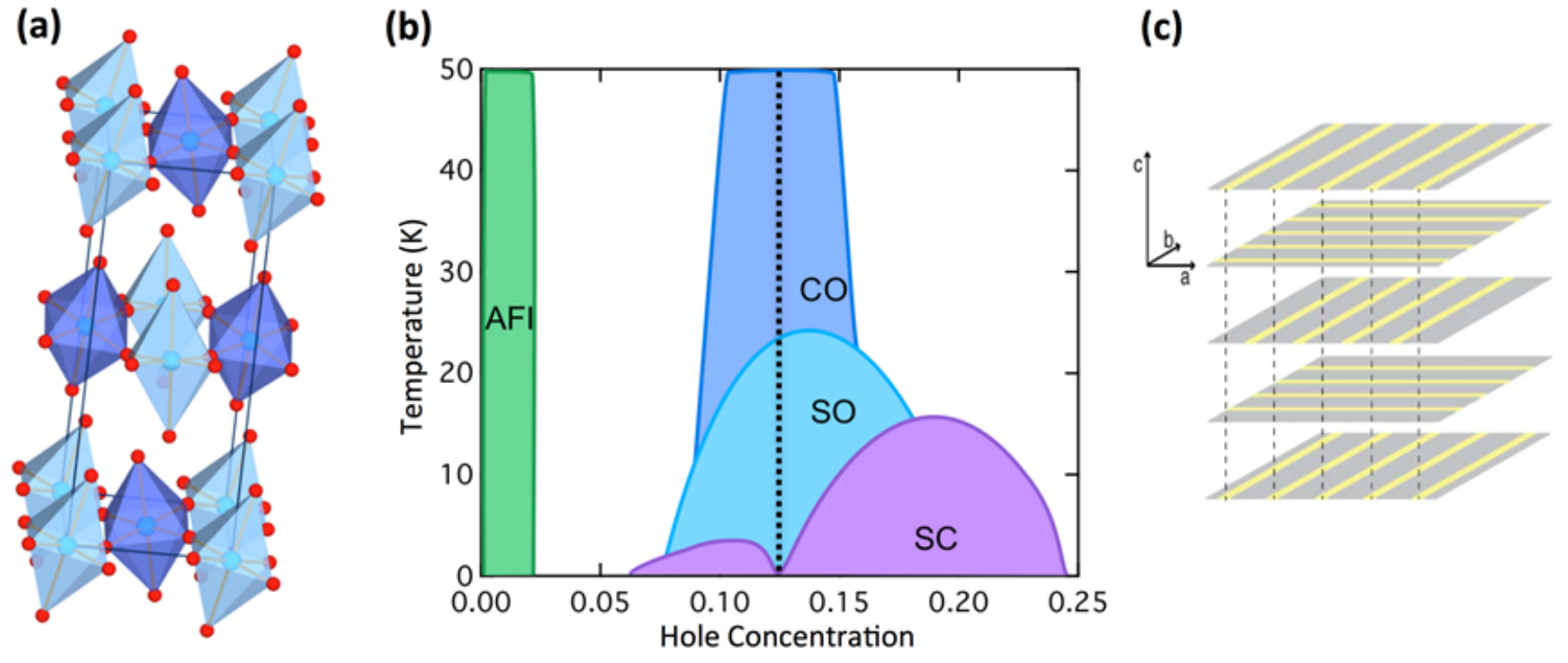

Fig. 9. (a) Schematic crystal structure and (b) phase diagram for $La_{1.8-x}Eu_{0.2}Sr_xCuO_4$. This compound is an antiferromagnetic insulator at $x$ = 0. A low-temperature structural distortion (LTT), associated with buckling of the $CuO_2$ planes, quenches superconductivity at all doping levels below $x$ = 1/8 (vertical dashed line). At this doping, a one-dimensional modulation of charges (CO) and spins (SO), the stripe state, emerges in the planes. At $x$ > 1/8, the compound is superconducting. (c) Periodic stacking of $CuO_2$ planes in the stripe phase. The stripe orientation rotates by 90° between layers, thus preventing interlayer superconducting phase coherence. Panel (b) adapted from [19]. Panel (c) reprinted with permission from [96]. Copyright 2014 by the American Physical Society.

The first vibrational control experiment on high-$T_c$ cuprates was performed on the 1/8-doped, non-superconducting $La_{1.675}Eu_{0.2}Sr_{0.125}CuO_4$ system, where the crystal lattice was dynamically perturbed by selectively driving an IR-active phonon mode with femtosecond MIR pulses [20]. The sample, held at a base temperature of 10 K, was excited at 16-μm wavelength, resonant with an in-plane Cu-O stretching mode comparable to that driven in the manganites (see Section 2).

This vibrationally-driven state was shown to be superconducting by time-resolved THz spectroscopy. At equilibrium, superconductivity in layered cuprates is reflected in the appearance of a so-called Josephson plasma resonance (JPR) in the *c*-axis THz optical properties. This resonance is a general feature observed in layered superconductors [97], well understood by considering that Josephson coupling between stacks of two-dimensional superconducting $CuO_2$ layers gives rise to the three-dimensional superconducting state [98,99,100] (see also Section 6.1). Figure 10(a) shows the equilibrium *c*-axis reflectivity of $La_{1.84}Sr_{0.16}CuO_4$ ($T_c$ = 35 K) below and above the critical temperature. The JPR appears as an edge near 60 $cm^{-1}$ (~2 THz). Above $T_c$, where interlayer transport is incoherent, the response becomes flat. The equilibrium THz reflectivity of non-superconducting $La_{1.675}Eu_{0.2}Sr_{0.125}CuO_4$ at 10 K, shown in Fig. 10(b), is also featureless.

Figure 10(c) reports the THz reflectivity of $La_{1.675}Eu_{0.2}Sr_{0.125}CuO_4$ measured at τ = +5 ps after excitation of the Cu-O stretching mode. A clear reflectivity edge was measured at the same 2-THz frequency reported for $La_{1.84}Sr_{0.16}CuO_4$ at equilibrium, suggesting the possibility of emergent non-equilibrium superconducting transport. Note however that the measured transient edge is only 0.1% in amplitude, due to the pump-probe penetration depth mismatch. At THz frequencies, the probe pulse interrogates a volume that is between $10^2$ and $10^3$ times larger than the transformed region beneath the surface, with this mismatch being a function of frequency. Such mismatch was taken into account by modeling the response of the system as that of a photo-excited thin layer on top of an unperturbed bulk (which retains the optical properties of the sample at equilibrium). By calculating the coupled Fresnel equations of such multi-layer system [101], one could then retrieve the full transient optical response of the photo-excited layer, *e.g.* its complex-valued dielectric function $\varepsilon_1(\omega,\tau) + i\varepsilon_2(\omega,\tau)$ or optical conductivity $\sigma_1(\omega,\tau) + i\sigma_2(\omega,\tau)$ [22]. Note that artifacts at early time delays, found sometimes in the optical properties extracted from THz pump-probe data, have been shown to be negligible for the experiments reported here [102].

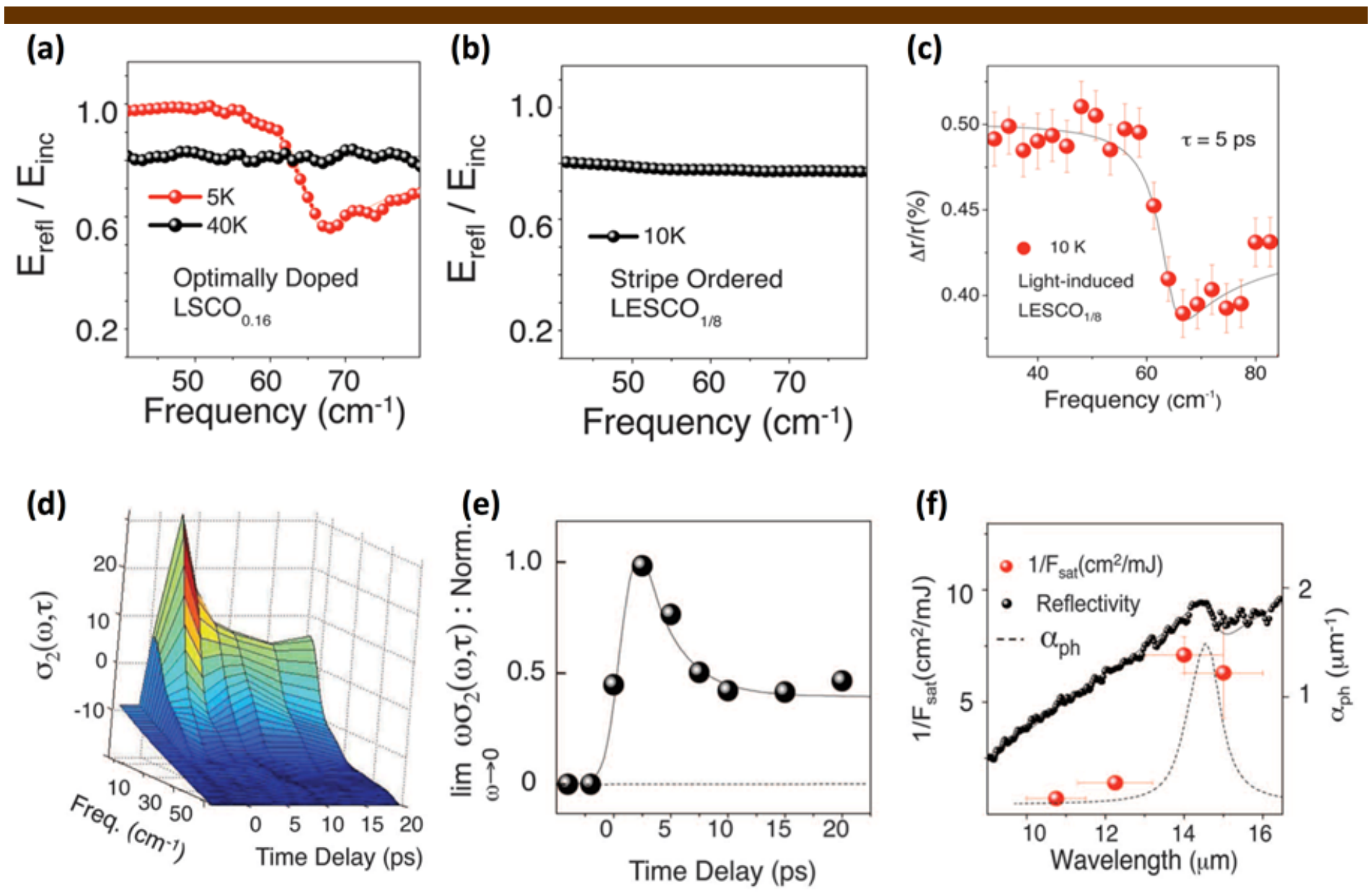

Fig. 10. (a) Equilibrium *c*-axis reflectivity of $La_{1.84}Sr_{0.16}CuO_4$ ($T_c$ = 35 K). In the superconducting state, the appearance of a JPR reflects coherent interlayer transport. Above $T_c$, incoherent ohmic transport is identified by a flat and featureless spectrum. (b) Equilibrium *c*-axis reflectivity of $La_{1.675}Eu_{0.2}Sr_{0.125}CuO_4$ at 10 K, showing the response of a non-superconducting compound. (c) Reflectivity changes induced in $La_{1.675}Eu_{0.2}Sr_{0.125}CuO_4$ by MIR excitation at 10 K, displaying a light-induced JPR. (d) Corresponding frequency- and time-delay-dependent imaginary conductivity (showing a 1/ω-like divergence) and (e) extracted superfluid density. (f) Inverse threshold (photo-susceptibility) as a function of pump wavelength (red dots), tuned in the spectral region of the Cu-O stretching mode. Data are shown along with the equilibrium *ab*-plane reflectivity and absorption coefficient (dashed) of $La_{1.675}Eu_{0.2}Sr_{0.125}CuO_4$. From [20]. Reprinted with permission from AAAS.

Importantly, in a superconductor at equilibrium the quantity $\lim_{\omega\to 0}\omega\sigma_2(\omega)$ is proportional to the superfluid density [103]. The same quantity can be dynamically traced by evaluating the low-frequency limit of the measured transient imaginary conductivity at different time delays $\tau$. As displayed in Figs. 10(d)-10(e), a finite superfluid density appears promptly after excitation and relaxes then into a plateau on a few picoseconds timescale, indicating the formation of a metastable, three-dimensional superconducting state. Figure 10(f) plots the pump-wavelength dependence of the inverse fluence threshold for light-induced superconductivity, used as a measure of photo-susceptibility. These data were compared to the equilibrium absorption coefficient $\alpha$ near the phonon energy (dashed line). The observed resonance of the photo-susceptibility at the Cu-O stretching mode frequency proves that direct coupling of the light electric field to the crystal structure triggers formation of the superconducting phase.

Further insights into the temperature dependence and relaxation dynamics of the light-induced superconducting state in $La_{1.675}Eu_{0.2}Sr_{0.125}CuO_4$ were provided in a later work [104]. By making use of an enhanced signal-to-noise ratio in the THz time-domain measurement, the authors could identify a photo-induced dynamics up to a temperature scale of ~80 K, which corresponds in $La_{1.675}Eu_{0.2}Sr_{0.125}CuO_4$ to the charge-ordering temperature $T_{CO}$. These measurements were taken under the very same MIR excitation of Ref. [20], where, due to the different experimental conditions, only a signal up to ~15 K was found above the noise. As shown in Fig. 11(a), a reflectivity edge, indicative of light-induced superconductivity, could be found all the way up to $T_{CO} \simeq 80$ K, a very large temperature for this family of cuprates. Furthermore, the edge size exhibited a discontinuous decrease when crossing $T_{SO} \simeq 25$ K (the spin-ordering temperature).

By analyzing the relaxation dynamics and comparing the measured THz conductivities with previous observations at equilibrium [105], it was identified that the transient JPR relaxes through a collapse of its coherence length, rather than of the superfluid density.

In addition, all time dependent data could be well fitted by a double exponential decay and the two relaxation time constants, $\tau_1$ and $\tau_2$, were plotted as a function of temperatures (Fig. 11(b)). This analysis revealed two different kinetic regimes: The lifetime of the photoinduced state remains essentially independent of base temperature below $T_{SO}$, while it follows an Arrhenius-like behavior (with the lifetime dropping exponentially with increasing temperature) in the charge-ordered regime, up to $T_{CO}$.

The exponential relaxation at $T_{SO} < T < T_{CO}$ can be reconciled with the expected kinetic behavior for a transition between two distinct thermodynamic phases separated by a free-energy barrier. Interestingly, the extracted energetic regime (~meV) corresponds to the energy scale of spin fluctuations measured in other single-layer cuprates [106,107,108], suggesting that the transition between the two phases may be regulated by spin rearrangements. This is consistent with the observation that strong spin fluctuations develop at equilibrium above $T_{SO}$ and survive up to $T_{CO}$ [109,110].

The departure from activated behavior for $T < T_{SO}$ may therefore be related to the freezing out of these spin fluctuations, being compatible with quantum coherent tunneling between two states, for instance between superconductivity and a pair-density-wave phase at constant carrier density. In this picture, in-plane Cooper pairing would already be present at equilibrium and superimposed or intertwined [111,112,113] with the stripe phase. The dynamical destruction of stripes would then allow a finite Josephson current along the *c* axis, giving rise to a transient, three-dimensional superconductor.

The fate of stripe order and LTT distortion during the vibrationally-induced transition into the transient superconducting state was investigated by femtosecond resonant soft x-ray diffraction in the strongly related compound $La_{1.875}Ba_{0.125}CuO_4$ [96]. Both static stripe order and the LTT distortion were measured through resonant diffraction near the oxygen K-edge. Here, static charge stripes can be observed at the $\boldsymbol{q}$ = (0.24 0 0.5) wave vector [88,114,115], while the LTT distortion can be directly measured through the (001) diffraction peak that is structurally forbidden in the high-temperature phases [116].

MIR pulses, tuned to the IR-active in-plane Cu-O stretching phonon [117], were used to excite the sample at T = 15 K. The pump fluence was kept at a similar value (~2 mJ/cm$^2$) as that used in Refs. [20,104].

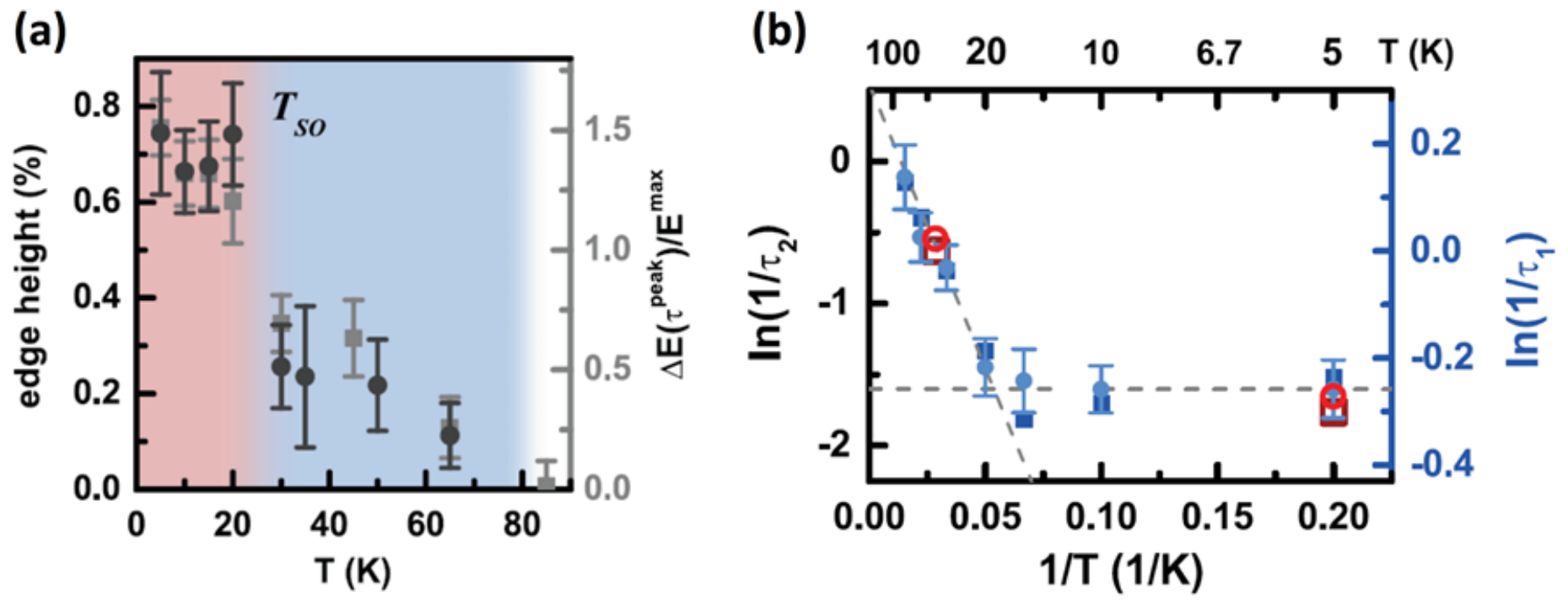

Fig. 11. (a) Pump-induced change in the THz field amplitude and corresponding size of the Josephson plasma edge measured in $La_{1.675}Eu_{0.2}Sr_{0.125}CuO_4$ at 1.8 ps time delay after MIR excitation, at different base temperatures. The JPR size remains approximately constant below the spin-order transition temperature $T_{SO}$, but drops rapidly between $T_{SO}$ and $T_{CO}$ (dark gray circles). (b) Arrhenius plot of the relaxation rate as a function of temperature. The fast component $\tau_1$ (dark blue squares) and the slow component $\tau_2$ (light blue circles) track the decay of the coherence length of the transient plasma. The lifetime of the transient state remains temperature independent (horizontal gray dashed line) below the spin-order transition temperature, $T_{SO} \simeq 25$ K. Above this transition, the lifetime exhibits an exponential temperature dependence, with an energy scale of 4 meV for $\tau_2$ and 0.8 meV for $\tau_1$ (gray dashed line). Reprinted with permission from [104]. Copyright 2015 by the American Physical Society.

The time-dependent integrated scattering intensity of the (0.24 0 0.5) stripe order diffraction peak in response to the MIR excitation is plotted in Fig. 12(a), showing a prompt decrease of ~70%. This result shows that stripe order is melted within less than 1 ps, a timescale similar to that observed for the onset of superconductivity in $La_{1.675}Eu_{0.2}Sr_{0.125}CuO_4$ [20]. This finding points towards a tight connection between stripe melting and light-induced superconductivity.

In contrast, the evolution of the LTT phase, probed by the (001) diffraction peak, is very different. The integrated scattered intensity of this structural peak drops by only 12%, and with a much longer time constant of ~15 ps, likely determined by acoustic propagation.

These combined experiments demonstrate that MIR lattice excitation in striped cuprates triggers the ultrafast formation of a non-equilibrium state, in which stripe order has disappeared while the LTT distortion still exists. As also stated above, the melting of charge order removes the periodic potential that suppresses interlayer coupling at equilibrium [94,95], thus inducing coherent coupling between the two-dimensional superconducting condensates of the $CuO_2$ planes on the few-hundred femtosecond time scale of the JPR. Noteworthy, the LTT distortion plays a minor role in this process, if at all, as also demonstrated by the observation of light-enhancement of superconductivity in striped compounds of the $La_{2-x}Ba_xCuO_4$ family upon non-resonant excitation at near-infrared and optical wavelengths [118,119].

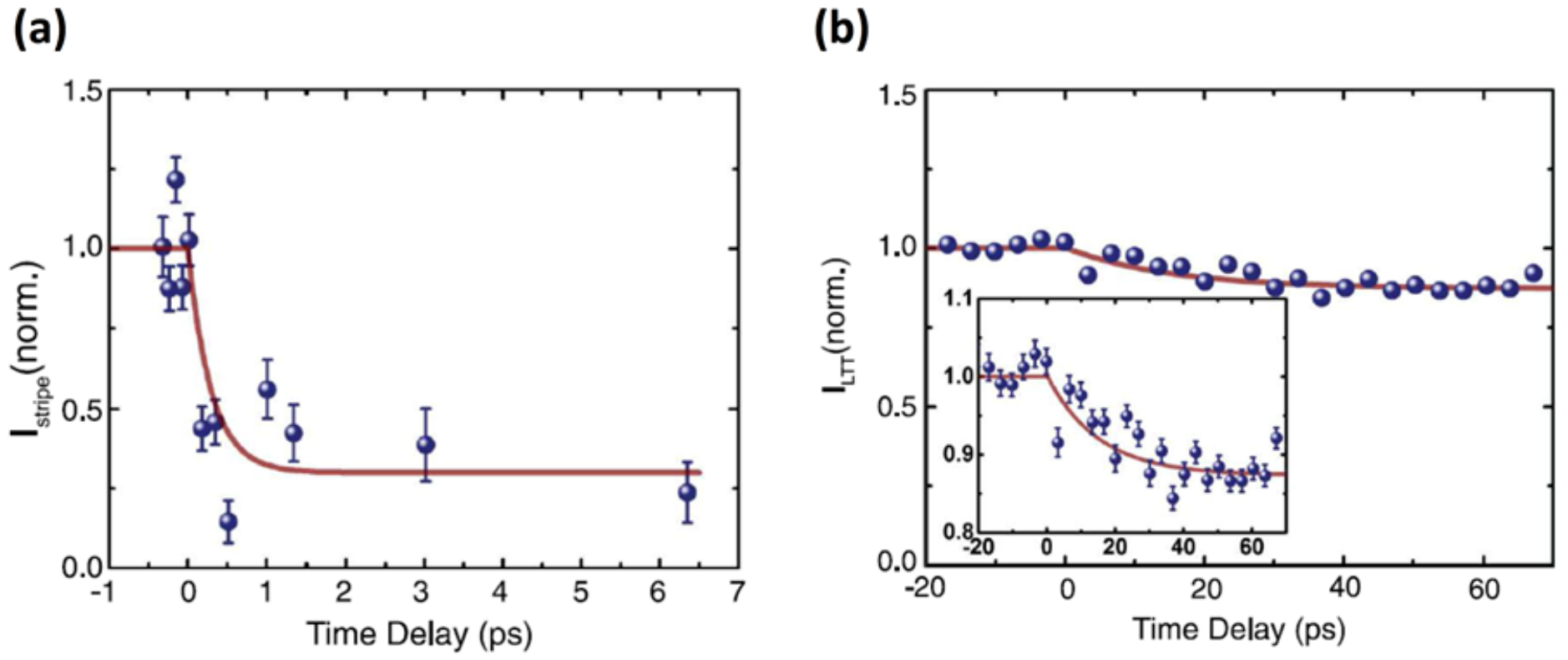

Fig. 12. (a) Transient intensity of the charge stripe order diffraction peak in $La_{1.875}Ba_{0.125}CuO_4$ measured at the (0.24 0 0.5) wave vector. MIR excitation results in a prompt sub-ps decrease of the scattered intensity. The red solid line is an exponential fit with a time constant of 300 fs, *i.e.* the resolution of the experiment. (b) Corresponding changes in the intensity of the (001) diffraction peak (reflecting the LTT distortion) in the same crystal under same excitation conditions. The red solid line is an exponential fit yielding a time constant of 15 ps. Reprinted with permission from [96]. Copyright 2014 by the American Physical Society.

### 3.2. Control of superconductivity in $YBa_2Cu_3O_x$

Light-control of superconductivity via phonon excitation has been recently applied also to the bilayer high-$T_c$ cuprate $YBa_2Cu_3O_x$ [22,120]. This compound crystallizes in an orthorhombic structure, made of bilayers of $CuO_2$ planes, separated by insulating Yttrium layers, as shown in Fig. 13(a). The bilayers are further spaced by Cu-O chains, which control the hole doping in the $CuO_2$ planes.

In analogy with single-layer compounds (see Section 3.1), the coherent inter-bilayer tunneling of Cooper pairs in the superconducting state of $YBa_2Cu_3O_x$ results in the appearance of a JPR in the *c*-axis THz optical response. This is visualized as an edge in the reflectivity, as shown in Fig. 13(b) for three hole concentrations [120] ($x$ = 6.45, 7% holes, $T_c$ = 35 K; $x$ = 6.5, 9% holes, $T_c$ = 50 K; $x$ = 6.6, 12% holes, $T_c$ = 62 K), all belonging to the "underdoped" region of the phase diagram, *i.e.* doping levels lower than that corresponding the maximum $T_c$ (~16%).

$YBa_2Cu_3O_x$ single crystals were photo-excited with MIR pulses tuned to be resonant with a phonon mode of $B_{1u}$ symmetry, which involves oscillations of the so-called "apical" oxygen atoms (see Fig. 13(a)). Note that the static atomic position of these apical oxygen atoms, and in particular their distance from the $CuO_2$ planes, is known to be an important parameter in high-$T_c$ cuprates, which affects the in-plane band structure and correlates strongly with the maximum achievable $T_c$ within each cuprate family [121]. It is therefore to be expected that large-amplitude oscillations of these atoms driven by the MIR pulses may have dramatic effects on the superconducting behavior of $YBa_2Cu_3O_x$.

The transient response measured after excitation below $T_c$ (*i.e.* in the superconducting state) is summarized in Fig. 14(a)-14(c). At all measured dopings and all $T < T_c$, a pump-induced increase in the imaginary conductivity was measured, indicative of enhanced inter-bilayer superconducting coupling. This $\Delta\sigma_2 > 0$ corresponded in reflectivity to the appearance of a new, stiffer JPR, at the expenses of that measured at equilibrium [120]. Note that this response is qualitatively different and opposite in phase from that measured in the same material after non-resonant electronic excitation at near-infrared and optical frequencies [122,123], which resulted in a depletion of superfluid density due to Cooper pair breaking.

When the base temperature was raised to 100 K, *i.e.* above the highest equilibrium $T_c$ of the $YBa_2Cu_3O_x$ family, in analogy with the observations reported in Section 3.1 for single-layer cuprates, a light-induced reflectivity edge was measured at ~1 ps time delay after excitation (see Fig. 14(g)-14(i)). Remarkably, as in the case of the equilibrium JPR (Fig. 13(b)), this edge was located in the 1-2 THz frequency range and was found to stiffen with increasing hole doping. Correspondingly, an increase in imaginary conductivity was also extracted (see Fig. 14(d)-14(f)), compatible with a possible light-induced inter-bilayer superconducting transport above equilibrium $T_c$.

Similarly to what found for $La_{1.675}Eu_{0.2}Sr_{0.125}CuO_4$ (Section 3.1), the relaxation of the transient reflectivity edge occurred with a double exponential decay. However, the time scales $\tau_1 \simeq 0.5$ ps and $\tau_2 \simeq 5$ ps were significantly faster and the ground state was fully recovered within ~10 ps, possibly suggesting a different driving mechanism for the observed dynamics.

The very same experiment was repeated at progressively higher base temperatures and a temperature scale T' for the disappearance of light-induced superconductivity ($\Delta\sigma_2 = 0$) was determined for each doping (see Fig. 15(a)-15(c)). T' was found to increase with decreasing hole concentration and to exceed room temperature at $x$ = 6.45, with a trend very similar to the well-known "pseudogap" transition line T* [124] (see phase diagram in Fig. 15(d)).

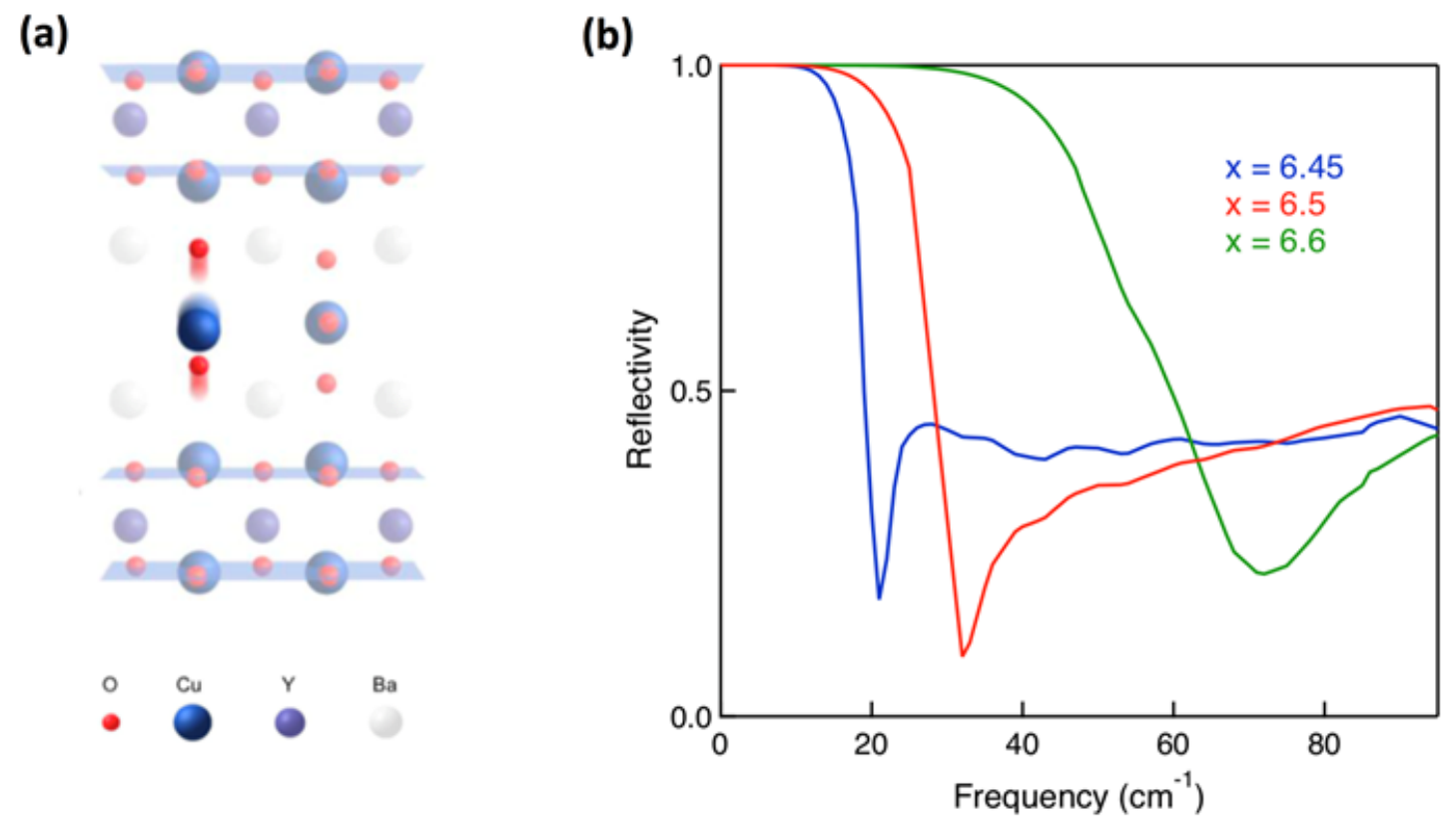

Fig. 13. (a) Crystal structure of orthorhombic $YBa_2Cu_3O_{6.5}$ and sketch of the resonantly excited $B_{1u}$-symmetry IR-active phonon mode, comprising *c*-axis motions of the apical oxygen atoms between bilayers. (b) Equilibrium reflectivity spectra of $YBa_2Cu_3O_{6.45}$, $YBa_2Cu_3O_{6.5}$, and $YBa_2Cu_3O_{6.6}$ measured along the *c* axis, showing clear edges at the JPR frequencies. Panel (a) Reprinted with permission from [18]. Copyright 2015 American Chemical Society. Panel (b) adapted from [120].

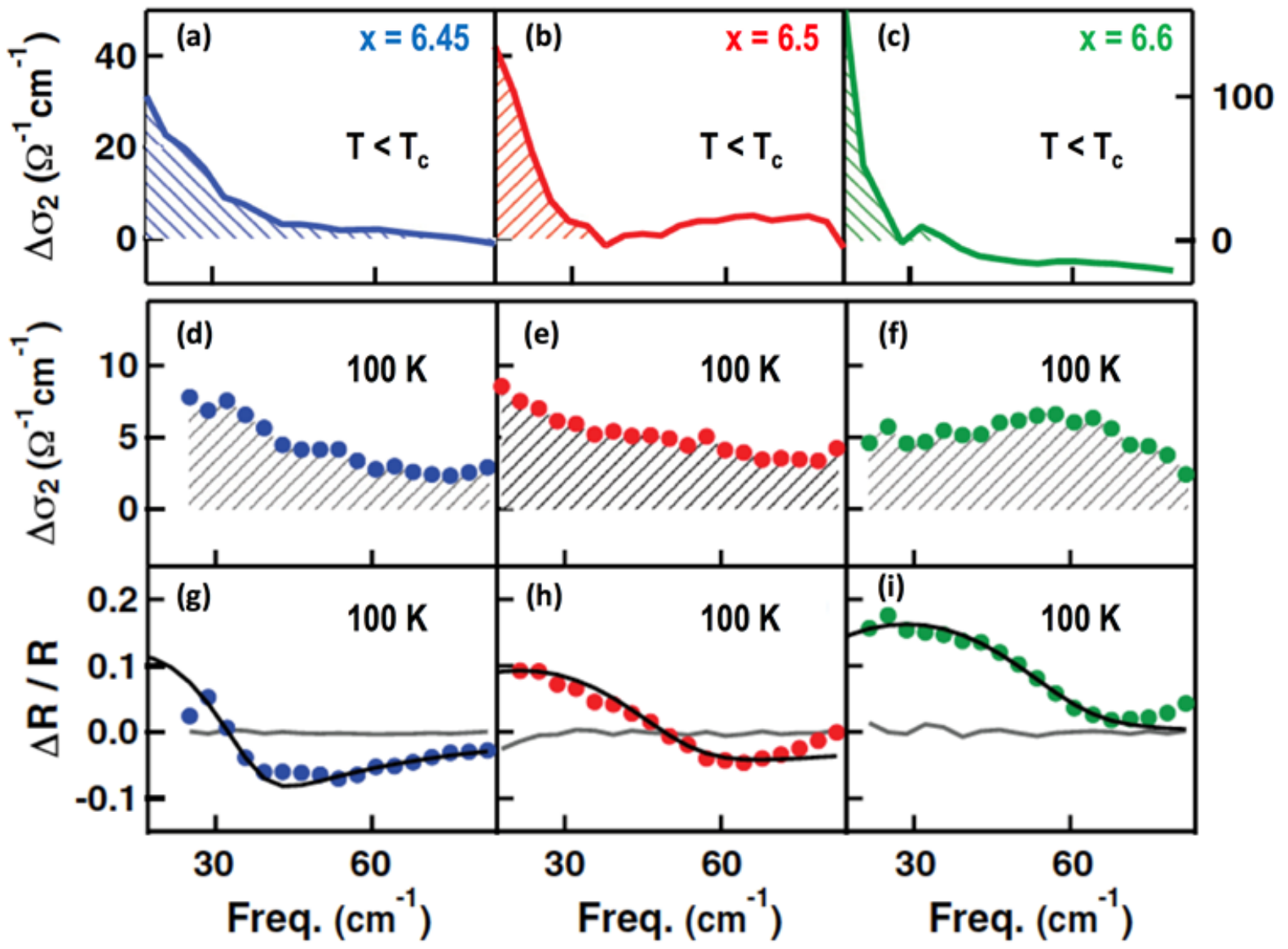

Fig. 14. (a)-(c) Transient changes in the *c*-axis imaginary conductivity of $YBa_2Cu_3O_{6.45}$, $YBa_2Cu_3O_{6.5}$, and $YBa_2Cu_3O_{6.6}$ measured below $T_c$ (*i.e.* in the superconducting state), 1 ps after MIR phonon excitation. Data in (a)-(b) refer to the right $\Delta\sigma_2$ scale, while those in (c) to that on the left. (d)-(f) Light-induced imaginary conductivity changes measured in the same samples, under the same excitation conditions, and at the same time delay as in (a)-(c), at a base temperature of 100 K (above $T_c$). (g)-(i) Light-induced reflectivity changes corresponding to the conductivity curves in (d)-(f). Reprinted with permission from [120]. Copyright 2014 by the American Physical Society.

The pseudogap phase in the equilibrium phase diagram of high-$T_c$ cuprates defines a region ($T < T^*$) characterized by anomalous electronic properties, where the possible presence of pre-formed Cooper pairs without phase coherence has been discussed in the past [125,126,127]. The experimental observations reported above may therefore suggest that phonon excitation transfers phase coherence to such preformed pairs, thus giving rise to a transient superconductor.

In order to get further insights into the driving mechanism of light-enhanced superconductivity in $YBa_2Cu_3O_x$, the dynamics of another Josephson plasmon was also investigated under the same excitation conditions. This mode is the so-called intra-bilayer plasma mode, located at ~15 THz and related to the Josephson tunneling within the bilayer units of the $YBa_2Cu_3O_x$ crystal structure [128,129] (see Fig. 16(a)). As the low-frequency JPR, also the intra-bilayer plasmon appears as a reflectivity edge or as a peak in the loss function, while another "mixed" mode (the so-called transverse JPR) gives rise to a peak in the real part of the optical conductivity, $\sigma_1(\omega)$ (Fig. 16(b)).

At equilibrium, the intra-bilayer JPR is observed at temperatures much higher than $T_c$ (up to ~150 K) and its presence has been used to define a region of the $YBa_2Cu_3O_x$ phase diagram with precursors of superconductivity in the normal state, possibly characterized by coherent tunneling of preformed Cooper pairs within the bilayer units [130,131].

In order to map out the response of this mode to MIR excitation of the apical oxygen phonon, time-resolved ultrabroadband (1 - 15 THz) spectroscopy was employed. This experiment was carried out at a single doping level ($x$ = 6.5) [22]. The measured transient optical properties of $YBa_2Cu_3O_{6.5}$ could be well fitted at all temperatures and time delays by assuming an effective medium with two components: The first one ($\gtrsim$80%) was substantially not affected by MIR excitation, while the second, photo-susceptible one ($\lesssim$20%) showed a stiffened (photo-induced) low-frequency JPR below (above) $T_c$ (see also Fig. 14), concurrently with a red-shifted intra-bilayer plasmon. Representative results of these fits are reported in Fig. 16(c)-16(d).

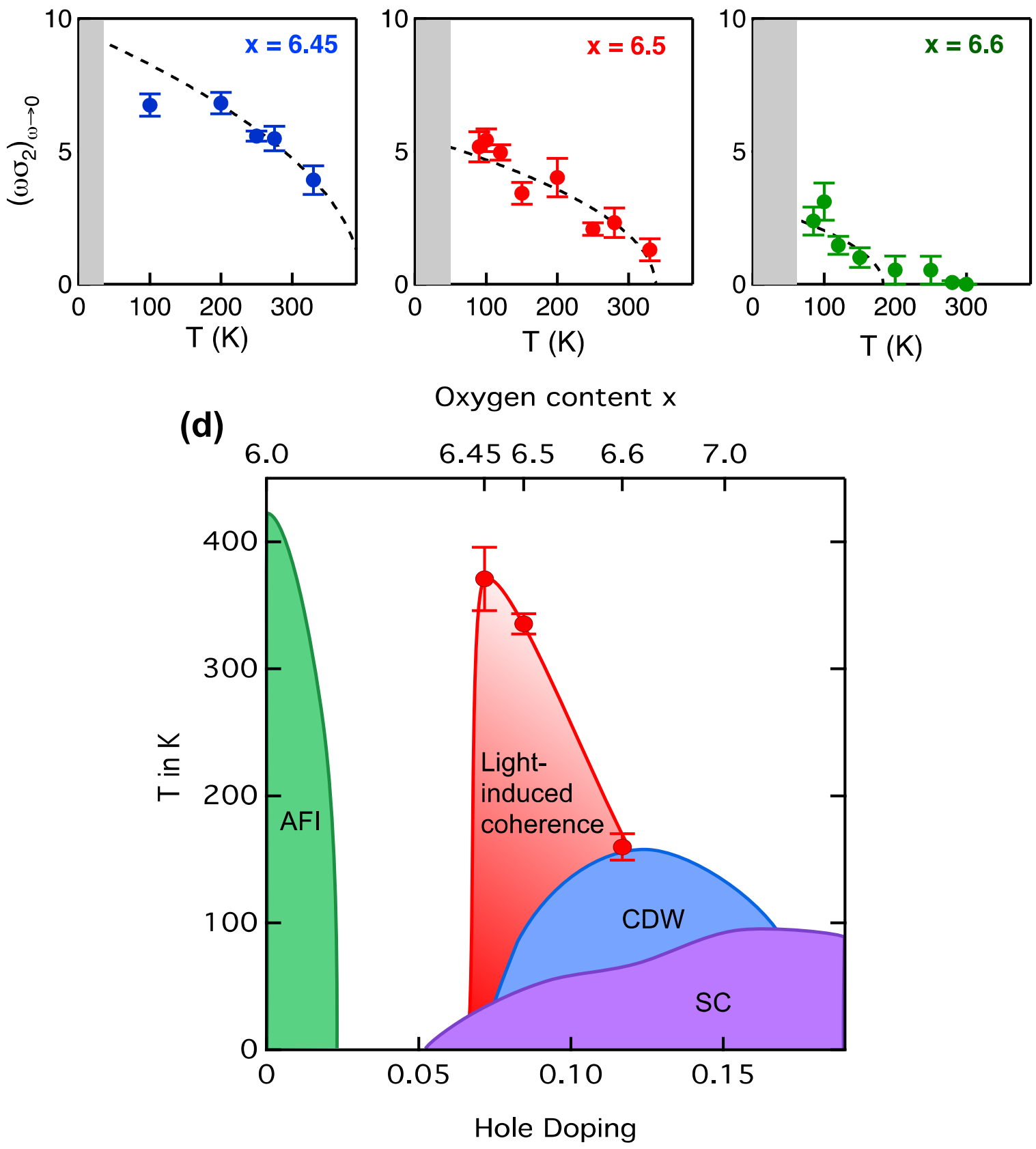

Fig. 15. (a)-(c) Strength of the light-induced inter-bilayer coupling measured in $YBa_2Cu_3O_{6.45}$, $YBa_2Cu_3O_{6.5}$, and $YBa_2Cu_3O_{6.6}$ at +1 ps pump-probe delay as a function of base temperature and quantified by the zero-frequency extrapolation of the enhancement in imaginary conductivity. Grey shaded regions refer to the equilibrium superconducting state. (d) Phase diagram of $YBa_2Cu_3O_x$. AFI, CDW, and SC refer to the equilibrium antiferromagnetic insulating, charge-density-wave, and superconducting phase, respectively. The red circles, estimated from the data in (a)-(c), delimit the region where signatures of possible light-induced superconductivity were measured.

All these observations define a scenario in which inter-bilayer Josephson coupling is enhanced (or induced) at the expenses of intra-bilayer tunneling, while the number of Cooper pairs is conserved. This redistribution of coherent coupling takes place only in a fraction ($\lesssim$20%) of the material, throughout the pseudogap phase of $YBa_2Cu_3O_x$. Notably, high-$T_c$ cuprates in this region of their phase diagram are known to be intrinsically inhomogeneous already at equilibrium [132] and this feature may also play a role in explaining the non-equilibrium physics reported here.

It should also be mentioned that the observation of transient superconductivity is not unambiguous. Most conservatively, the experimental data can also be fitted by the optical properties of a metal with a very long scattering time ($\tau_S \simeq 7$ ps, corresponding to the lifetime of the state). This value of $\tau_S$ implies a d.c. mobility of $10^3$ – $10^4$ cm$^2$ V$^{-1}$ s$^{-1}$, which would be highly unusual for incoherent transport in oxides. Furthermore, as the position of the reflectivity edge does not move with the number of absorbed photons and only the fraction of material that is switched is a function of laser field, the results are hardly compatible with above-gap photoconductivity. If one also excludes more exotic effects, like conduction by a sliding one-dimensional charge density wave [133], photo-induced transient superconductivity is the most plausible scenario.

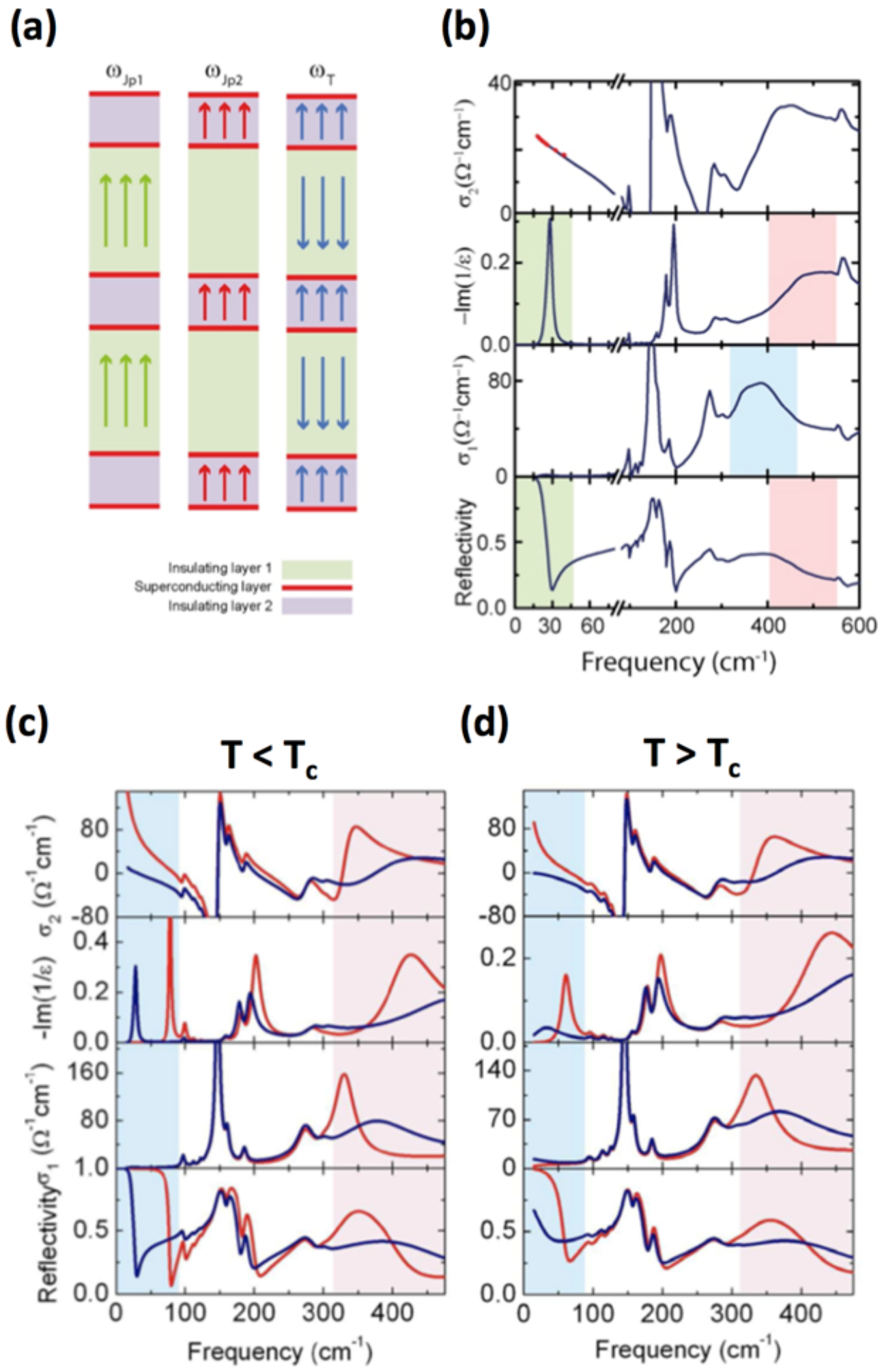

Fig. 16. (a) In the superconducting state, the structure of $YBa_2Cu_3O_{6.5}$ can be viewed as two Josephson junctions in series, giving rise to two longitudinal modes ($\omega_{Jp1}$, $\omega_{Jp2}$) and a transverse mode ($\omega_T$) (arrows indicate the direction of the current). (b) Equilibrium *c*-axis optical properties of $YBa_2Cu_3O_{6.5}$. Superconductivity is evidenced by the $1/\omega$ divergence (red dashed) in the imaginary conductivity. Both longitudinal JPRs appear as peaks in the loss function and edges in reflectivity (green and red shaded). The transverse JPR appears as a peak in $\sigma_1$ (blue shaded). (c)-(d) Effective medium fits used to reproduce the transient broadband optical response after MIR excitation. Blue curves refer to the unperturbed component ($\gtrsim$80%) (equilibrium superconductor below $T_c$ and normal insulator above $T_c$). Red curves indicate the photo-susceptible component ($\lesssim$20%), which has a stiffer inter-bilayer JPR and red-shifted high-frequency plasmons. Reprinted from [22].

While in single-layer cuprates the observed non-equilibrium superconductivity was explained by the vibrationally-driven melting of the competing charge ordered phase (Section 3.1), the framework in $YBa_2Cu_3O_x$ is more complicated. The recent observation of a charge-density-wave (CDW) phase in the underdoped region of the equilibrium phase diagram of bilayer cuprates [134,135] suggested that also in this case phonon excitation might simply be removing a CDW phase, which was preventing superconductivity at higher temperatures at equilibrium.

As for $La_{1.875}Ba_{0.125}CuO_4$ (see Section 3.1), femtosecond resonant soft x-ray diffraction was employed also in underdoped $YBa_2Cu_3O_{6.6}$ to measure the ultrafast changes of CDW correlations under the same MIR excitation conditions used in Refs. [22,120]. It was found that when coherent inter-bilayer transport is enhanced by optical excitation of the apical oxygen distortions, at least 50% of the in-plane CDW order is melted [136]. However, while this result indicates that charge ordering and superconductivity are competing in $YBa_2Cu_3O_{6.6}$ up to the CDW transition temperature (~150 K), the observation of enhanced coherent inter-bilayer transport up to 300 K and above at a doping concentration ($x$ = 6.45) where no CDW is present at equilibrium (see Fig. 15(d)) [135] clearly indicates that additional physics must be invoked to explain light-induced superconductivity in $YBa_2Cu_3O_x$.

Among the possible driving mechanisms, one should mention for example dynamical stabilization [137,138]. As the 15-μm modulation used here occurs at frequencies that are high compared with plasma excitations between planes, one could envisage a dynamically stabilized stack of Josephson junctions, by direct coupling of the oscillatory field to the order parameter.

Alternatively, within a nonlinear phononics scenario (see Section 2.1), excitation of the lattice may create a displaced crystal structure [18,19] with atomic positions more favorable to high-temperature superconductivity [121,139,140]. This possibility, along with experimental evidences, is described in the following Section.

### 3.3. Nonlinear lattice dynamics in $YBa_2Cu_3O_x$

In search of a microscopic picture for the transient response observed in $YBa_2Cu_3O_x$ after MIR excitation, the underlying nonlinear lattice dynamics was investigated in the $x$ = 6.5 sample using femtosecond hard x-ray diffraction at the LCLS free electron laser [23].

As the direct product $B_{1u} \otimes B_{1u}$ is of $A_g$ symmetry, the resonantly driven $B_{1u}$ phonon mode of Fig. 13(a) can only couple to Raman-active phonons of $A_g$ symmetry. By means of Density Functional Theory calculations four strongly coupled $A_g$ Raman modes were identified, all involving $c$-axis motion of the apical oxygen and planar copper atoms. According to the theory of nonlinear phononics (Section 2.1), the crystal lattice is then promptly distorted into a non-equilibrium structure along the linear combination of the atomic motions associated with these Raman modes.

In Fig. 17(a) we report the demonstration of nonlinear lattice dynamics occurring in the transient superconducting state of $YBa_2Cu_3O_{6.5}$. Hard x-ray diffraction data were taken at 100 K, under the same excitation conditions of Refs. [22,120], which gave rise to coherent inter-bilayer transport. The dynamics of two representative diffraction peaks, the (–2 –1 1) and (–2 0 4), are displayed. A prompt change in diffracted intensity was observed, as expected from the rearrangement of the atoms in the unit cell predicted by the model of nonlinear phononics. The decay to the equilibrium structure happened on the same timescale as the relaxation of light-induced superconductivity (see Section 3.2), indicating an intimate connection [22,120].

The exact amplitude and sign of the changes in diffraction intensity carry fundamental information on the light-induced lattice rearrangement. In order to reconstruct the transient crystal structure, the computed coupling strengths to the $A_g$ Raman modes were combined with structure factor calculations, predicting the changes in diffraction intensity of all measured Bragg peaks for any given $B_{1u}$ amplitude. The dynamics of all measured Bragg peaks could then be fitted simultaneously with the driving amplitude as only free parameter [23] (fits shown as red curves in Fig. 17(a)).

The reconstructed transient crystal structure of Fig. 17(b) shows an increase in the in-plane O-Cu-O bond buckling and a decrease in apical oxygen to planar copper distance. Furthermore, the intra-bilayer distance increases, whereas the $CuO_2$ planes within the bilayers move closer together, effectively enhancing the inter-bilayer coupling. This last observation is intuitively consistent with the ultrabroadband THz spectroscopy measurements reported in Section 3.2 [22], which indicated that the appearance of inter-bilayer superconducting coupling comes at the expenses of intra-bilayer tunneling strength.

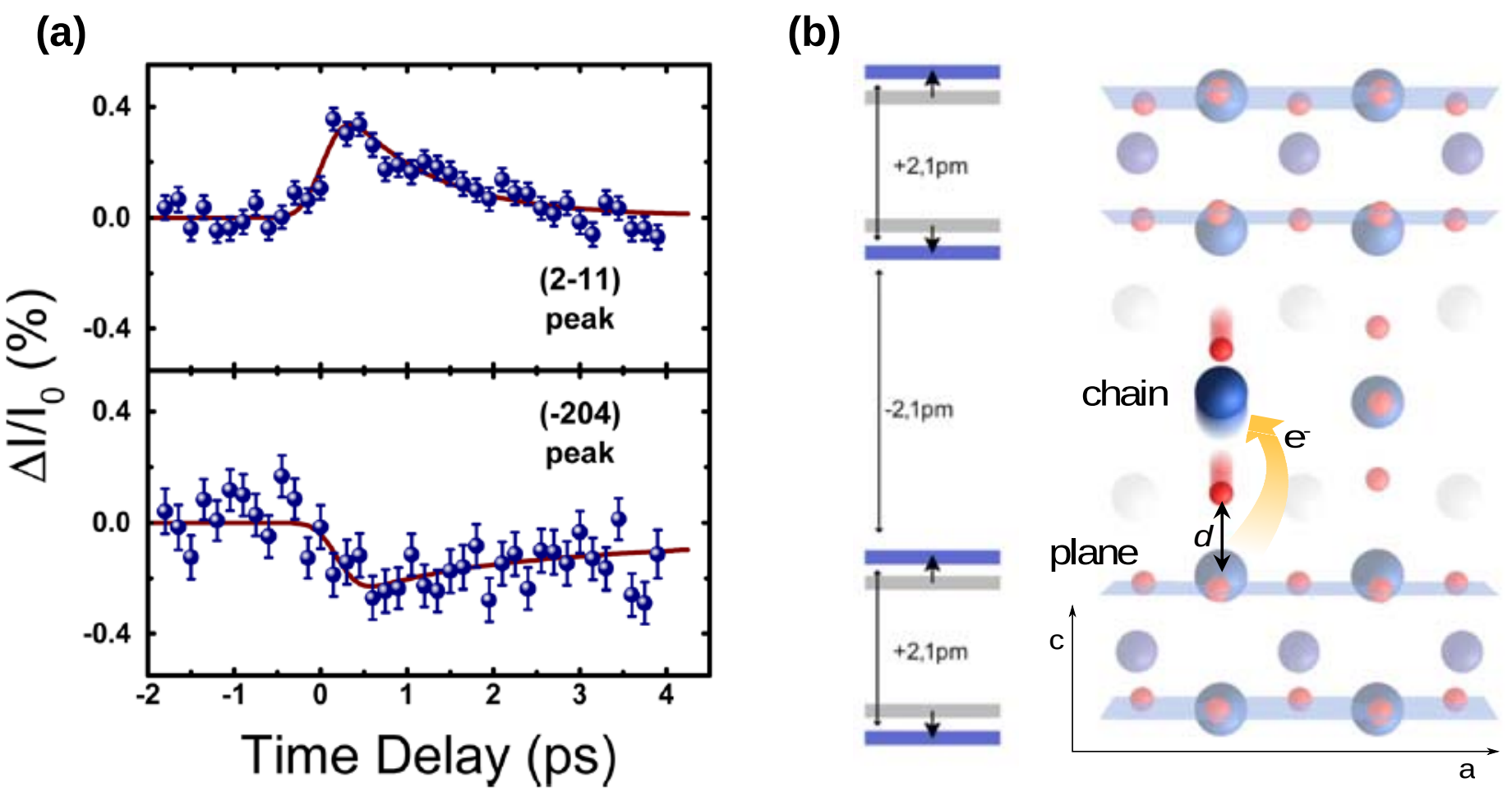

Fig. 17. (a) Relative changes in diffracted intensity of the (−2 −1 −1) and the (−2 0 4) Bragg peaks measured in $YBa_2Cu_3O_{6.5}$ after MIR excitation at 100 K. Results of simultaneous fits to all data are shown as red lines. (b) Sketch of the reconstructed transient crystal structure at the peak signal. The atomic displacements from the equilibrium structure involve a decrease in inter-bilayer distance, accompanied by an increase in intra-bilayer distance. Panel (a) reprinted with permission from [18]. Copyright 2015 American Chemical Society.

Density Functional Theory calculations using the transient crystal structure predict also a charge transfer from the $CuO_2$ planes to the Cu-O chains, effectively increasing the hole doping of the planes. Such self-doping effect was recently found to accompany the temperature-driven superconducting transition at equilibrium in $YBa_2Cu_3O_{6.9}$ and might be of key importance to explain the formation of the superconducting phase [141].

In addition, a recent theoretical study analyzed the effects of the discussed lattice dynamics on superconductivity and charge order [142]. An increase in $T_c$ by even 100% was predicted, along with a suppression of charge order, also consistent with the experimental observations of Ref. [136].

Such nonlinear lattice dynamics is expected not only to induce a directional displacement in the crystal lattice along all coupled Raman mode coordinates. The theory of nonlinear phononics (Section 2.1) predicts in addition that all displaced modes which have a long eigenperiod compared to the excitation pulse duration will exhibit coherent oscillations about the displaced atomic positions. The coherent lattice response of $YBa_2Cu_3O_{6.55}$ was measured using an optical probe, aiming at providing a complete picture of the structural dynamics in the light-induced superconducting state [143].

An atomic motion along Raman coordinates modulates the polarizability tensor and can be observed as an oscillation in the sample reflectivity. This coherent structural response was measured by probing the transient reflectivity at 800 nm wavelength using 35-fs pulses. MIR pulses were used to resonantly excite the IR-active $B_{1u}$ mode as in Refs. [22,23,120]. Under these conditions, all $A_g$ Raman modes with finite coupling to the $B_{1u}$ mode and of up to $\sim$6 THz in frequency are expected to be coherently driven.

Oscillations in the transient reflectivity were found by probing both in the $ab$ (in-plane) and $c$ (out-of-plane) directions, as shown in Fig. 18(a). Three dominant frequency components appeared in the Fourier transform spectrum, attributed to the lowest-frequency $A_g$ modes of $YBa_2Cu_3O_{6.55}$ (Fig. 18(b)). Consistent with the Raman tensor for the $A_g$ modes, the phase of the coherent response was the same for the two orthogonal probe polarizations [144].

The oscillation amplitudes in real space could not be quantified from these data alone, as the changes in the 800-nm polarizability depend on the unknown Raman tensor elements [145]. However, an estimate of the oscillatory amplitude could be obtained by combining these data with the previous x-ray diffraction experiment. According to this analysis, the coherent atomic motion is dominated by a change in distance of planar Cu atoms $d$ along the crystallographic $c$ axis. Following excitation, the Cu atoms of bilayers move apart from each other by ~3 pm, corresponding to a relative change in distance of 1%. This motion is accompanied by coherent oscillations with an estimated amplitude of ~0.9 pm, which decay within 3 ps after excitation.

According to Density Functional Theory calculations [143], this oscillatory motion would induce periodic charge redistributions between the Cu atoms of planes and chains, which may cause a dynamical stabilization of interlayer fluctuations by a modulation of electronic properties [146].

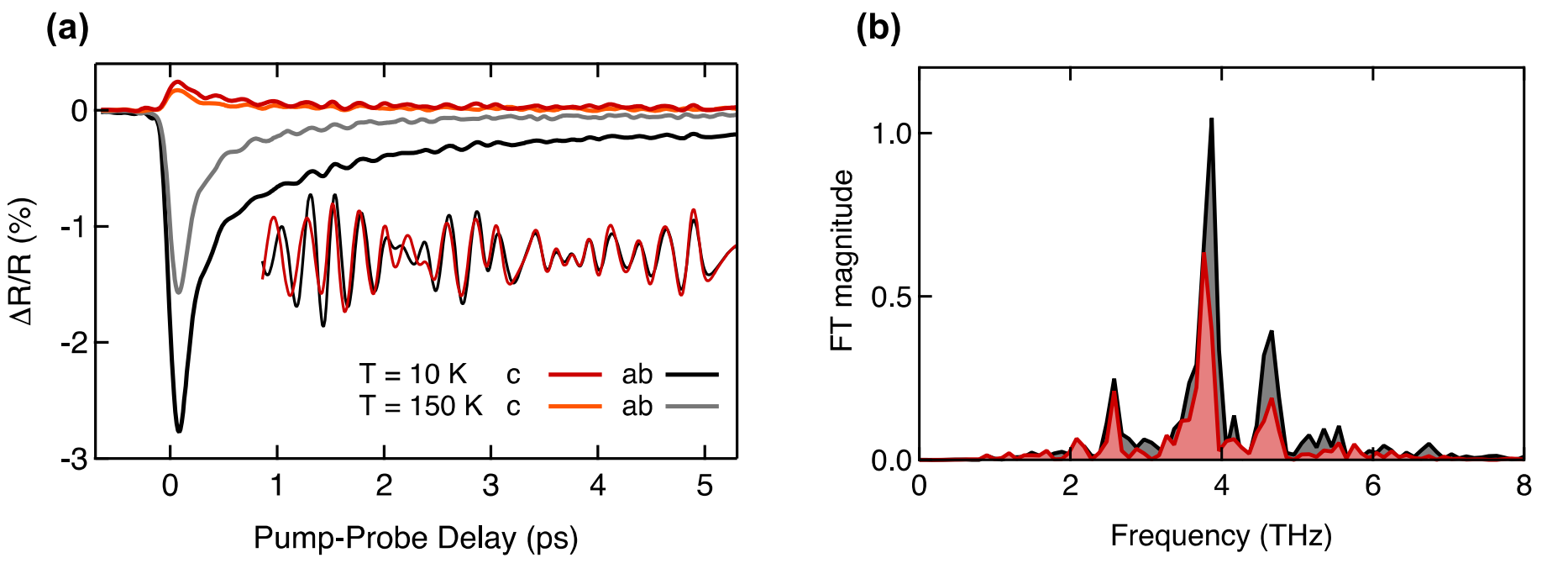

Fig. 18. (a) Time-resolved changes in the reflection of 800 nm pulses polarized in-plane (*ab*) and out-of-plane (*c*) following MIR excitation. The oscillatory components shown in the inset are extracted from the 10 K measurement. (b) Fourier transform of these oscillations, showing three components, attributed to different $A_g$ Raman modes. Adapted from [143].

## 4. Excitation of local vibrations in molecular solids

As discussed in the previous Sections, coherent excitation of anharmonically coupled lattice modes can be used to create transient crystal structures with new electronic properties. Examples of such nonlinear phonon control are ultrafast insulator-to-metal transitions, melting of magnetism, and light-induced superconductivity.

Nonlinear phononics has been realized and understood in the limit of cubic anharmonic coupling [23,57], which results in rectification of an excited lattice oscillation and in the net displacement of the atomic positions along a second phonon mode. Changes in the electronic properties result then from the altered bond angles and atomic distances, which perturb hopping amplitudes and exchange interactions.

However, local parameters like Mott correlations, of importance in many complex materials, are not addressed by nonlinear phononics. Conversely, these can be modulated in molecular solids by driving local molecular degrees of freedom to large amplitudes [147]. The excitation of local modes at MIR frequencies is different from the case of nonlinear phononics in that the molecular orbital and, consequently, the on-site charge density are controlled [148,149].

In this Section we show how resonant excitation of such local molecular vibrational modes has been used to affect the on-site Coulomb interaction in the one-dimensional Mott insulator ET-$F_2$TCNQ [24,25] and to control superconductivity in the alkali-fulleride $K_3C_{60}$ [26].

### 4.1. Modulation of local Mott correlations in ET-$F_2$TCNQ

A Mott insulator is a half-filled solid in which electrons are localized due to their mutual Coulomb repulsion. This physics is typically described by the so-called extended Hubbard Hamiltonian:

$$\hat{H} = -t\sum_{\ell,\sigma}(c^{\dagger}_{\ell\sigma}c_{(\ell+1)\sigma} + \mathrm{h.c.}) + U\sum_{\ell}\hat{n}_{\ell\uparrow}\hat{n}_{\ell\downarrow} + V\sum_{\ell}\hat{n}_{\ell}\hat{n}_{\ell+1}$$

where $c^{\dagger}_{\ell\sigma}(c_{\ell\sigma})$ is the creation (annihilation) operator for an electron at site $\ell$ with spin σ, and $\hat{n}_{\ell\sigma}$ is the associated number operator, $U$ ($V$) denotes the on-site (nearest-neighbor) Coulomb repulsion, and $t$ the hopping amplitude [12,13].

The key features of the Hubbard model are well reproduced in some molecular solid, for example in the one-dimensional charge transfer salt ET-$F_2$TCNQ. The crystal structure of this material is shown in Fig. 19(a). One-dimensional Mott physics is observed along the crystallographic $a$ axis, where the half-filled chain of ET molecules is characterized by small intersite tunneling amplitude ($t$ ~ 40 meV) and large Coulomb repulsion ($V$ ~ 120 meV and $U$ ~ 840 meV). Hence, although the system is fractionally filled, it shows a large insulating gap in its excitation spectrum, followed by a "charge-transfer" band centered at 5500 cm$^{-1}$ (~700 meV, see black curve in Fig. 19(b)) [150].

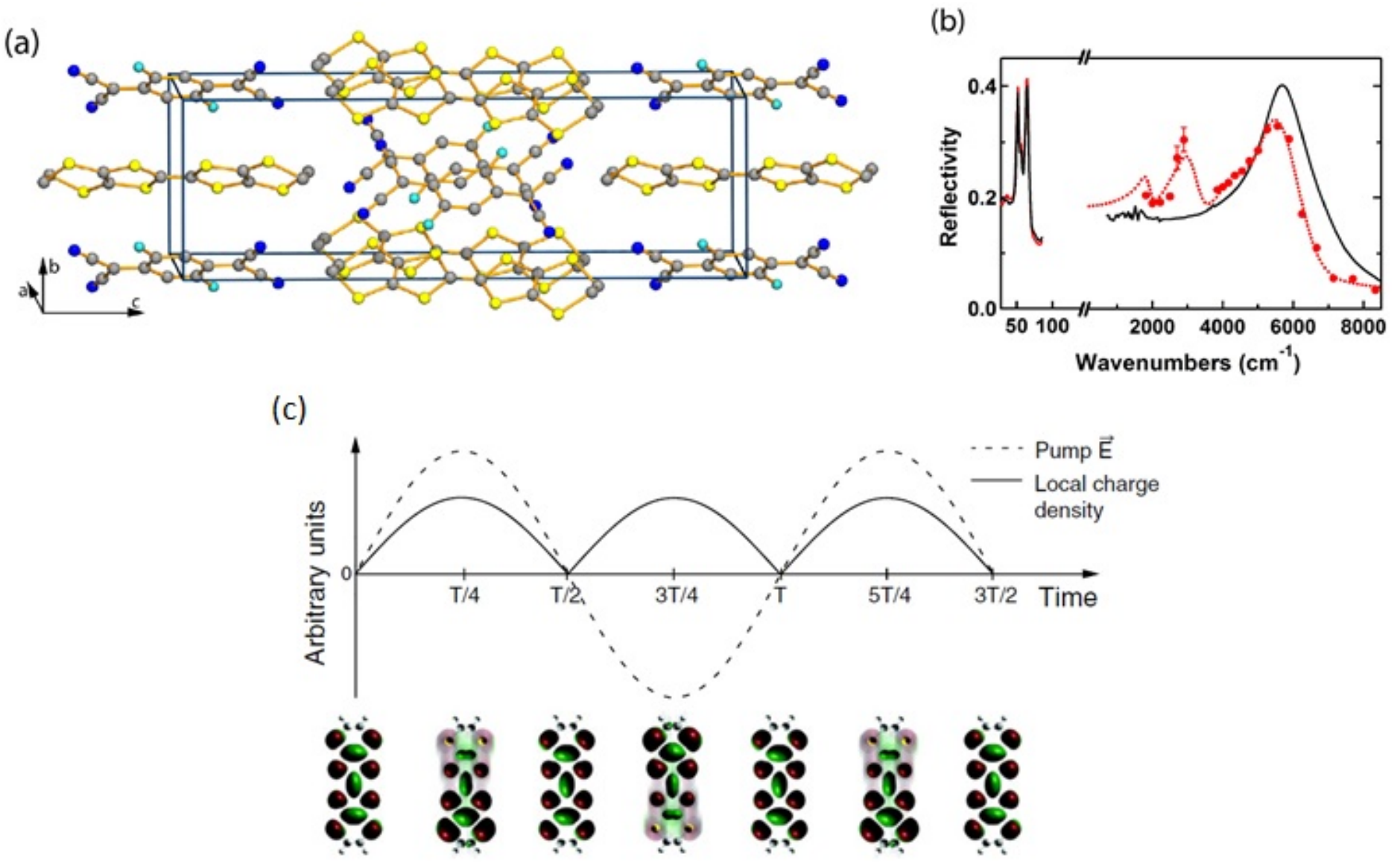

Fig. 19. (a) Crystal structure of ET-$F_2$TCNQ. The ET molecules form a one-dimensional, half-filled Mott-insulating chain along the $a$ axis. (b) Frequency-dependent reflectivity at equilibrium (black) and at the peak of the modulating infrared field (red). Full circles indicate experimental data, whilst the dashed line is a lorentzian fit to the data. In the low-frequency range, full lines indicate the equilibrium (black) and transient (red) reflectivity, displaying two unscreened phonon modes of the molecular crystal. (c) Temporal evolution of the pump electric field (dashed black line) together with the resultant change in the local charge density (solid black line) and the corresponding orbital motion of the vibrationally excited ET molecule over time. Panels (a) and (b) reprinted from [24]. Panel (c) reprinted with permission from [25]. Copyright 2015 by the American Physical Society.

What differentiates molecular solids from complex oxides (like cuprates or manganites) is the presence of a wider variety of IR-active structural modes, which are all accessible with nonlinear optical techniques, thus offering unprecedented possibilities for control experiments. These modes are located in well-separated spectral ranges: Collective phonons are found only at very low frequencies (~1 THz), while localized molecular vibrations are observed in the MIR.

In two subsequent experiments [24,25] one of these localized vibrations of the ET molecule was resonantly driven in ET-$F_2$TCNQ with strong MIR fields (up to 10 MV/cm) tuned to ~10 μm wavelength. This vibrational excitation applied a time-dependent deformation of the valence orbital wave function, thus acting on the local charge densities and modulating the Hamiltonian parameters introduced above. This situation is depicted schematically in Fig. 19(c).

The pump-induced reflectivity changes along the ET molecule chains ($a$ axis) were measured in the near-, mid-infrared, and terahertz range, with a time resolution of ~300 fs. A prompt shift of the charge transfer band to lower frequency was observed, along with new distinct peaks inside the gap (see Fig. 19(b)). No reflectivity change was detected for frequencies above the charge transfer resonance and no metallic response was found in the THz range, at variance with the Drude-like metallic response observed after above-gap optical excitation [12]. The excited state relaxed then rapidly (on a sub-picosecond timescale), most likely following the vibrational relaxation time.

A detailed understanding of the observed dynamics was achieved by considering additional terms in the Hubbard Hamiltonian, assuming in first instance that the classical vibrational mode coordinate $q_\ell$ only couples to the local charge density. This gives terms of the form $\hat{n}_{\ell\sigma} f(q_\ell) + \hat{n}_{\ell\uparrow}\hat{n}_{\ell\downarrow} g(q_\ell)$, where $f(q_\ell)$ and $g(q_\ell)$ are two functions not known a priori. Expanding the functions $f$ and $g$ as a series, one obtains:

$$\hat{H}_{e-vib} = \sum_\ell \hat{n}_\ell \left(A_1 q_\ell + \mathrm{A}_2 q_\ell^2 + \cdots\right) + \sum_\ell \hat{n}_{\ell\uparrow}\hat{n}_{\ell\downarrow}\left(B_1 q_\ell + \mathrm{B}_2 q_\ell^2 + \cdots\right)$$

Since the molecule is centrosymmetric and the vibrational mode has odd symmetry, linear terms in $q_\ell$ [151,152] vanish ($A_1 = B_1 = 0$). Every molecule is coherently driven with its coordinate in time τ described as $q_\ell(\tau) = C\ \sin(\Omega_{IR}\tau)$, being $C$ the driving amplitude and $\Omega_{IR}$ the vibrational mode's eigenfrequency. This implies that the $A_2$ term couples to the total density, resulting in an irrelevant global phase shift. One is then left with a quadratic coupling to the on-site Coulomb interaction of the form:

$$\hat{H}_{e-vib} = B_2 q_\ell^2(\tau)\hat{n}_{\ell\uparrow}\hat{n}_{\ell\downarrow} = (C/2)B_2[1 - \cos(2\Omega_{IR}\tau)]\hat{n}_{\ell\uparrow}\hat{n}_{\ell\downarrow}$$

Importantly, the coefficient $B_2 < 0$ because the vibration will, in general, cause the valence orbital to spatially expand. Thus, two effects are expected as long as an odd molecular mode is driven: a time-averaged reduction of the on-site repulsion together with its modulation at $2\Omega_{IR}$.

The former of the two effects, *i.e.* the reduction in $U$, was fully reproduced experimentally, as evidenced by the photo-induced red-shift of the charge transfer peak in Fig. 19(b). On the other hand, the frequency modulation at $2\Omega_{IR}$, which should appear within a classical model as sidebands at multiples of $\pm 2\Omega_{IR}$ on each side of the shifted charge transfer resonance, is only partially observed in the experiment, as sidebands are only seen to the red of the main peak. Only by considering a strong asymmetry between empty and doubly occupied sites, along with a fully quantum treatment of finite mass oscillators, it was possible to fully explain this observation and all experimental line shapes [24].

The $2\Omega_{IR}$ frequency modulation was also investigated directly in time domain in a second experiment [25]. Here, vibrational oscillations were excited by MIR pump pulses at 10 μm with a stable carrier-envelope phase, in which the temporal offset between the electric field and the intensity envelope is constant and locked over consecutive laser shots. Hence, every laser pulse drove the molecule with identical phase, allowing measurements with sub-cycle resolution over many pump pulses. The evolution of these oscillations was then monitored by a delayed probe pulse of ~10 fs duration, thus providing the necessary time resolution to observe coherent oscillations of the charge transfer resonance at ~1.7 μm wavelength.

Figure 20(a) reports the spectrally integrated time-resolved reflectivity changes, showing fast oscillations at approximately 70 THz (corresponding to $\sim 2\Omega_{IR}$) with an amplitude of ~1%. Note that these oscillations are convolved with the 10 fs duration of the probe pulse, longer than their 8-fs half period, thus reducing their amplitude. In the same figure, the response after deconvolution, now exhibiting stronger oscillations, is also shown.

Further insight was possible by analyzing the spectrally resolved signal (Fig. 20(b)). This evidenced a reduction in the total reflectivity, a red shift of the charge transfer peak (consistent with the observations reported in [24]), as well as oscillations with the same 70 THz frequency detected in the spectrally integrated measurements of Fig. 20(a). The spectra were then modeled with Kramers-Kronig (KK) consistent variational dielectric function fits [153], yielding the time- and frequency-dependent optical conductivity.

A detailed analysis of the conductivity line shapes [154,155] crucially allowed to identify the presence of $2\Omega_{IR}$ oscillations only in the time dependent on-site Coulomb repulsion $U(\tau)$, and not in the nearest-neighbor term $V(\tau)$. This finding implies that the on-site Hubbard $U$ can be selectively modulated by the excitation of the local molecular vibration, at variance with what reported to date for excitation of collective phonon modes. This result opens up new routes for coherent control of interactions in many-body system, a task to date only possible with cold-atom optical lattices [156].

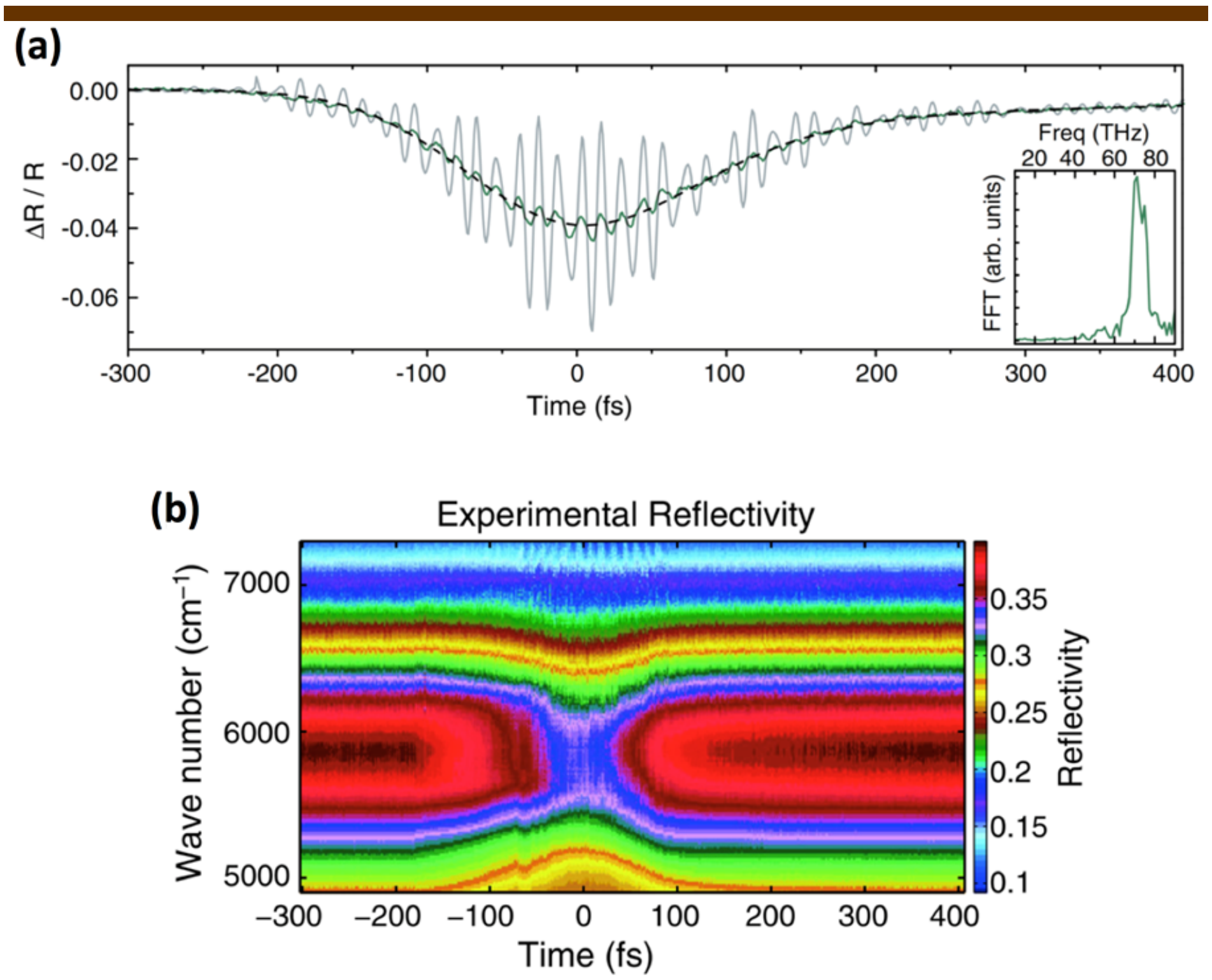

Fig. 20. (a) Spectrally integrated time-dependent reflectivity changes measured in ET-$F_2$TCNQ with a pump fluence of 0.9 mJ/cm$^2$ at room temperature. Data are shown along with a double exponential fit (dashed black) and the deconvolved reflectivity changes (solid gray). The inset shows a Fourier transformation of the measured oscillations peaking at 70 THz. (b) Corresponding frequency-resolved reflectivity as a function of pump-probe time delay. Reprinted with permission from [25]. Copyright 2015 by the American Physical Society.

### 4.2. Possible light-induced superconductivity in $K_3C_{60}$

A similar driving of local molecular vibrations has also been applied to $K_3C_{60}$, a molecular solid crystallized in a face-centred cubic structure (Fig. 21(a)), where each $C_{60}^{3-}$ ion contributes three half-filled $t_{1u}$ molecular orbitals to form narrow bands [157]. These electronic states give rise to superconductivity at equilibrium below $T_c$ = 20 K [158], mediated by a combination of electronic correlations [159] and local molecular vibrations [160].

In this experiment [26], femtosecond MIR pulses were tuned between 6 μm and 15 μm (a spectral region where different local molecular vibrations are present) and used to excite $K_3C_{60}$. The resulting changes in THz-frequency reflectivity and optical conductivity were determined at different pump-probe time delays and normalized to the absolute equilibrium optical properties measured on the same sample.

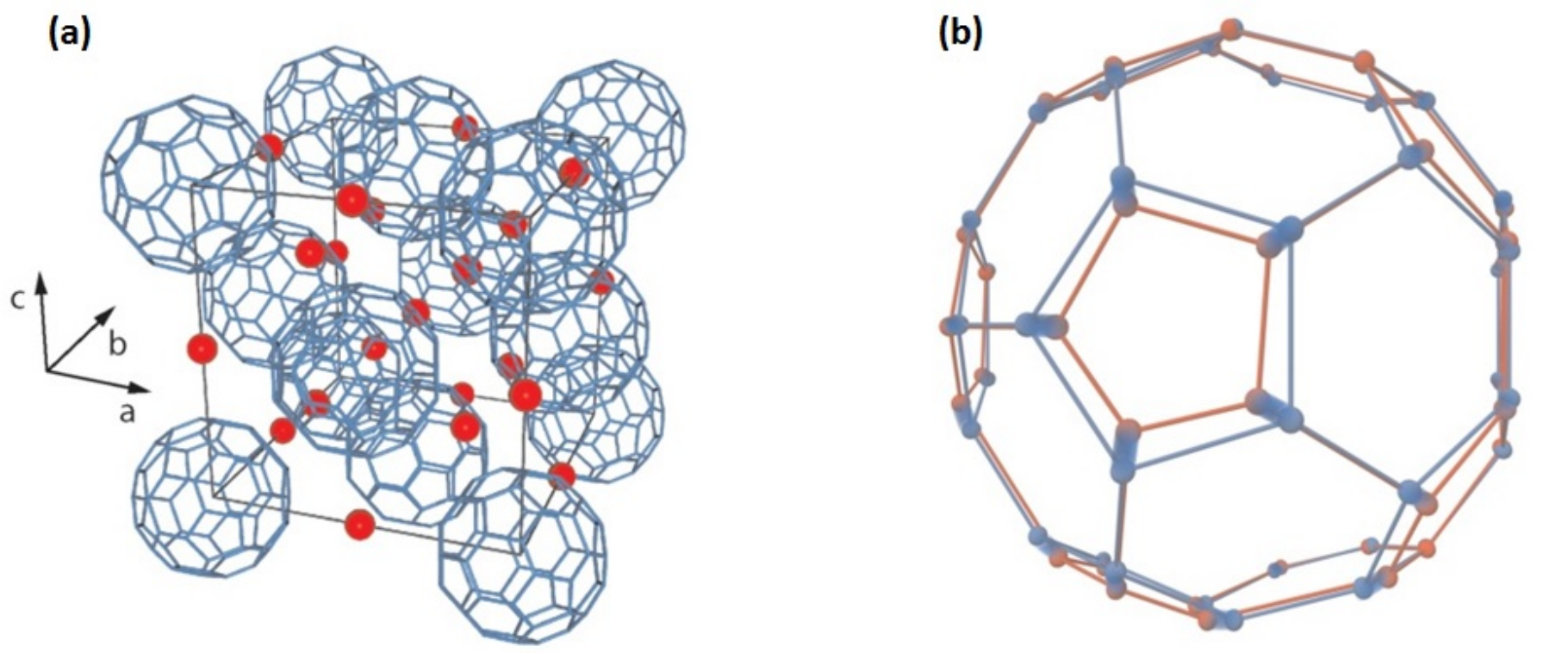

Fig. 21. (a) Face-centered cubic unit cell of $K_3C_{60}$. Blue bonds link the C atoms on each $C_{60}$ molecule. K atoms are represented as red spheres. (b) $C_{60}$ molecular distortion (red) along the $T_{1u}$ vibrational mode coordinates. The equilibrium structure is displayed in blue. Reprinted from [26].

In Fig. 22 we show the transient response (reflectivity and complex conductivity) measured in metallic $K_3C_{60}$ at high temperature after MIR excitation at 7 µm (resonant with the $T_{1u}$ molecular mode displayed in Fig. 21(b)), along with the superconducting response measured at equilibrium below $T_c$.

For all temperatures between $T_c$ = 20 K and T = 100 K, photo-excitation induced transient changes in the optical properties very similar to those observed when cooling at equilibrium: At 1 ps pump-probe time delay, a saturated reflectivity at R=1, a gapped $\sigma_1(\omega)$, and divergent $\sigma_2(\omega)$ were observed. Remarkably, all transient optical spectra could be fitted with the same function used for the low-temperature superconductor at equilibrium [26].

The non-equilibrium fits evidenced a gap in $\sigma_1(\omega)$ nearly twice as large as the superconducting gap at equilibrium. When the same experiment was repeated at higher base temperatures, the effect progressively disappeared, although a sizeable enhancement in carrier mobility could be measured up to room temperature.

One possible interpretation for the data above posits that excitation of molecular vibrations can coherently stimulate a transition from the high-temperature metallic state into a non-equilibrium superconducting phase. For the strong optical fields used in the experiment (~1.5 MV/cm) and from the polarizability of the $T_{1u}$ molecular vibrations, one can estimate large oscillatory distortions of the C-C bonds amounting to several percents of their equilibrium bond lengths. Hence, in analogy with the nonlinear phononics scenario introduced in Section 2.1, the large amplitude excitation is expected to extend beyond linear response and to deform the structure of the molecule along other, anharmonically coupled coordinates. To lowest order, these couplings are described by $q_{T1u}{}^2Q$ terms in the nonlinear lattice Hamiltonian [19,57], where $q_{T1u}$ is the directly driven mode coordinate, and $Q$ is the coordinate of any distortion contained in the irreducible representation of $T_{1u} \times T_{1u}$. A sizeable distortion is expected along normal mode coordinates of $H_g$ symmetry. Because $H_g$ modes are thought to assist pairing at equilibrium [157], it is possible that a quasi-static $H_g$ molecular deformation might favour stronger superconductivity, for example by causing an increase in the electron-phonon coupling [26].

In analogy with what reported for ET-$F_2$TCNQ in Section 4.1, excitation of local $T_{1u}$ vibrations is also expected to modulate local electronic correlations. An order-of-magnitude estimate for this effect has been extracted from frozen atomic motions for the $T_{1u}$ mode, predicting an increase in $U$ as high as ~10% for the orbitals orthogonal to the vibrational [26]. Such large asymmetric change would unbalance the occupancy of the three $t_{1u}$ orbitals and possibly interplay with the dynamical Jahn-Teller coupling, which is known to contribute to superconductivity near equilibrium [157].

However, as in the case of cuprates reported in Section 3, the interpretation proposed above is not unique. Firstly, some other transient non-superconducting states would also be consistent with the optical properties revealed by this experiment (for example a large reduction in carrier scattering rate without condensation or a sliding, non-commensurate charge density wave).

Secondly, the MIR pulses used to excite the sample couple here not only to the vibrations but also to the metallic carriers. Therefore, the observed dynamics might also be originated by a direct modulation of the local charge density, without involving molecular vibrations.

In the balance, regardless of the specific mechanism, these data are indicative of striking emergent physics away of equilibrium and open many new opportunities and challenges for the use of local vibrations to control molecular solids.

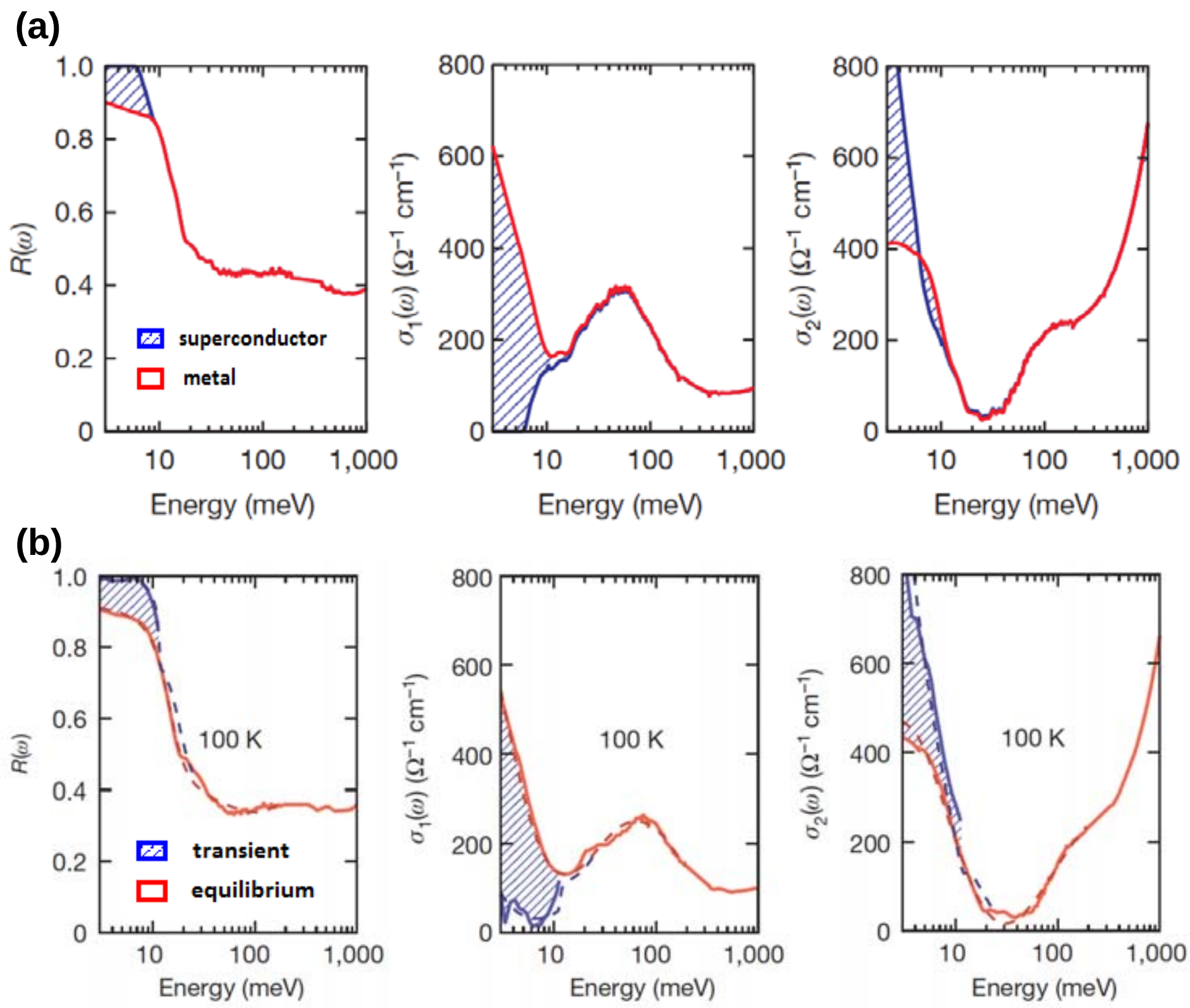

Fig. 22. (a) Equilibrium reflectivity and complex optical conductivity of $K_3C_{60}$ measured at T = 25 K (red) and T = 10 K (blue). (b) Same quantities measured at a base temperatures T = 25 K at equilibrium (red) and 1 ps after MIR excitation (blue). Reprinted from [26].

## 5. Material control with single- and sub-THz pulses

Strong transient fields can nowadays be routinely generated at single-THz and sub-THz frequencies using table-top femtosecond laser sources [17] or Free-Electron Lasers [161]. These pulses have typical duration between 1 and 10 ps, they can have constant carrier-envelope phase, and their electric- and magnetic-field amplitudes can exceed 1 MV/cm and 0.3 T, respectively [162]. When such transient fields impinge on complex matter, they interact with various degrees of freedom, thus providing a versatile handle for control. For instance, they can couple to lattice [163] or spin [164] degrees of freedom, they can non-resonantly excite or accelerate electrons (possibly driving phenomena like switching of orbital order [165] or insulator-to-metal transitions [166]), or they can directly couple to collective excitations of superconducting condensates [28,29]. In this Section we focus on some of the most recent highlights achieved in the control of complex materials using such single- or sub-THz pulses.

### 5.1. Terahertz control of magnetism

The magnetic component of intense THz transients has been shown to enable ultrafast control of magnetic degrees of freedom through the so-called "Zeeman interaction". When a THz pulse impinges on a material surface, the magnetic field $\vec{B}(t)$ associated with it exerts a torque $\vec{m} \times \vec{B}(t)$ on any permanent magnetic dipole moment $\vec{m}$ in the solid, as for instance that carried by the electron spins. In materials displaying magnetic ordering, the spins are oriented along a direction determined by the effective internal magnetic field $\vec{B}_{int}$. An external transient magnetic field $\vec{B}(t)$ can deflect the spins from their equilibrium orientation, thus determining an additional torque $\vec{m} \times \vec{B}_{int}$, which leads then to spin precession [27].

In most ferromagnetic materials, collective spin modes are typically found at frequencies well below 100 GHz. For this reason, THz magnetic field transients have allowed so far only non-resonant spin control [167,168]. In such experiments, in order to maximize the Zeeman torque, the THz magnetic field was set perpendicular to the unperturbed sample magnetization. The induced spin dynamics was then probed by magneto-optical Kerr effect (MOKE), *i.e.* by measuring the induced rotation in the polarization of an optical probe pulse.

A representative magnetization trace, measured on a 10-nm thick Co film [168] is displayed in Fig. 23(a). Before THz excitation, an external magnetic field was used to drive the macroscopic magnetic moment to saturation. The excitation pulse, with a frequency spectrum centered at 2.1 THz, carried ~1.5 optical cycles with a field amplitude of ~40 mT. Apart from a quarter period phase shift, corresponding to a delay of ~50 fs, the MOKE signal (red curve) exhibited a variation that was phase-locked and close to identical to the driving THz magnetic field (blue dots). This magnetization dynamics, fully dominated by the phase of the THz stimulus, does not rely on resonant excitation and can be very well understood by simple classical precession (see simulated black curve in Fig. 23(a)).

At variance with ferromagnets, spin control in antiferromagnetic materials is far more challenging, as these typically consist of two spin-ordered sublattices with opposite magnetization, resulting in a zero net magnetic moment. Despite this cancellation persists also when a fast external magnetic field modulates the sublattice magnetization, THz spin control of antiferromagnets can still rely on the enhanced spin-light coupling at magnetic resonances, as for instance spin waves (magnons).

A first experiment of resonant excitation of spin waves was performed on the prototypical antiferromagnet NiO [164]. A 45-μm-thick NiO crystal was illuminated with an intense THz transient covering a spectral range from 0.1 to 3 THz (with an estimated peak magnetic field of 40 mT), which excited resonantly a magnon at 1-THz frequency. The subsequent spin dynamics was then monitored by an optical probe pulse, which measured the Faraday rotation, *i.e.* the transient circular birefringence driven by the induced magnetization. As shown in Fig. 23(b), the THz magnetic pulse triggered a spin precession with a period of 1 ps, corresponding exactly to the 1-THz magnon. Due to the relatively long lifetime of the spin wave (~30 ps), control of the dynamics was possible over an extended time window. For example, the spin precession could be switched off almost completely by using the torque of a second THz magnetic pulse arriving precisely 6.5 oscillation cycles after the first [164].

In both experiments discussed above, the electromagnetic pulses were shown to couple directly to magnetism via the Zeeman torque induced by the magnetic field [164,168,169]. Such direct excitation has the advantage of minimal excess heat deposition but is on the other hand very challenging due to the low magnetic field strengths of currently realizable THz sources.

Thanks to the coexistence of different ferroic orders, multiferroics offer new interesting possibilities for controlling magnetism [170], for example by application of electric fields. While optical pulses have already been shown in the past to affect the magnetic structure of multiferroics on femto- and picosecond time scales [171], recently strong few-cycle THz pulses, tuned to resonance with an *electromagnon*, were successfully used to transiently modify the magnetic structure of multiferroic $TbMnO_3$ [172]. Electromagnons are electric-dipole active spin excitations directly connected to the magnetoelectric coupling [173] and show up as a broad peak in the THz spectrum of $TbMnO_3$ (with the strongest feature at 1.8

THz). The spin motion resulting from resonant THz excitation (peak electric fields up to 300 kV/cm), was investigated with time-resolved resonant soft x-ray diffraction at the Mn $L_2$ edge by measuring a specific $(0q0)$ reflection.

At temperatures deep in the multiferroic phase, the x-ray signal showed oscillations resembling the shape of the THz pump-pulse electric field (see Fig. 23(c)). As expected in case of spin motion driven by the electric field, the delay between THz pump and x-ray signal corresponded exactly to half of a single oscillation cycle. In addition, a sign inversion of the pump electric field resulted in an opposite sign of changes in the diffraction intensity transients.

Remarkably, the observed modulation of the diffraction peak intensity was over an order of magnitude larger than that expected for unconstrained spin precession driven directly by the magnetic field component of the THz pulse [172], revealing the great potential of multiferroic materials for this kind of application.

A recent progress in the field of ultrafast control of magnetism has been achieved by generalizing the concept of nonlinear phononics (see Section 2) to drive atomic rotations in the rare-earth orthoferrite $ErFeO_3$. At variance with all other experiments described in this Section, here pump pulses at shorter wavelengths (in the MIR range) were used to excite pairs of non-degenerate phonon vibrations [174]. The induced breaking of time-reversal invariance through this lattice rotation was shown to mimic the application of a magnetic field and manifested itself in the excitation of large amplitude spin precession [174].

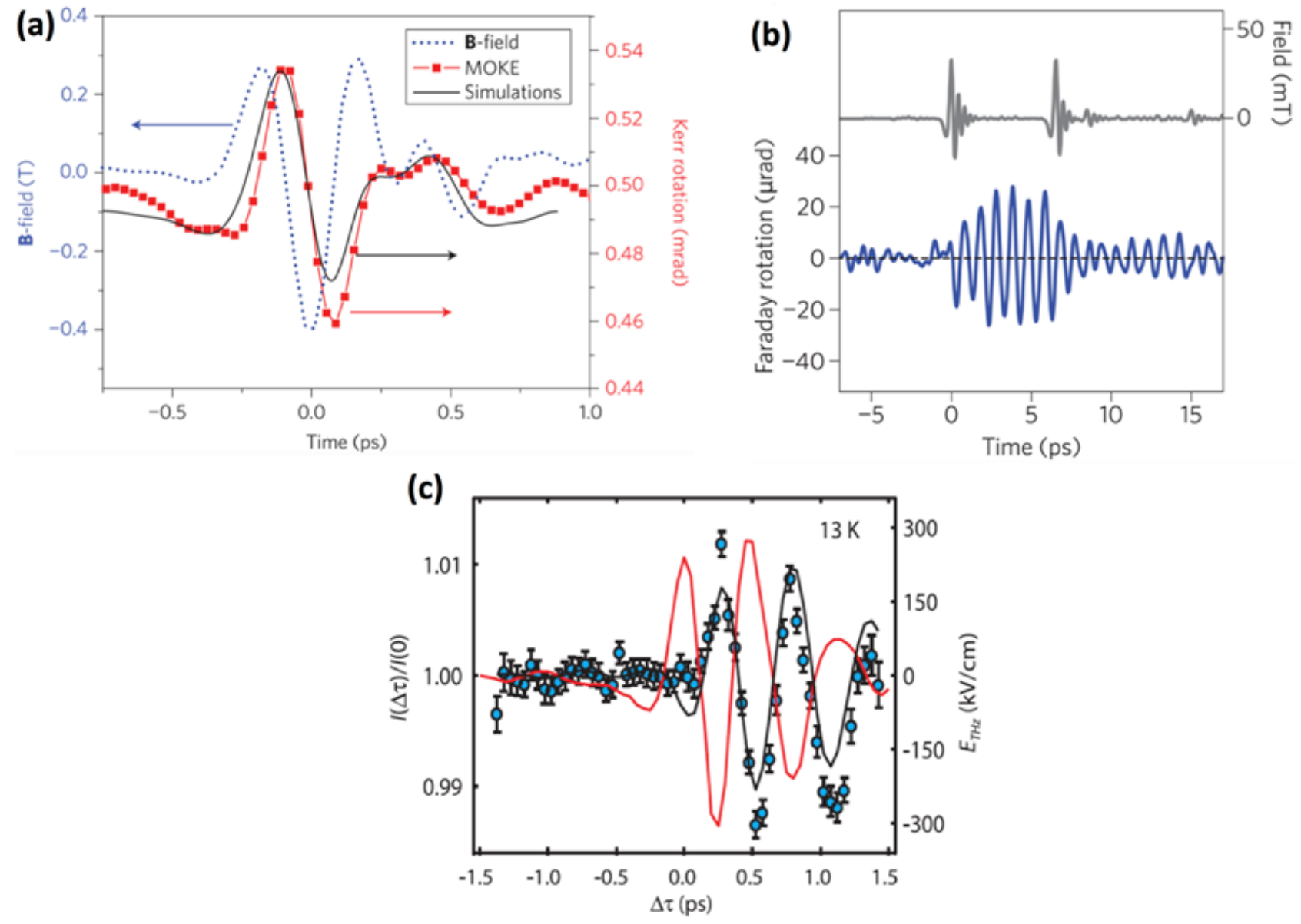

Fig. 23. (a) Femtosecond magnetization dynamics in 10-nm thick Co, represented by the MOKE signal (red curve) and initiated by single-cycle THz magnetic field (blue dotted curve). The black line indicates the simulated response [168]. (b) Magnetization dynamics (blue curve) measured in antiferromagnetic NiO after excitation with a pair of 0.5-ps, 40-mT THz transients (grey curve). The applied pulse sequence is shown to switch a 1-THz spin wave on and off [164]. (c) Magnetic diffraction intensity of the $(0q0)$ peak of $TbMnO_3$ (blue symbols, left axis) compared with the electric field of the pump trace (red solid line, right axis) as a function of time delay, measured in the multiferroic phase (T = 13 K) [172]. Panel (a) reprinted by permission from Macmillan Publishers Ltd: Nature Photonics [168], copyright 2013. Panel (b) reprinted by permission from Macmillan Publishers Ltd: Nature Photonics [27], copyright 2013. Panel (c) from [172]. Reprinted with permission from AAAS.

## 5.2. Non-resonant strong field control

Most of the examples of THz control presented above take advantage of resonant coupling to some material degrees of freedom, thus allowing for coherent accumulation of the energy deposited by the driving THz field in a selected mode. In addition to such resonant manipulation, high-field THz sources allow even non-resonant nonlinear control, as they can accelerate electrons to very high energies over just half a cycle of their driving field. For instance, for free electrons shone with a 1-THz pulse of 0.3 MV/cm peak field, the *ponderomotive* energy reaches values as high as 1 eV, which is comparable with typical bandgap energies [27]. In a certain amount of experiments, such non-resonant THz excitation has been successfully applied to a variety of semiconducting materials, being able to drive field-dependent processes including nonlinear manipulation of excitons [175], electron mass renormalization [176], and high-harmonic generation by dynamical Bloch oscillations [177].

More recently, non-resonant THz field control has also been applied to strongly correlated materials, being able, for example, to induce an insulator-to-metal transition in vanadium dioxide by enhanced THz fields in metamaterial structures [166]. Vanadium dioxide ($VO_2$) displays a temperature-driven insulator-to-metal transition at T = 340 K in which both the lattice and on-site Coulomb repulsion play an important role. In this experiment, aiming at intense THz excitation, metamaterial structures in the form of split-ring resonators [178] (essentially sub-wavelength LC circuits) were deposited on a $VO_2$ film, serving as local resonant THz concentrators (see Fig. 24(a)). It was shown that the THz electric field inside the 1.5-μm $VO_2$ capacitive gap could be enhanced by more than an order of magnitude, from the ~300 kV/cm free-space value to ~4 MV/cm (see Fig. 24(b)) [166].

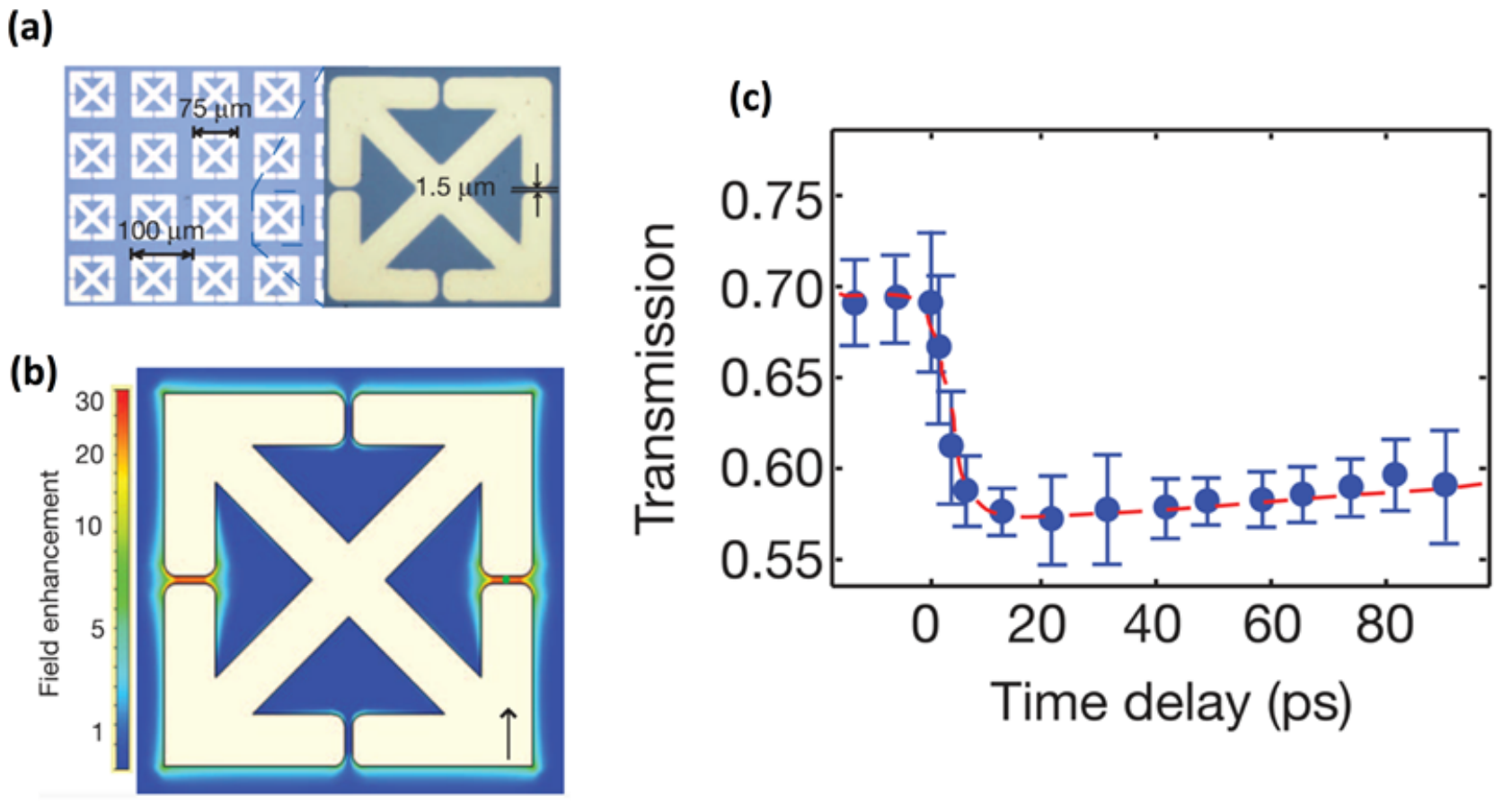

Fig. 24. (a) Optical image of metamaterial split-ring resonators deposited on $VO_2$/sapphire. (b) Simulated resonant THz field enhancement as a function of position in the split-ring resonator structure. (c) THz transmission pump-probe scan taken at ω = 0.8 THz (*i.e.*, at the metamaterial dipole resonance) for a sample temperature T = 324 K and a THz pump field of 1 MV/cm. Red curve is an exponential fit. Reprinted by permission from Macmillan Publishers Ltd: Nature [166], copyright 2012.

Figure 24(c) shows the time-dependent transmission of a weak-field THz probe pulse that was variably delayed with respect to the high-field THz pump pulse at an in-gap field strength of ~1 MV/cm. The observed changes are compatible with an increase in the $VO_2$ conductivity towards that of the metallic phase. The measured buildup time of the response (~8 ps) is comparable with that observed in optical-pump THz-probe experiments and consistent with a percolative phase transition [179,180]. The pump-induced changes persisted with little or no relaxation for at least 100 ps. In addition, when the same experiment was repeated at the maximum achievable in-gap THz fields (~4 MV/cm), irreversible damages to the $VO_2$ metamaterials were observed.

This work demonstrates that metamaterial-enhanced high-field THz pulses can be used to study correlated-electron materials in a non-perturbative regime. The technique is extremely versatile, and can potentially be used to study THz-induced phase transitions in other correlated materials and transition metal oxides.

### 5.3. Control over superconducting order

Resonant THz control has also been applied to the weakly bound electronic states of superconducting condensates, the so-called Cooper pairs, which are characterized by typical binding energies in the single-THz range. Selective breaking of these pairs by intense THz transients has been recently shown for the classical BCS superconductor NbN [181].

In this experiment, the ultrafast dynamics of the BCS gap, $2\Delta$ = 1.3 THz was mapped out after an impulsive excitation of quasiparticles. By using intense THz pulses with peak electric fields in excess of 50 kV/cm, the photoinjection of high-density quasiparticles as much as $10^{20}$ /cm$^3$ was realized. A measurement of the transient THz optical conductivity revealed that, after irradiation, superconductivity is rapidly switched off within the duration of the single-cycle THz pulse. This observation is in stark contrast with what measured upon excitation in the near-infrared or visible, where much longer buildup times (up to 20 ps) are found.

Indeed, pumping at ~eV photon energies ($\omega_{pump} \gg 2\Delta$) generates hot electrons, whose excess energy is transferred to a large amount of high-frequency phonons. These phonons cause in turn pair breaking, resulting in the suppression of superconductivity [182]. In contrast, the THz pump pulse injects low-energy quasiparticles directly, in absence of phonon emission from hot electrons, thus driving the observed ultrafast change of the BCS state.

A similar excitation scheme has been also applied to the doped superconducting compound $Nb_{1-x}Ti_xN$, with a gap $2\Delta \lesssim 1$ THz [28]. Remarkably, a transient oscillation was found to emerge in the THz response of this material (see Fig. 25(a)), whose frequency coincided with the value of the BCS gap energy [28].

As already theoretically predicted [183], when a BCS ground state is non-adiabatically excited by a short laser pulse, the coherence between different quasiparticle states should lead to oscillations of the order parameter (the so-called Higgs amplitude mode). Such non-adiabatic excitation requires a short pump pulse with the duration $\tau_{pump}$ small enough compared to the response time of the BCS state, which is characterized by the BCS gap $\Delta$ as $\tau_\Delta = \Delta^{-1}$ [28]. Here femtosecond optical pulses at ~eV photon energies are not ideal, because of the huge excess energies of photoexcited hot electrons. On the other hand, the single-cycle THz pulses used in the experiment, being resonant to the BCS gap and having a duration of $\tau_{pump}$~1.5 ps (corresponding to a ratio $\tau_{pump}/\tau_\Delta \sim 0.6$), are indeed suitable to observe such coherent dynamics.

The measured oscillation frequency was systematically mapped out as a function of pump fluence (Fig. 25(a)), displaying a scaling with the asymptotic value $2\Delta_\infty$ reached by the gap energy after its initial reduction due to THz-induced quasiparticle excitation [181] (Fig. 25(b)).

The coherent nonlinear interplay between collective modes and THz light fields were further investigated in the prototypical BCS superconductor NbN by measuring the real-time evolution of the order parameter under the driving field of multi-cycle (as opposed to single-cycle), narrowband THz pulses [29]. Photon energies of 0.3, 0.6, and 0.8 THz were used for excitation, all located below the NbN superconducting gap in the low temperature limit (1.3 THz).

The ultrafast dynamics of the superconducting order parameter driven by the multi-cycle pump pulse was then probed through the transmittance of a single-cycle THz probe pulse. An oscillatory signal at twice the frequency of the driving field emerged during the pump pulse irradiation (Fig. 25(c)). This $2\omega_{pump}$ oscillation of the order parameter was interpreted in terms of precession of Anderson's pseudospins [29,184].

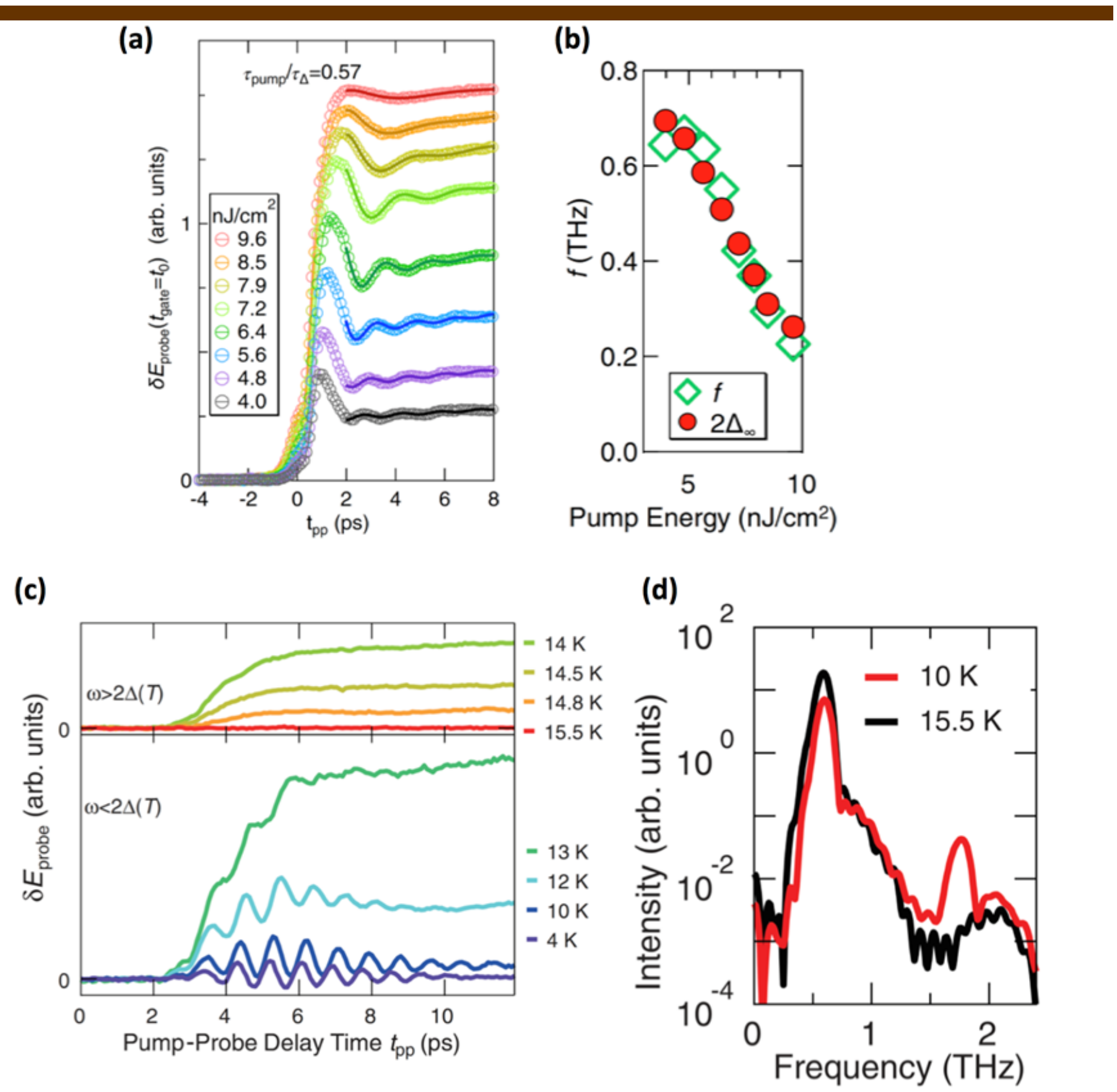

Fig. 25. (a) Temporal evolution of the change of the probe electric field in $Nb_{1-x}Ti_xN$ after excitation with single-cycle THz pulses at 4 K for different pump fluences. Oscillations with a ~ps period are apparent. Solid curves are fits to the data [28]. (b) Extracted oscillation frequency *f*, scaling with the asymptotic gap energy $2\Delta_\infty$ as a function of pump fluence [28]. (c) Time evolution of the change of the probe electric field in NbN after excitation with multi-cycle narrowband THz pulses ($\omega_{pump}$ = 0.6 THz), displaying clear oscillations at $2\omega_{pump}$ = 1.2 THz [29]. (d) Corresponding power spectrum of the transmitted THz pump pulse measured below (10 K) and above (15.5 K) $T_c$ = 15 K. The below-$T_c$ trace shows a strong third harmonic generation at 1.8 THz [29]. Panels (a) and (b) reprinted with permission from [28]. Copyright 2013 by the American Physical Society. Panels (c) and (d) from [29]. Reprinted with permission from AAAS.

Because the coherent interaction between the superconductor and the THz electromagnetic radiation resulted in nonlinear oscillations of the order parameter, generation of higher odd-order harmonics in the transmitted pump THz pulse was also expected to occur. This was verified by nonlinear THz pump transmission on the same material, which indeed showed the appearance of a strong $3\omega_{pump}$ component (Fig. 25(d)). As reported in Section 6.3, analogous third order nonlinearities have also been recently observed high-$T_c$ cuprates driven with strong THz fields made resonant with Josephson plasma modes.

Importantly, the mechanism responsible for the coherent oscillations and third harmonic generation reported in [28,29] is currently under debate. While in [28,29,185] it was speculated that the Higgs mechanism was responsible for such phenomena, more recently it has been discussed that this dynamics is most likely triggered via a Raman process, whose cross section is far larger then that related to the Higgs mechanism itself [186,187,188].

Another remarkable result achieved upon THz driving is the observation of radiation-enhanced superconductivity [189]. This phenomenon had been already discovered in a few type-I superconductors (Al, Sn) upon excitation at microwave frequencies [190,191], between the inelastic scattering rate and the superconducting gap frequency 2Δ. Utilizing intense, narrow-band, picosecond terahertz pulses from a Free-Electron Laser, tuned to just below and above 2Δ of the BCS superconductor NbN, it was demonstrated that the superconducting gap can be transiently increased also in a type-II dirty-limit superconductor. The effect was found to be particularly pronounced at higher temperatures and was attributed to radiation-induced non-thermal electron distribution, persisting on a 100 ps time scale [189].

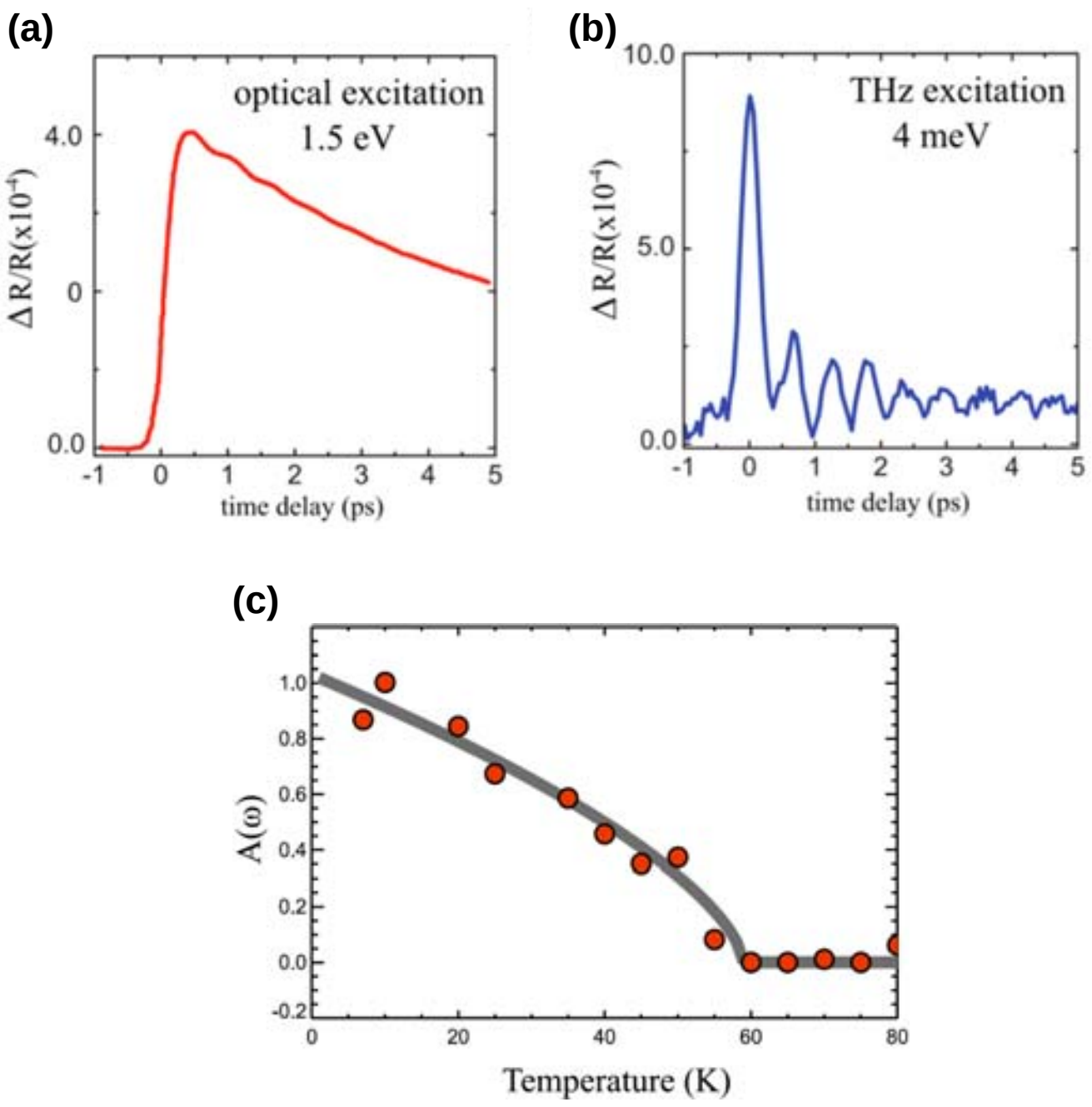

Fig. 26. (a) Time evolution of the 800-nm reflectivity change of YBCO after optical excitation at 15 K [194]. (b) Time trace of the 800-nm reflectivity change due to a THz excitation at 10 K [192]. Optical excitation was performed with a fluence of ~1.5 μJ/cm², resulting in a large number of photoinduced quasiparticles (large background in figure). On the contrary, THz stimulation used a fluence of ~40 μJ/cm², showing large oscillations with minimal quasiparticle generation. (c) Amplitude of the Fourier mode at 1.8 THz, $A(\omega = 1.8\,\text{THz})$, measured upon THz excitation as a function of temperature. Reprinted from [192].

As last example, we report here an experiment in which intense THz pulse excitation has been employed to coherently manipulate the interplay between superconductivity and charge-density-wave (CDW) order in the high-$T_c$ cuprate $YBa_2Cu_3O_x$ (YBCO) [192]. As already discussed in Section 3.2, the presence of an incommensurate CDW state in bulk underdoped YBCO was recently experimentally verified [134,135]. The observed enhancemement of CDW order when superconductivity was suppressed by a strong magnetic field [193] indicated that the CDW and the superconducting state are competing.

Optical pump experiments on CDW-ordered YBCO have demonstrated the excitation of coherent dynamics with a frequency of ~1.8 THz, corresponding to a soft phonon related to the CDW order [194]. However, the coherent oscillations were obscured by a large incoherent background (due to scattering of "hot" electrons), which decayed then on a time scale of a few picoseconds (Fig. 26(a)). In the same material, excitation with strong single-cycle THz pulses with peak fields up to ~400 kV/cm allowed to directly access electronic states near the Fermi energy at the nodal point of the $d$-wave superconductor, thus inducing large ~1.8 THz oscillations with almost negligible incoherent background (Fig. 26(b)) [192].

Remarkably, the temperature-dependent dynamics was also found to be very different from that induced by optical pump. Firstly, while the collective oscillations induced by optical excitation persisted to temperatures well above the superconducting $T_c$, all the way to $T_{CDW}$, the THz-induced collective dynamics disappeared slightly below $T_c$ (Fig. 26(c)). Secondly, unlike optical excitations, neither the frequency nor the phase of the oscillations shifted as a function of temperature. These results demonstrate how THz excitation provides a unique mechanism for driving this superconductor out of equilibrium via interaction with the CDW state. The observed coherent dynamics may be ubiquitous among the underdoped cuprates and reveal perspectives for investigating the interplay between superconductivity and CDW.

In the next Section we will discuss extensively how strong THz pulses can be used not only to couple to the amplitude of the superconducting condensate wave function (*i.e.* to the density of Cooper pairs), but also to its macroscopic phase. This is achieved by stimulation of Josephson plasma modes in high-$T_c$ cuprates.

## 6. Josephson Plasmonics

In the normal state of high-$T_c$ cuprates, charge transport is metallic within the $CuO_2$ planes, while the electrodynamics along the *c* axis (*i.e.* along the direction perpendicular to the planes) is highly incoherent, displaying an insulating behavior with absence of any plasma response from the normal carriers. Below $T_c$, *c*-axis coherent transport is established by the superconducting condensate, which supports interlayer excitations of the superconducting order parameter phase in terms of collective plasma oscillations of the Cooper pairs. These oscillations occur at a specific frequency and become visible in the form of the so-called Josephson Plasma Resonance (JPR) in the 0.1 – 3 THz frequency range (see also Section 3). In this Section, after a general introduction to the physics of Josephson Plasma Waves (JPWs), we review a series of experiments in which high control capability of interlayer transport in high-$T_c$ cuprates has been achieved by driving these Josephson plasma modes at single- and sub-THz frequencies.

### 6.1. Josephson physics in high-$T_c$ cuprates

High-$T_c$ cuprates can be modeled as quasi two-dimensional materials with superconducting $CuO_2$ planes weakly coupled via Josephson tunneling through insulating barriers (Fig. 27(a)). The superconducting order parameter $\psi(x, y, z, t)$ is appropriately described by considering a constant order parameter amplitude throughout the material $\Delta(x, y, z, t) = \Delta$ and a set of phase parameters $\{\varphi_n(x, y, t)\}_{n=1,\dots,N}$, each of them representing the phase difference of a Josephson junction made of two neighboring superconducting layers [195].

To achieve a formal description of the phase dynamics in a stack of Josephson junctions, one should start from the case of a single junction made of two bulk superconductors separated by a thin insulating layer. The two Josephson relations are given by: $\frac{\partial\varphi}{\partial t} = \frac{2e}{\hbar}V(t)$ and $I = I_c \sin(\varphi)$, $V(t)$ being the voltage drop between the two superconductors and $I_c$ being the critical current of the junction. In *long* Josephson junctions, where the spatial dependence of the phase has to be taken into account as well, the two Josephson relations are supplemented with the Maxwell's equations and $\varphi(x, t)$ evolves according the so-called sine-Gordon equation, which reads in one dimension [196]:

$$\frac{1}{\omega_p^2}\frac{\partial^2\varphi}{\partial t^2} - \lambda_J^2\frac{\partial^2\varphi}{\partial x^2} + \sin(\varphi) = 0$$

Here $\lambda_J$ is the Josephson penetration depth, $\omega_p$ the JPR frequency, and $c_0 = \lambda_J\omega_p$ is the so-called Swihart velocity (the propagation velocity of JPWs).

For a stack of $N$ Josephson junctions, a similar derivation can be performed, and, consequently, the dynamics of the phases $\{\varphi_n(x, t)\}_{n=1,\dots,N}$ is described by a set of coupled sine-Gordon equations with appropriate boundary conditions [197,198,199].

In the following, we outline most of the key physical phenomena encompassed by the sine-Gordon equation and we discuss how these manifest in the case of a stack of Josephson junctions.

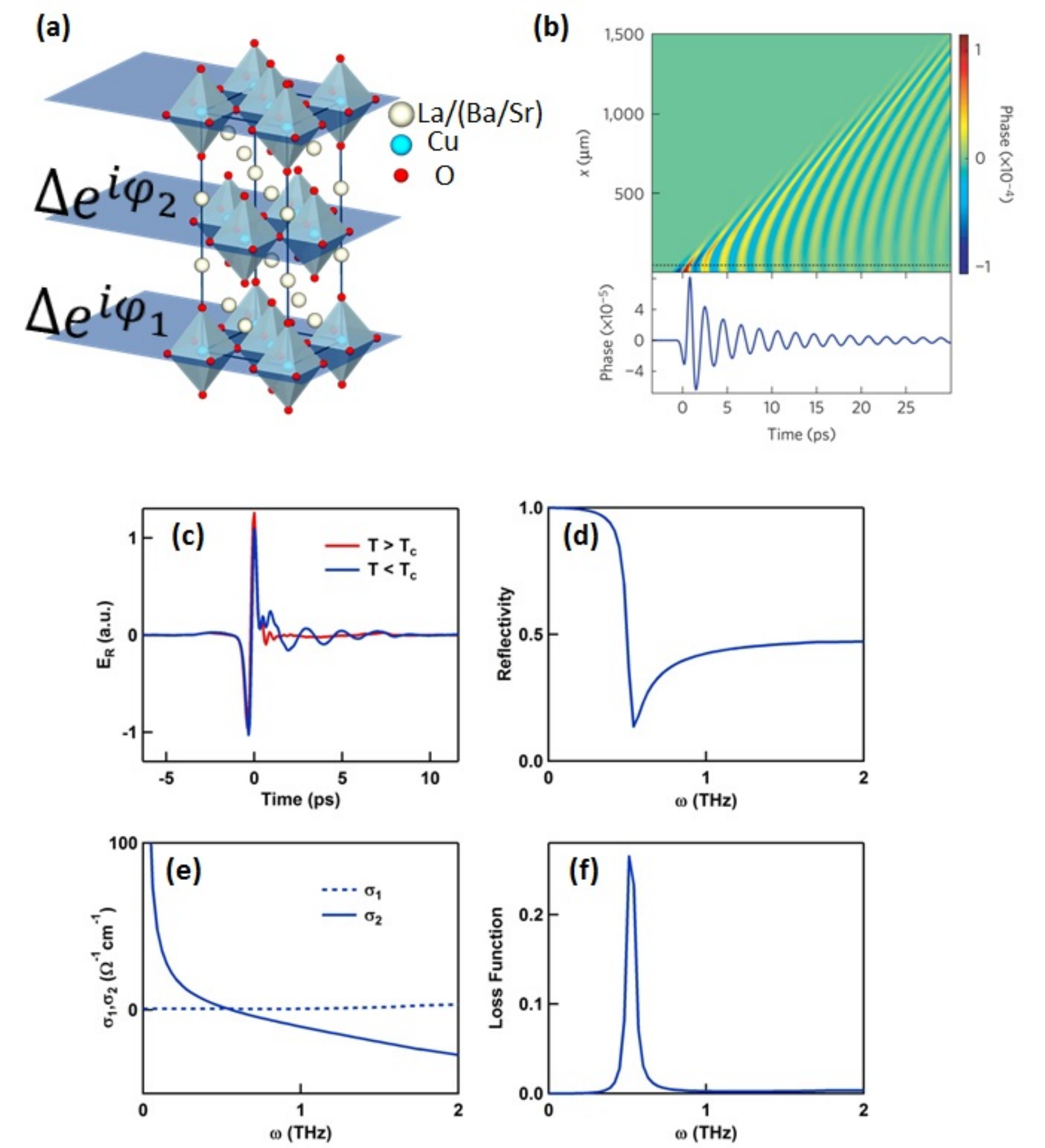

Fig. 27. (a) Crystal structure of a high-$T_c$ cuprate represented as a stack of Josephson junctions. The superconducting order parameter is described by considering a constant amplitude Δ and a phase $\varphi_i$ which is different for each $CuO_2$ layer. (b) Simulation of the Sine-Gordon equation for a time dependent probe field at the surface of the superconductor. The top panel is a color plot of the Josephson phase as a function of time and depth into the material, while the lower panel is the corresponding time evolution for a line cut close to the surface boundary. (c) Reflected THz probe field, (d) absolute reflectivity, (e) complex optical conductivity, and (e) loss function of $La_{1.905}Ba_{0.095}CuO_4$, as determined via time-domain THz spectroscopy. Panel (b) reprinted from [32].

If one ignores in a first instance the time dependence, the Sine-Gordon equation writes:

$$-\lambda_J^2 \frac{\partial^2 \varphi}{\partial x^2} + \sin(\varphi) = 0$$

This expression describes the screening of an external magnetic field by the superconducting Josephson currents. For long junctions, in the linear regime where $\sin(\varphi) \sim \varphi$, the solution takes the form $\varphi \sim e^{-x/\lambda_J}$. In this case, the applied magnetic field is expelled from the junction. For a stack of junctions, this corresponds to the well-known Meissner effect. In nonlinear regime, when a sufficiently strong magnetic field is applied, the equation above describes the magnetic flux penetration in the form of Josephson vortices.

The dynamics of the Josephson effect in short junctions is also well described within a RLC circuit equivalent model [200]. This is derived by neglecting the spatial dependence in the Sine-Gordon equation, and by applying a DC bias current $I$ and a damping term $\beta$:

$$\frac{1}{\omega_p^2}\frac{\partial^2\varphi}{\partial t^2} + \beta\frac{\partial\varphi}{\partial t} + \sin(\varphi) = \frac{I}{I_c}$$

Here the junction is modeled by a nonlinear inductance $L$ (describing the Josephson coupling between the superconducting layers), a capacitance $C$ (the capacitive coupling between the superconducting layers) and a resistance $R$ (due to the finite conductivity of residual quasiparticles), all in parallel.

For $I < I_c$, the equation has a time-independent solution, $\varphi(t) = \varphi_0$, and the junction stays in the zero resistance state, where superconducting carriers "short" the normal carriers. On the other hand, for $I > I_c$, time dependent states appear, corresponding to the so-called AC Josephson effect. Here the junction develops a finite voltage $V_{DC}$ and, according to the first Josephson relation, the phase evolves as $\varphi(t) = \frac{2e}{\hbar}V_{DC}t$. Consequently, $I(t) = I_c\sin(\frac{2e}{\hbar}V_{DC}t)$ oscillates at a frequency $\frac{2e}{\hbar}V_{DC}$.

The JPR at $\omega_p$ can be detected with a probe electric field polarized along the $c$ axis, with its wave vector parallel to the $CuO_2$ layers. In this case, the different Josephson phases of the stack are excited all in phase ($\varphi_n(x,t) = \varphi(x,t)$ for all $n$) and their dynamics behaves as in the case of a single Josephson junction:

$$\frac{1}{\omega_p^2}\frac{\partial^2\varphi}{\partial t^2} - \lambda_c^2\frac{\partial^2\varphi}{\partial x^2} + \varphi = 0$$

The first two terms of this equation describe wave propagation at a velocity given by $\omega_p\lambda_c = c/\sqrt{\epsilon_\infty}$, *i.e.* the speed of light in the material ($\epsilon_\infty$ is the dielectric constant of the insulating barrier at THz frequencies). The last term introduces a cut-off frequency, $\omega_p$, for the wave propagation.

Indeed, for plane waves of the type $\varphi \sim e^{i(k_x x-\omega t)}$, the relation dispersion takes the form: ${k_x}^2 = \frac{\epsilon_\infty}{c^2}(\omega^2 - \omega_p^2)$. At $\omega < \omega_p$, $k_x$ becomes imaginary and wave propagation inside the material is forbidden. On the other hand, at $\omega > \omega_p$ JPWs can propagate inside the material.

As already discussed in Section 3 the response of JPWs can be probed by optical spectroscopy at THz frequencies [201]. The excitation of linear JPWs in response to a broadband THz pulse, whose frequency content overlaps with the JPR, is shown in Fig. 27(b). Therein, the evolution of the Josephson phase $\varphi(x,t)$ is shown on a color scale as a function of time $t$ and depth $x$ inside the material. We observe that JPW propagation is strongly dispersive and is bounded in the $(x,t)$ plane by an upper line, corresponding to the light cone (*i.e.* the speed of light in the material, $c/\sqrt{\epsilon_\infty}$).

Experimentally, a quasi-single-cycle THz probe pulse (polarized along the $c$ axis), measured after reflection from the superconductor, shows clear oscillations in its trailing edge, which are absent above $T_c$ (Fig. 27(c)). This reflects the excitation of JPWs inside the material. The corresponding reflectivity spectrum (Fig. 27(d)) exhibits a well-defined edge at $\omega_p$. As already discussed above, for all $\omega < \omega_p$ THz waves are fully reflected, while for $\omega > \omega_p$ they are partially transmitted.

As for the complex optical conductivity (Fig. 27(e)), its real part $\sigma_1(\omega)$ is fully gapped, whereas the imaginary part $\sigma_2(\omega)$ diverges at low frequencies as $\sim S/\omega$, due to the inductive nature of the response of superconducting carriers. The parameter $S$ quantifies the strength of the interlayer superconducting coupling and is proportional to the superfluid density. This quantity will be used in Section 6.2 to determine how excitation of a high-$T_c$ cuprate with intense single-cycle THz pulses allows, under certain conditions, to modulate in time on ultrafast time scales the dimensionality of the superconductor.

Another way to visualize the JPR response is in term of the loss function $L(\omega) = -\mathrm{Im}(1/\varepsilon(\omega))$, $\varepsilon(\omega)$ being the dielectric permittivity (Fig. 27(f)). $L(\omega)$ peaks when the real part of the permittivity crosses zero (at $\omega = \omega_p$) and typically shows a lorentzian lineshape, whose width at quantifies the damping of the JPR due to quasi-particle excitations. Its departure from the exact lorentzian shape can reflect static or dynamic inhomogeneities of the Josephson coupling [202]. In Section 6.4 this concept will be used to describe Josephson plasma excitations in highly nonlinear regime.

### 6.2. Bi-directional gating of interlayer charge transport

Josephson coupling in cuprate superconductors can be tuned by means of static external perturbations, such as DC electric bias currents [203,204] or magnetic fields [205,206]. Here we report on the recent demonstration [30] of control of interlayer superconducting transport on an ultrafast time scale by means of strong THz field transients, in absence of injection of incoherent quasiparticles.

For this purpose, the external perturbation of the Josephson coupling had to be quasi-static (adiabatic), *i.e.* slow compared to the time scale of the JPR ($\sim 1/\omega_p$). This condition was achieved by applying a non-resonant THz driving, with its spectral content limited to frequencies lower than $\omega_p$. Such experiment was performed by exciting optimally doped $La_{1.84}Sr_{0.16}CuO_4$ ($T_c$ = 36 K, $\omega_p \approx$ 2 THz) with intense single-cycle THz pulses having a spectrum peaked at ~450 GHz and limited below 1.5 THz (Fig. 28), generated with the tilted pulse front technique in $LiNbO_3$ [17].

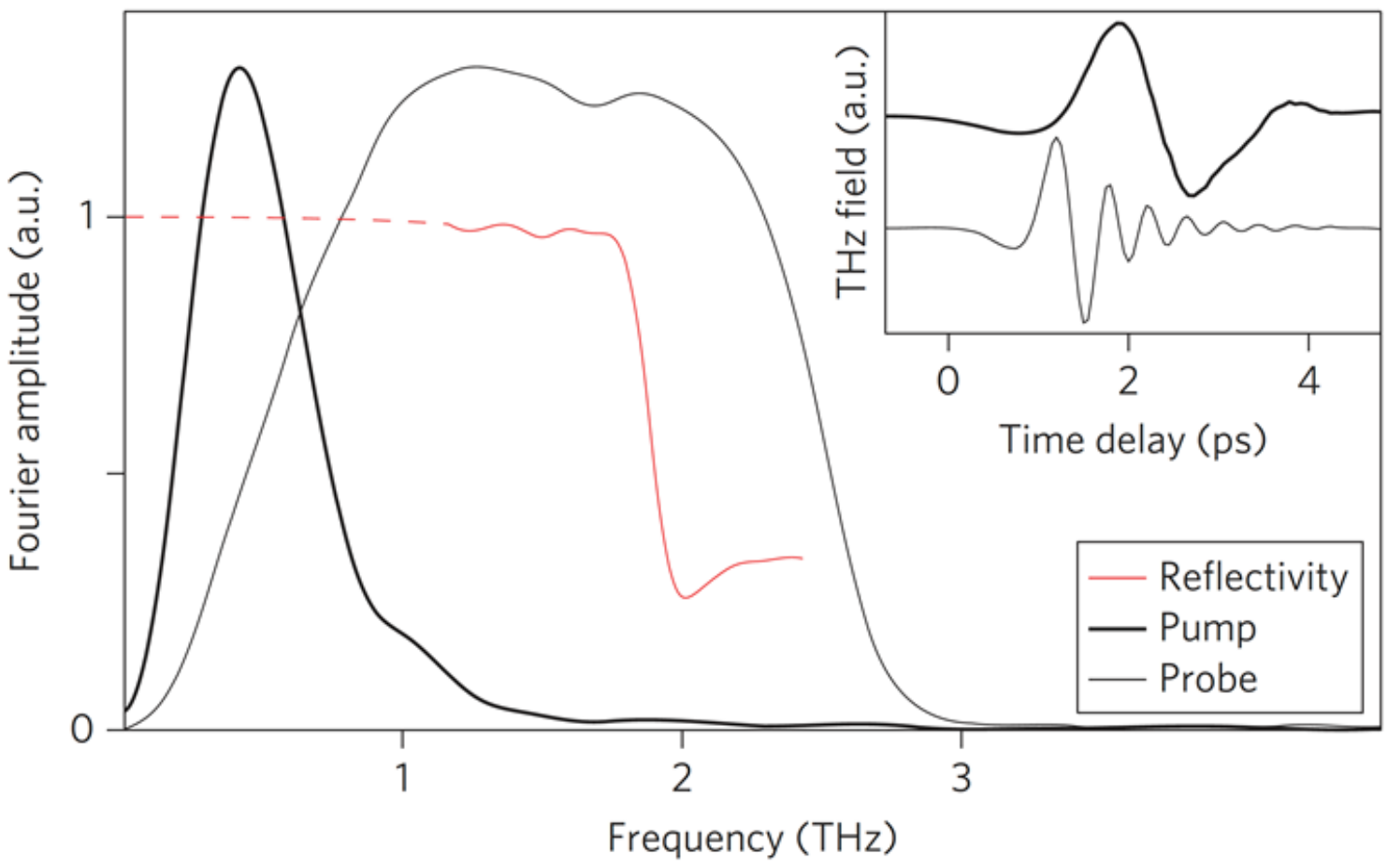

Fig. 28. THz pump (thick black) and probe (thin black) spectra, shown along with the reflectivity of $La_{1.84}Sr_{0.16}CuO_4$ (red curve). Corresponding time domain traces are displayed in the inset. Reprinted from [30].

The pump pulse, polarized along the $c$ axis, was fully reflected off the superconductor, penetrating only within the Josephson penetration depth $\lambda_c \sim 5$ μm as an evanescent wave. The transient complex conductivity of the perturbed material was determined as a function of pump-probe delay by time-resolved THz spectroscopy, for both in-plane and out-of-plane polarizations. As displayed in Fig. 29, $\sigma_{ab}(\omega, \tau)$ was almost unaffected by the excitation. This demonstrated that, in stark contrast with what observed upon optical [207] or in-plane THz pumping [192], no Cooper pair breaking and quasiparticle formation was observed under these excitation conditions.

On the other hand, the out-of-plane optical conductivity, $\sigma_c(\omega, \tau)$, appeared to be strongly modulated by the excitation. As a function of pump-probe time delay $\tau$, the low-frequency divergence of $\mathrm{Im}[\sigma_c(\omega, \tau)]$ was completely depleted and then recovered back, showing fast oscillations on a sub-picosecond time scale. For each delay $\tau$ where $\mathrm{Im}[\sigma_c(\omega, \tau)]$ was depleted, a metallic Drude peak appeared in $\mathrm{Re}[\sigma_{ab}(\omega, \tau)]$ from the flat and featureless equilibrium spectrum. Importantly, the observed conductivity modulation revealed a state in which the dimensionality of the superconductor became time dependent on an ultrafast time scale. This modulation occurred only for the duration of the pump pulse, in absence of any long-lived response.

The physics of the time-dependent Josephson coupling observed in this experiment is well captured by the two Josephson relations. From $\frac{\partial\varphi(\tau)}{\partial\tau} = \frac{2e}{\hbar}V(\tau)$ and $I(\tau) = I_c \sin(\varphi(\tau))$, one derives $\frac{\partial I(\tau)}{\partial t} = I_c \cos(\varphi(\tau))\frac{\partial\varphi(\tau)}{\partial t} \propto I_c \cos(\varphi(\tau))\, V(\tau)$. Out of equilibrium, the effective critical current of the junction is modulated as $I_c \cos(\varphi(\tau))$. Consequently, the Josephson coupling strength, defined as $\omega_p^2 \propto I_c$, is renormalized by the excursion of the phase $\varphi(\tau)$ as $\omega_p^2(\tau) = \omega_{p,0}^2 \cos(\varphi(\tau))$ ($\omega_{p,0}$ being the equilibrium JPR frequency).

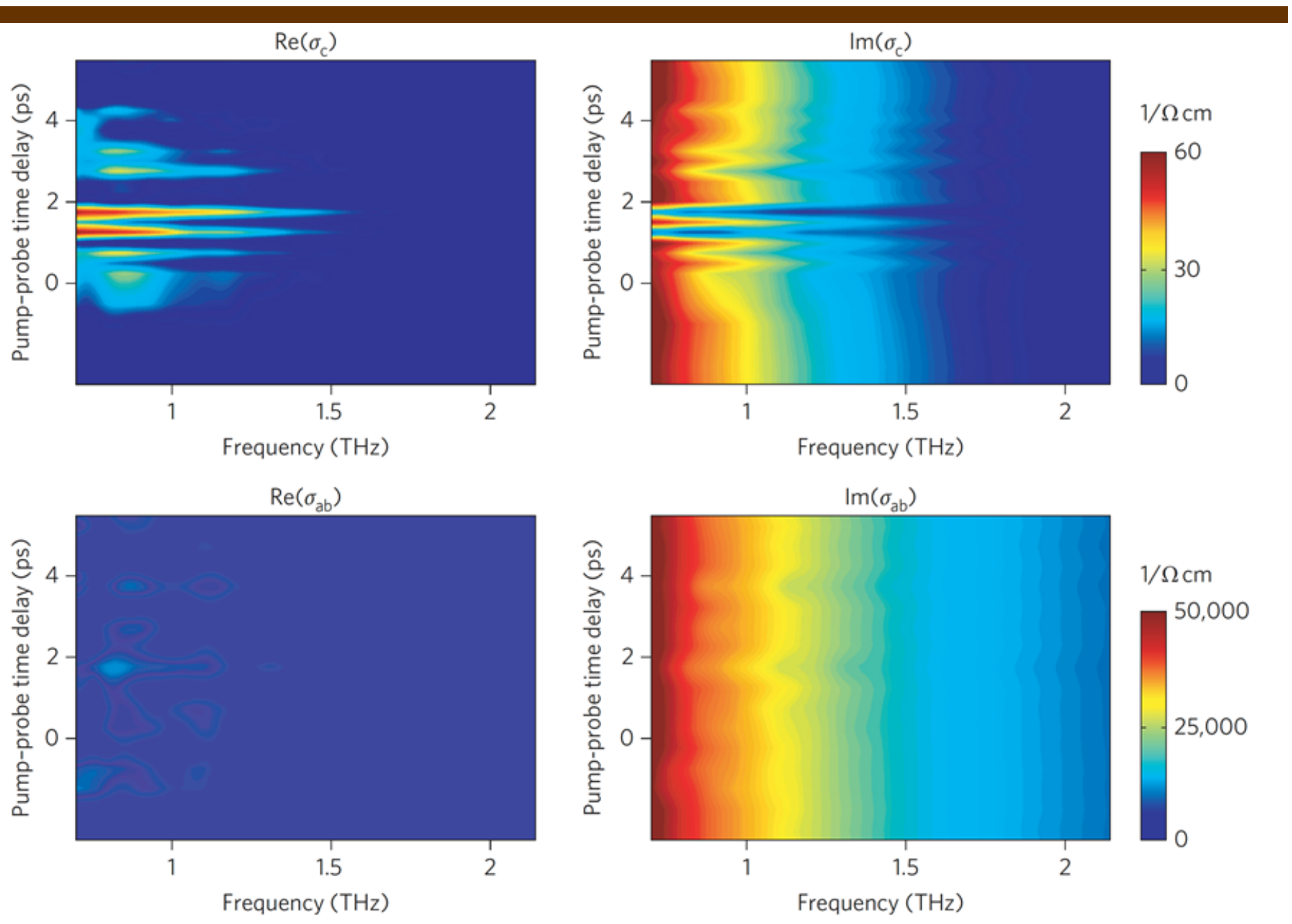

Fig. 29. Time-delay- and frequency-dependent complex optical conductivity of $La_{1.84}Sr_{0.16}CuO_4$ measured after THz driving, for polarization parallel and perpendicular to the $CuO_2$ planes. Reprinted from [30].

For low enough phase gradients ($\varphi(\tau) \approx 0$), $c$-axis transport occurs via the inductive response of the condensate (divergence in $\mathrm{Im}[\sigma_c(\omega,\tau)]$) and the Josephson coupling stays unaltered ($\omega_p^2(\tau) = \omega_{p,0}^2 \cos(\varphi(\tau)) \approx \omega_{p,0}^2$). On the other hand, for the strong phase gradients driven in the experiment ($\varphi(\tau) \approx \frac{\pi}{2}$), the Josephson coupling vanishes ($\omega_p^2(\tau) \rightarrow 0$). The inductive response is consequently "switched-off" and charge transport happens through normal quasiparticles (Drude peak in $\mathrm{Re}[\sigma_c(\omega,\tau)]$).

Figure 30 displays the advance in time of the integral of the electric field of the THz pump $\int_0^\tau E_{pump}(t)dt \sim \varphi(\tau)$ along with the measured strength of the inductive response $S(\tau)/S(\tau < 0)$, where $S(\tau) = \lim_{\omega\rightarrow 0} \omega\, Im[\sigma_c(\omega,\tau)] \propto \omega_p^2(\tau)$ . This quantity is well fitted by a function of the form $\left|\cos(\alpha \int_0^\tau E_{pump}(t)dt)\right|$ ($\alpha$ being a free fit parameter), thus demonstrating the modulation of the Josephson coupling strength introduced above. As the Josephson phase driven by the THz pump pulse reaches $\frac{\pi}{2}$, superconducting coupling vanishes, being then recovered when the phase is driven closer to $\pi$. This model is however not well suited to describe the dynamics at low field strengths, where the phase stays well below $\frac{\pi}{2}$ and superconducting transport dominates the optical properties.

The modulation frequency of the transport properties of the superconductor was been mapped out as a function of THz pump field strength [30], showing a linear scaling above a critical threshold field of ~75 kV/cm. This voltage-to-frequency conversion is a new demonstration of nonlinear THz physics, which is reminiscent of the AC Josephson effect.

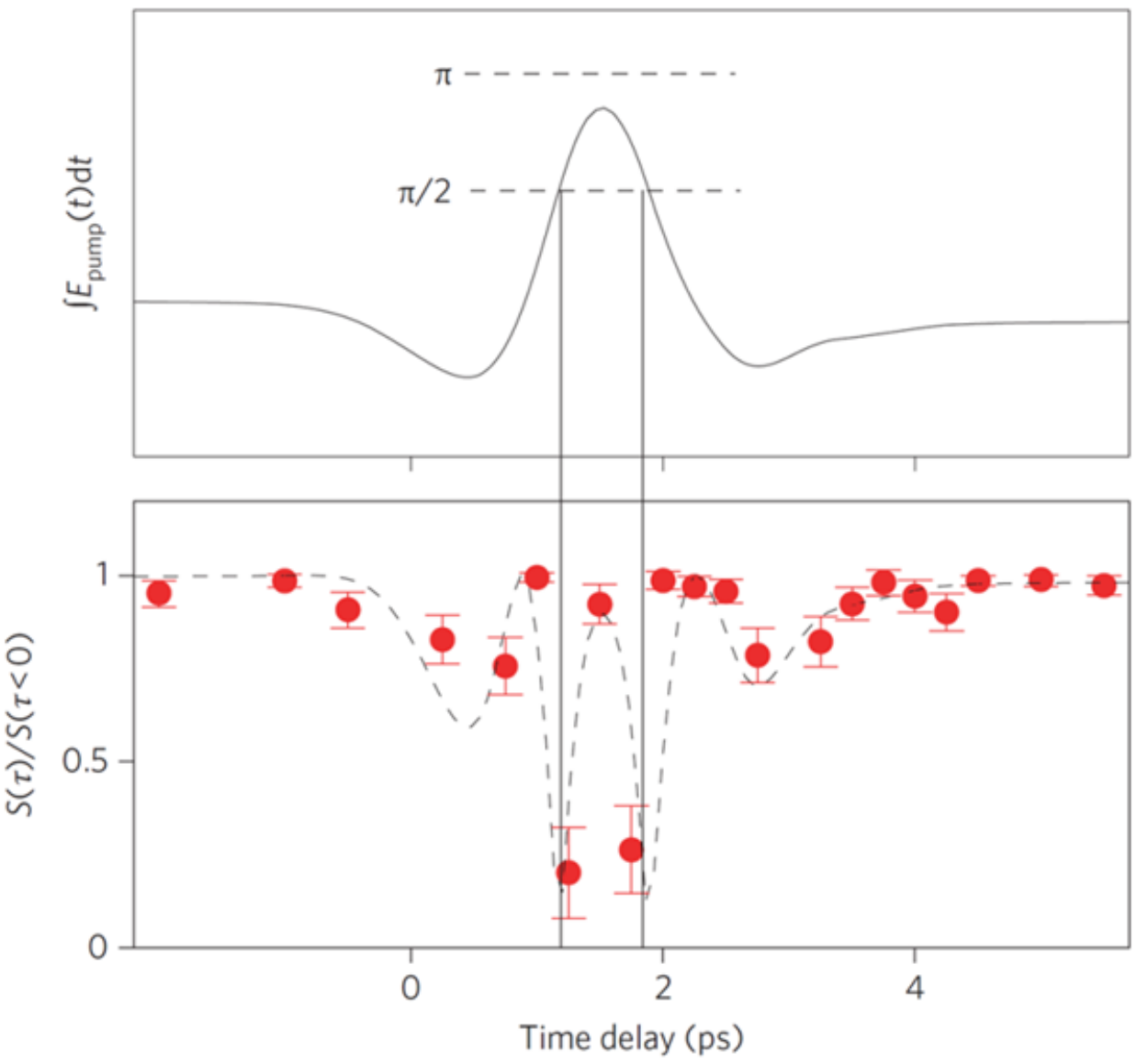

Fig. 30. Top panel: Time dependence of the integral of the experimentally measured THz electric field transient. Bottom panel: Corresponding evolution of the interlayer superconducting coupling strength, quantified by $S(\tau)/S(\tau < 0)$, where $S(\tau) = \lim_{\omega\to 0} \omega\, Im[\sigma_c(\omega,\tau)]$ (red dots). The dashed line is a fit to the data with the function $\left|\cos(\alpha \int_0^\tau E_{pump}(t)dt)\right|$. Reprinted from [30].

### 6.2. Parametric amplification of Josephson Plasma Waves

In the previous Section, we reported on the possibility to gate the interlayer superconducting transport in a quasi-static fashion. This was achieved by driving the Josephson phase with an intense, non-resonant THz pulse, whose frequency content was set to be entirely below the JPR frequency. Consequently, the THz field was screened within the Josephson penetration depth and propagation effects were neither expected nor observed.

When the frequency of the driving THz field comes into resonance with the JPR frequency, this adiabatic picture is no longer valid and new aspects related to the time and space dynamics of the phase are expected to come into play. Such scenario was explored recently [32] by stimulating with strong THz pulses another single-layer cuprate: $La_{1.905}Ba_{0.095}CuO_4$ ($T_c$ = 32 K), with $\omega_p \approx 500$ GHz (see optical properties reported in Fig. 27(c)-27(f)). The THz pump pulse spectrum (see Fig. 28), was in this case in resonance with the JPR.

In the weakly-driven regime, the THz pulse excites JPWs that propagate into the material (see Fig. 27(b)). These waves are linear because the phase excursion $\varphi(x,t) \ll \frac{\pi}{2}$ corresponds to the linear regime of the Josephson coupling: $\sin(\varphi) \sim \varphi$. Josephson plasma oscillations with a period of ~2 ps are therefore expected (Fig. 27(b)), corresponding to $\omega_p \approx 500$ GHz.

When the THz pump field is increased to few tens of kV/cm, thus accessing the strongly-driven regime, the phase excursion acquires values approaching $\sim\frac{\pi}{2}$, for which evident nonlinearities are expected. If in first instance one neglects the spatial dependence in the sine-Gordon equation, the dynamics of Josephson phase is equivalent to that of a classical pendulum in nonlinear regime, $\frac{1}{\omega_p^2}\frac{\partial^2\varphi}{\partial t^2} + \sin(\varphi) = 0$, with a resonance frequency $\omega_p$ that depends on the oscillation amplitude, red-shifting with increasing driving.

A full spatial and temporal simulation of the Josephson coupling strength in the strongly driven regimes is shown in Fig. 31(a). This was obtained by computing the relative Josephson coupling strength as $\omega_p^2(x,\tau)/\omega_{p,0}^2 = \cos(\varphi(x,\tau))$. While for weak driving this quantity remains almost unaffected, when the amplitude of the Josephson phase approaches values of the order of $\sim\frac{\pi}{2}$, the superconducting coupling almost vanishes.

This effect is similar to that reported in the previous Section, in which the superconducting layers could be decoupled at certain time delays $\tau$ such that $\omega_p^2(\tau) \approx 0$ when $\varphi(\tau) \approx \frac{\pi}{2}$. Nevertheless, the major difference here is represented by the long-lived nature of the modulation, which is due to the resonant excitation of the JPR.

In Fig. 31(b), the experimental proof of this phenomenon is reported. In response to the intense THz pump pulse, a modulation of the optical response around the JPR frequency was observed as a function of pump-probe delay. While the THz pump pulse was only ~1-2 ps long, the total duration of the induced dynamics was of the order of ~10 ps, hence showing a long-lived character.

The observed response presented two distinct types of signal: a fast oscillatory dynamics, superimposed to a rectified background. The oscillatory component could be better visualized in time domain after subtraction of the rectified part (Fig. 31(c)) and subsequent Fourier transforming (Fig. 31(d)). Both experiment and simulation clearly exhibited long-lived oscillations with a frequency of $2\omega_{p,0}$ ≈1 THz.

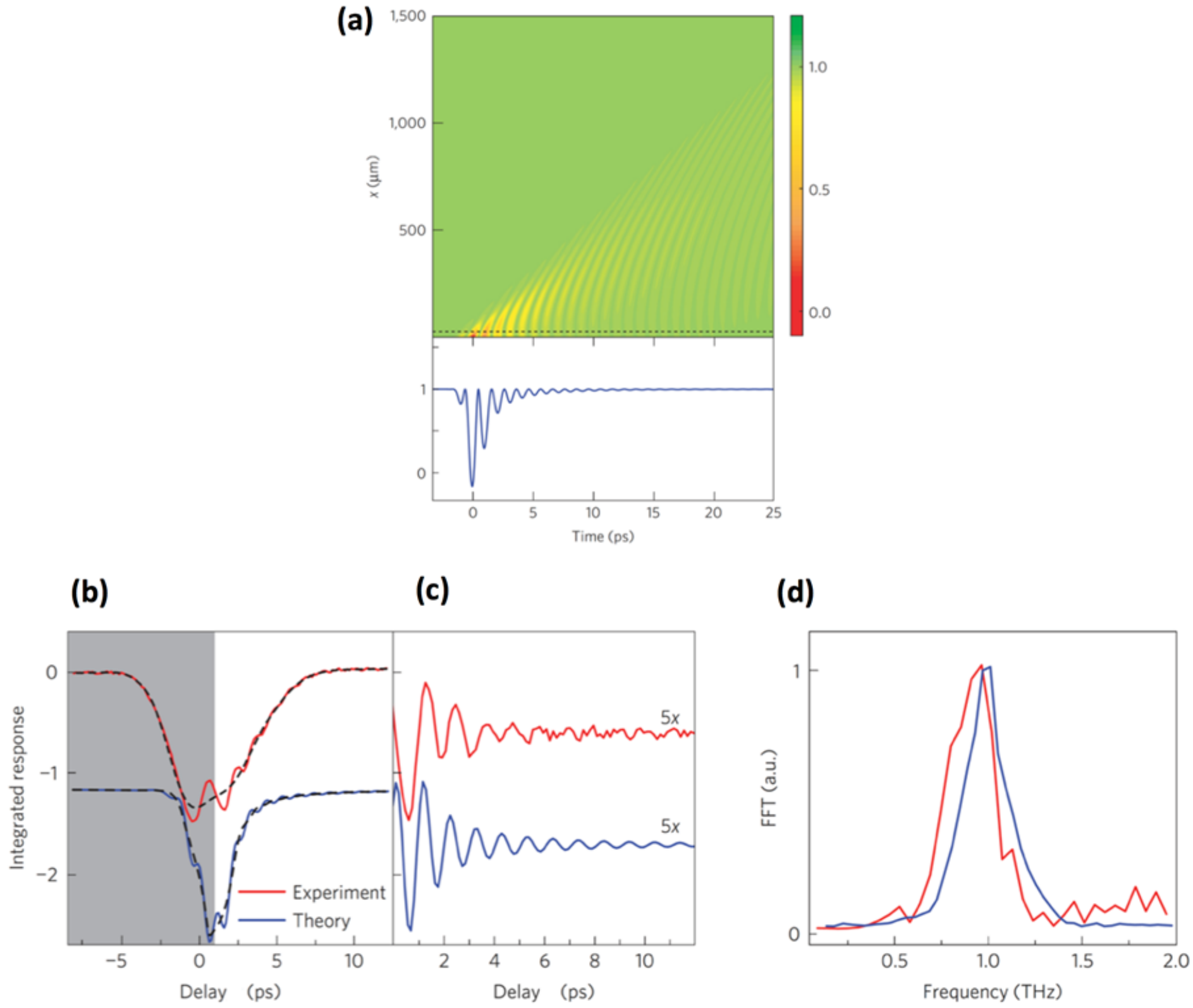

Fig. 31. (a) Change in the plasma oscillator strength, $\omega_p^2(x,\tau)/\omega_{p,0}^2$, induced by a strong THz pump field, as determined by numerically solving the sine-Gordon equation in nonlinear regime. The top panel is a color plot as a function of time and depth into the material, while the lower panel is the corresponding time evolution for a line cut close to the surface boundary. (b) Experimental (red curve) and simulated (blue curve) pump-probe response measured in $La_{1.905}Ba_{0.095}CuO_4$ as a function of pump-probe delay. The signal buildup region, affected by perturbed free induction decay, is shaded in grey. (c) Oscillatory component of the signals obtained by Fourier filtering the curves in (b). (d) Fourier transforms of the oscillatory signals shown in (c). Reprinted from [32].

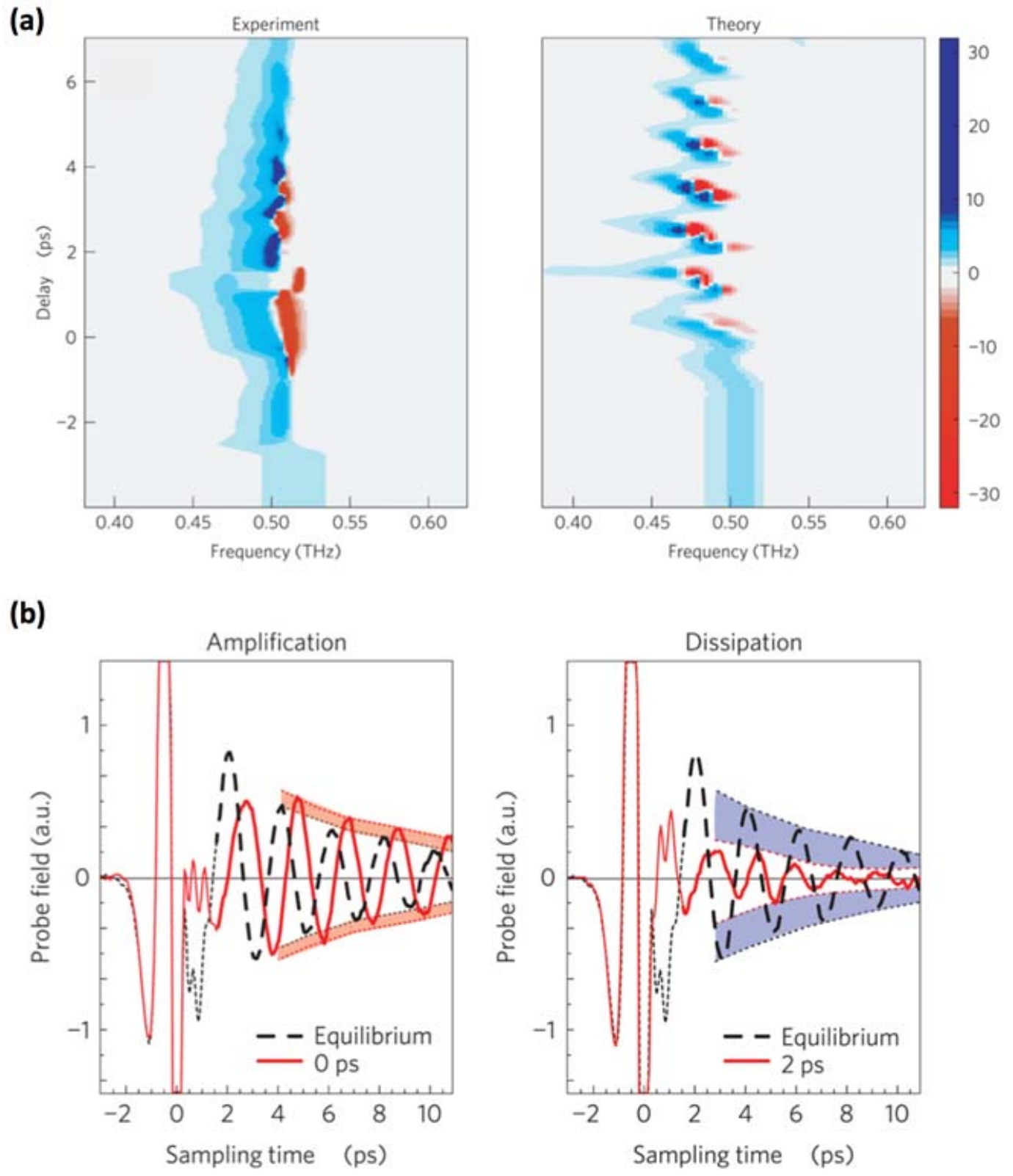

Fig. 32. (a) Time-delay and frequency dependent loss function of $La_{1.905}Ba_{0.095}CuO_4$ determined experimentally after THz excitation and by numerically solving the sine-Gordon equation in nonlinear regime. (b) THz probe field traces measured at selected pump-probe delays, shown along with the same quantity at equilibrium (pump off) to highlight amplification and dissipation. Reprinted from [32].

The nature of this response can be understood by noting that the modulation of the Josephson coupling in this nonlinear regime is such that $\omega_p^2(\tau) = \omega_{p,0}^2 \cos(\varphi_0 \cos(\omega_{p,0}\tau)) \approx \omega_{p,0}^2 \left[1 - \frac{\varphi_0^2}{4} - \frac{\varphi_0^2}{4}\cos(2\omega_{p,0}\tau)\right]$. This formula includes both the rectified component, $\omega_{p,0}^2 \left[1 - \frac{\varphi_0^2}{4}\right]$, and the $2\omega_{p,0}$ oscillations.

The most striking phenomenon associated with the parametric modulation of the JPR was observed after analyzing the frequency-resolved optical properties. In Fig. 32(a), the loss function $L(\omega, \tau)$ is displayed as a function of pump-probe delay $\tau$. In both experiment and simulations a strong reshaping, accompanied by fast oscillations between positive (blue) and negative (red) values, was found. Note that $L(\omega, \tau) < 0$ corresponds to a negative absorption coefficient and therefore to signal amplification. This effect is observed even more directly in the time domain. In Fig. 31(b), we report the THz probe pulses measured before and after excitation (at two selected pump-probe delays), showing either amplification or damping of the plasma oscillations.

These experimental observations could be fully understood within the model of Josephson phase dynamics outlined previously. If one considers the Josephson phase to be the sum of the phases driven by the pump and probe pulses, $\varphi(\tau) = \varphi_{pump}(\tau) + \varphi_{probe}(\tau)$ (with $\varphi_{pump}(\tau) = \varphi_0 \cos(\omega_{p,0}\tau)$), then, an effective equation of motion describing the dynamics of $\varphi_{probe}(\tau)$ can be derived [32]:

$$\frac{\partial^2 \varphi_{probe}(\tau)}{\partial t^2} + \omega_{p,0}^2 \left[1 - \frac{\varphi_0^2}{4} - \frac{\varphi_0^2}{4}\cos(2\omega_{p,0}\tau)\right] \varphi_{probe}(\tau) = 0$$

This equation of motion corresponds to a Mathieu equation. In this particular case where the natural frequency $\omega_{p,0}$ is driven at $2\omega_{p,0}$, it is expected to observe amplification or damping of the oscillator amplitude depending on the relative phase between the modulation and the oscillator.

The observations reported here demonstrate that terahertz JPWs can be parametrically amplified, exhibiting the expected sensitivity to the relative phase of strong and weak fields mixed in this process and the oscillatory dependence at twice the frequency of the drive. This effect is of interest for applications in photonics and may also potentially lead to control of fluctuating superconductivity [208,209], perhaps even over a range of temperatures above $T_c$ [105].

### 6.4. Optical excitation of Josephson plasma solitons

We report here on the observation of excitations of the Josephson plasma, which take full advantage of the spatial, temporal, and nonlinear properties of the dynamics allowed by the sine-Gordon equation [31]. Due to its sine type of nonlinearity, the sine-Gordon equation possesses particular solutions, which describe self-sustained wave packets preserving their shape while propagating into the medium, the so-called *solitons*. Solitons have been extensively investigated in conventional superconductors and stacks of Josephson junctions [210]. However, dynamical states involving solitons are in general created in the dissipative phase of the current-voltage characteristic or with quasi-static external perturbations.

A recent experiment demonstrated all-optical generation and detection of these modes, starting from the non-dissipative superconducting state and with rapidly varying electromagnetic fields impinging on the surface of the superconductor [31]. Close to the JPR frequency, the nonlinearity of the Josephson coupling is expected to manifest very strongly, even for modest driving fields. Experimentally, understanding how the sine nonlinearity manifests close to the JPR requires a narrowband THz pump pulse that can be tuned continuously in its vicinity. To achieve this condition, the experiment was performed at the Free Electron Laser FELBE (Helmholtz-Zentrum Dresden-Rossendorf). The superconductor under investigation was, as in the experiment described in Section 6.2, optimally doped $La_{1.84}Sr_{0.16}CuO_4$ ($T_c$ = 36 K, $\omega_p \approx 2$ THz).

For a narrowband THz pump pulse centered at a frequency $\omega_{FEL} = 1.1\omega_p$, one expects from simulation that linear (E = 9 V/cm) and nonlinear (E = 39 kV/cm) regimes do not differ significantly (see Fig. 33(a)). In both cases, the THz pump pulse would generate a wave packet that propagates into the material at the same group velocity (given by the slope of the wave packet in the $(x, t)$ panels of Fig. 33(a)). On the other hand, for $\omega_{FEL} = 1.05\,\omega_p$, tuned closer to the JPR, qualitative changes are more clearly observed (Fig. 33(b)). In the linear regime the wave packet propagates in a similar manner, while in nonlinear regime this excitation breaks into a train of more localized wave packets, which penetrate deeper into the material. This shows that the typical threshold electric fields for observing nonlinearities in the JPW propagation reduces as the THz excitation frequency is tuned closer to the JPR.

When the narrowband THz pump pulse is centered at frequencies $\omega_{FEL} = 0.97\,\omega_p$, that is below the JPR, for field strengths corresponding to the linear excitation regime, the impinging electromagnetic wave is evanescent and fully reflected off the superconductor (Fig. 33(c)). Closer to threshold, at E = 38 kV/cm, the wave is still reflected off after penetrating over longer distances into the superconductor, while right above threshold (E = 39 kV/cm), a propagating mode emerges, with a very low group velocity compared to the speed of light in the material. This mode is a Josephson plasma soliton, which concentrates the electromagnetic energy in space and time and carries it in a self-sustained fashion without any noticeable distortion of its shape. This soliton solution (called *breather*) corresponds to a vortex–antivortex pair that oscillates back and forth in time during its propagation, at a frequency $\omega_p$. The characteristics of this mode, such as its precise shape, speed, or oscillation frequency depend critically on the excitation conditions near this electric field threshold, as confirmed by the simulations run at E = 42 kV/cm, which show a soliton propagating at much higher speed. Also, multiple soliton modes can be obtained in a similar way if the pump intensity is increased to even higher values.

In order to detect experimentally the phase disturbance in the vicinity of the JPR frequency, the loss function $L(\omega, \tau)$ of $La_{1.84}Sr_{0.16}CuO_4$ was determined throughout the pump-induced dynamics under different excitation conditions. FEL THz pump pulses with ~25 ps duration and a narrow bandwidth of $\Delta\omega_{FEL}/\omega_{FEL} \approx 1-2\%$ were used for excitation, with their central frequency tunable in the vicinity of $\omega_p$.

Figure 34(a)-(b) shows a comparison between the experimental and numerical perturbed loss functions as a function of the pump-probe time delay $\tau$ for THz pump pulses tuned above the JPR ($\omega_{FEL} > \omega_p$). These consistently show a redshift of $L(\omega,\tau)$, as expected when THz pump pulses generate nonlinear Josephson plasma wave packets of the type shown in Fig. 33(a)-(b). This redshift occurs because, when the Josephson phase is forced to oscillate as $\varphi(t) = \varphi_0\cos(\omega_{FEL}t)$, the time averaged Josephson coupling strenght $\langle\omega_p^2(\tau)\rangle_\tau = \omega_{p,0}^2\langle\cos(\varphi_0\cos(\omega_{FEL}\tau))\rangle_\tau$ decreases compared to equilibrium. One should point out that oscillation of $\omega_p^2(\tau)$ as those reported in Section 6.3 should be present also in this case, but they would be too fast to be observed with the time resolution of the experiment. The phenomenon described here demonstrates, as already predicted theoretically [211,212], the experimental feasibility of using the JPR nonlinearities to open a transparency window for THz plasma waves in an otherwise electromagnetically opaque frequency region.

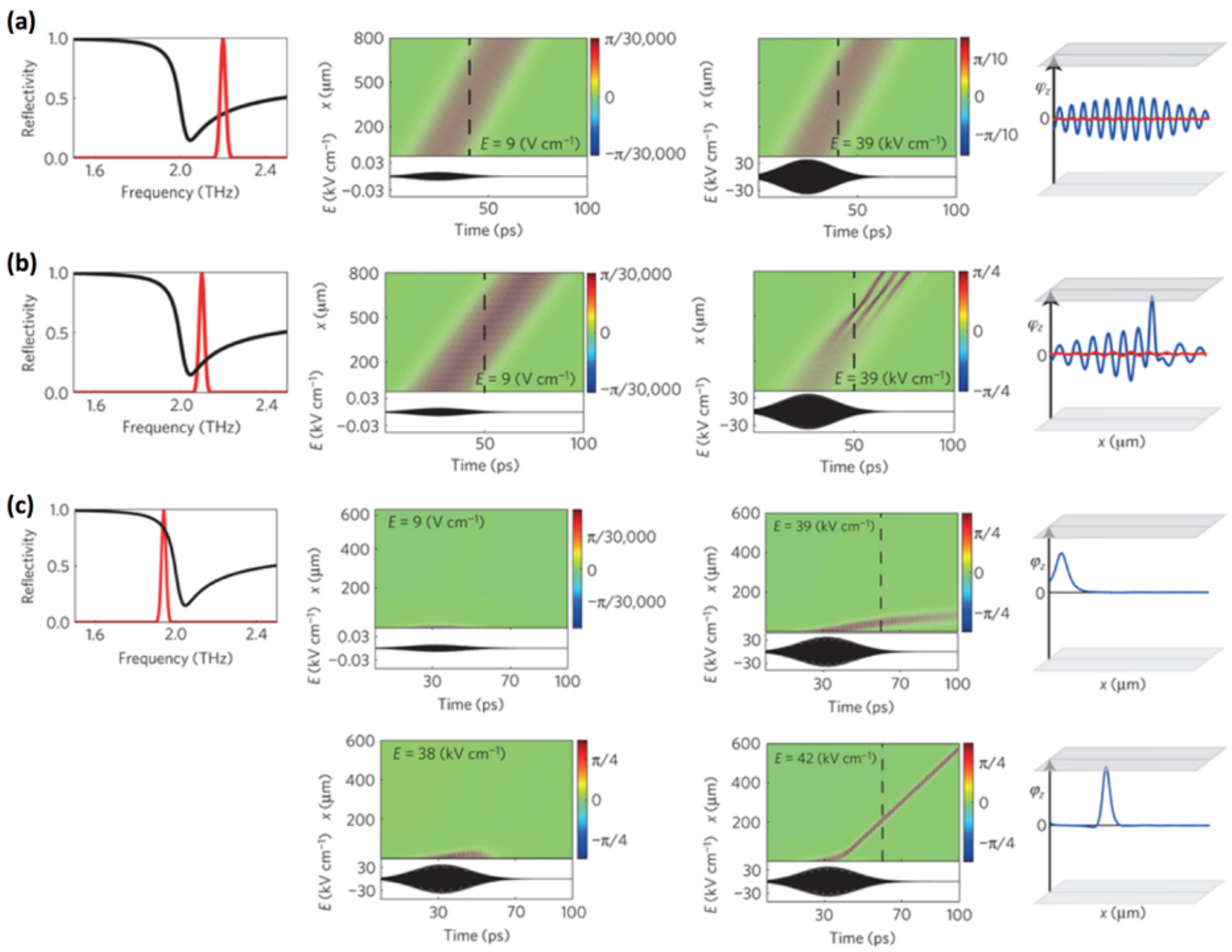

Fig. 33. Simulated space- and time-dependence of the Josephson phase $\varphi(x,\tau)$ for THz pump excitations with spectra centered at (a) $\omega_{FEL} = 1.1\ \omega_p$, (b) $\omega_{FEL} = 1.05\ \omega_p$, and (c) $\omega_{FEL} = 0.97\ \omega_p$. Left panels: THz pump spectrum (red curve) shown along with the JPR in the $La_{1.84}Sr_{0.16}CuO_4$ equilibrium reflectivity (black). Central panels: Two dimensional simulations of the Josephson phase $\varphi(x,\tau)$ as a function of time ($\tau$) and depth ($x$), showing wave propagation inside the superconductor for different THz pump fields. Right panels: Snapshots of the spatial profile of $\varphi(x,\tau)$ in linear (red) and nonlinear (blue) regime. Reprinted from [31].

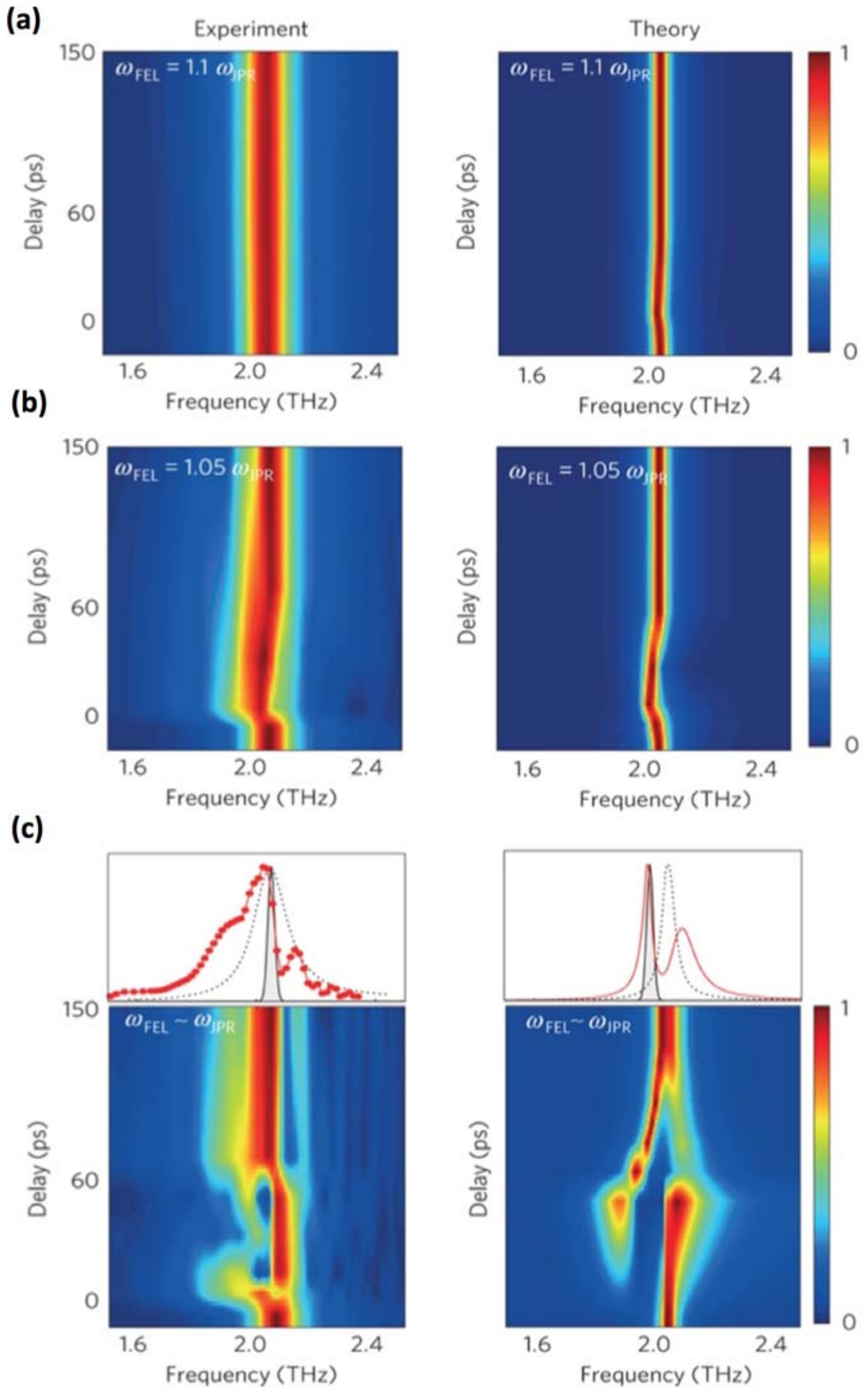

Fig. 34. Color plots representing the time evolution of the frequency resolved loss function of $La_{1.84}Sr_{0.16}CuO_4$, determined for different excitation frequencies: (a) $\omega_{FEL} = 1.1\ \omega_p$, (b) $\omega_{FEL} = 1.05\ \omega_p$, and (c) $\omega_{FEL} \approx \omega_p$. Experimentally measured (left) and simulated (right) data are both displayed. Line cuts shown in red in (c) are taken at $\tau$ = 80 ps and shown along with equilibrium loss functions (dashed) and pump spectra (solid lines). Reprinted from [31].

We now discuss the case of $\omega_{FEL}{\sim}\omega_p$, shown in Fig. 34(c). A long-lived perturbation of the Josephson phase, characterized by a strong reshaping of the loss function is apparent in both experiment and simulation. This peculiar spectral reshaping, exhibiting a splitted lineshape and a dip, can be intuitively understood as originating from the presence of a highly localized Josephson phase deformation within the unperturbed superconductor, *i.e.* a Josephson soliton. This is interacting with the continuum of linear Josephson plasma waves that are excited and probed by the THz probe pulse.

The propagation of these linear plasma waves is strongly perturbed by the presence of the soliton, acting as a dynamical barrier potential for Josephson wave propagation, thus causing multiple reflection and transmission events and leading to interference effects in the probe spectrum. Such interference phenomenon, where continuum excitations are coupled to discrete ones, is reminiscent of the physics at play in the Fano effect [213] or in electromagnetically induced tranparency [214,215].

Various applications to plasmonics [216], as well as new strategies for optical control of superconductivity, can be predicted from the present work. For instance the control of flux-carrying phase kinks, which has been considered in conventional Josephson junctions for information transport and storage, can now be driven and detected by light. Combining these ideas with terahertz coherent control techniques [217] may also open new opportunities, possibly allowing to generate, stop, accelerate, or slow down Josephson plasma solitons.

## 7. Conclusions

In this Review Article we reported on the state of the art of light control experiments on complex materials, with particular focus on high-$T_c$ superconductors, performed by means of intense THz field transients. Among the wide variety of striking dynamical phenomena we focused particularly on highly nonlinear lattice dynamics, modulation of electron correlations, insulator-to-metal transitions, and control and enhancement of superconductivity.

After having introduced the concept of mode-selective control of the crystal lattice, we discussed how, via the so-called *nonlinear phononics* mechanism, a net displacement of the atomic positions could be achieved, driving electronic and magnetic phase transitions in many complex oxides.

The implication that resonant lattice excitation has on the control of superconductivity in high-$T_c$ cuprates was also extensively discussed. In particular, it was shown how stimulation of Cu-O stretching modes in single-layer cuprates can lead to stripe melting and enhancement of superconductivity, or how nonlinear lattice excitation can drive a double-layer cuprate into a transient structure with a superconducting-like response up to room temperature.

In addition, we also reviewed the most recent results achieved by stimulating local vibrational modes in molecular solids, reporting prototypical examples of direct modulation of local electron correlations and control of superconductivity.

Finally, in the last part of this Review Article, we focused on experiments performed at longer excitation wavelengths, in the single- and sub-THz range. We summarized the main achievements in the control of magnetism and superconductivity, before turning our attention to the field of *Josephson plasmonics*, which revealed a completely new class of quantum nonlinear phenomena in complex matter.

All results reported here demonstrate the immense, and not yet fully exploited, potential of THz fields to control and steer matter. In the next years, further technological developments in the fabrication of even more powerful THz sources may allow to observe new classes of exciting and unexpected phenomena and increase exponentially our material control capabilities.

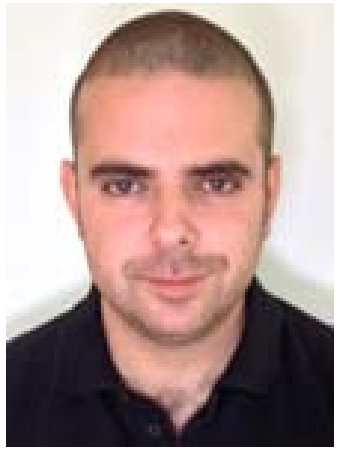

**Daniele Nicoletti** received his Ph.D. for his work on infrared spectroscopy of high-$T_c$ superconductors from "Sapienza" University of Rome, Italy, in 2010. Currently, he is staff scientist at the Max-Planck Institute for the Structure and Dynamics of Matter. His research is mainly focused on the ultrafast control of complex materials using strong terahertz fields.

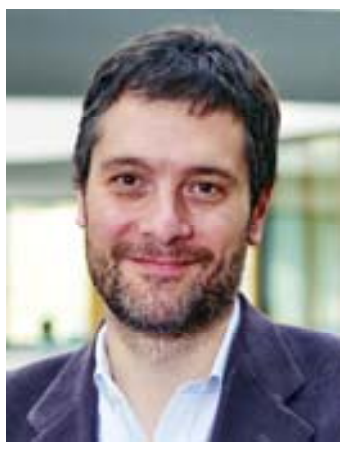

**Andrea Cavalleri** received his Ph.D. from the University of Pavia, Italy, in 1998. He held research positions at the University of Essen (Germany), at the University of California, San Diego, and at the Lawrence Berkeley National Laboratory. In 2005, he joined the faculty of the University of Oxford, where he was promoted to Professor of Physics in 2006. He became a Professor of Physics at the University of Hamburg in 2008 and is also the founding director of the Max Planck Institute for the Structure and Dynamics of Matter in Hamburg. Cavalleri has co-developed the field of ultrafast x-ray science and has been a pioneer in the optical control of emergent phases in quantum condensed matter.